\documentclass{jpsj3}
\usepackage{type1cm}

\title{
Spin-Atomic Vibration Interaction 
and Spin-Flip Hamiltonian 
of a Single Atomic Spin in a Crystal Field
}

\author{Satoshi~Kokado\thanks{E-mail address: tskokad@ipc.shizuoka.ac.jp}, 
Kikuo~Harigaya$^1$, and Akimasa~Sakuma$^2$
}
\inst{
Faculty of Engineering, Shizuoka University, Hamamatsu 432-8561, Japan\\
$^1$Nanosystem Research Institute, AIST, Tsukuba 305-8568, Japan\\
$^2$Graduate School of Engineering, Tohoku University, Sendai 980-8579, Japan
}
\abst{
We derive the spin-atomic vibration interaction $V_{\rm SA}$ 
and the spin-flip Hamiltonian $V_{\rm SF}$ 
of a single atomic spin in a crystal field. 
We here apply the perturbation theory to a model with 
the spin-orbit interaction and the kinetic and potential energies 
of electrons. 
The model also takes into account 
the difference in vibration displacement 
between an effective nucleus and electrons, $\Delta {\mbox{\boldmath $r$}}$. 
Examining 
the coefficients of $V_{\rm SA}$ and $V_{\rm SF}$, 
we first show that 
$V_{\rm SA}$ appears for $\Delta {\mbox{\boldmath $r$}}$$\ne$0, 
while $V_{\rm SF}$ is present 
independently of $\Delta {\mbox{\boldmath $r$}}$. 
As an application, we next obtain $V_{\rm SA}$ and $V_{\rm SF}$ of 
an Fe ion in a crystal field of tetragonal symmetry. 
It is found that 
the magnitudes of the coefficients of $V_{\rm SA}$ 
can be larger than those of 
the conventional spin-phonon interaction 
depending on vibration frequency. 
In addition, transition probabilities per unit time 
due to $V_{\rm SA}$ and $V_{\rm SF}$ are investigated 
for the Fe ion 
with an anisotropy energy of $-|D|S_Z^2$, 
where $D$ is an anisotropy constant 
and $S_Z$ is the $Z$ component of a spin operator. 
}

\kword{spin-atomic vibration interaction, spin-flip Hamiltonian, 
spin-phonon interaction, 
single atomic spin, spin-orbit interaction, 
crystal field, surface adsorbate, perturbation theory}

\begin{document}
\maketitle

\section{Introduction}
\label{intro}

Recently, the magnetic properties of 
molecular spin\cite{Caneschi,Sessoli,Mn12_1,Mn12_2,Politi,Awaga,Barra,Misiorny,Ardavan} 
and atomic spin\cite{Hirjibehedin,Hirjibehedin1,Park,Rugar,Tsukahara,Yayon,Etz,Serrate,Kokado,Kokado1} systems 
have been extensively studied 
to develop ultimate microscopic elements 
for mass-storage devices and quantum information devices. 
In particular, the spin relaxation\cite{Mn12_1,Misiorny,Ardavan,Kokado1} 
has attracted much attention 
from the viewpoints of data writing and storage. 
This spin relaxation\cite{Amasha} means that 
the spin relaxes from the excited state to the ground spin state 
via interactions between the spin and other degrees of freedom. 
The degrees of freedom are considered to be, for example, 
the atomic and lattice vibrations. 
Here, 
the vibrational states reflect 
structures of the systems. 
Typical structures of the above atomic spin systems are 
a magnetic ion adsorbed on a substrate\cite{Hirjibehedin} 
or the magnetic ion trapped between a tip and a substrate.\cite{Zhao,Huang} 
For such atomic spin systems, however, 
theoretical studies of the interactions 
between the spin and vibrations have scarcely been performed so far. 


A relevant interaction is considered to be spin-phonon interactions,
\cite{Van1,Van2,Mattuck,Orbach,Stoneham,Tucker,Stoneham_b,Hartmann} 
in which a localized spin 
interacts with the phonon of a lattice system. 
The expressions of the spin-phonon interaction
\cite{Van1,Van2,Mattuck,Orbach,Stoneham,Tucker,Stoneham_b} 
have been originally derived 
using 
the spin-orbit interaction 
and 
the modulation of crystal field potential energy 
due to the lattice vibration. 
In particular, 
Van Vleck\cite{Van1,Van2} obtained an expression 
applying a perturbation theory to the following model: 
the crystal field potential energy in an equilibrium state 
was included in the unperturbed Hamiltonian, 
while the spin-orbit interaction 
and the modulation of crystal field potential energy 
were the perturbed one.\cite{Stoneham_b} 
Mattuck and Strandberg\cite{Mattuck} also 
derived essentially the same expression as Van Vleck's one\cite{Van1} 
using a slightly different method. 
On the other hand, Hartmann-Boutron {\it et al}.\cite{Hartmann} 
proposed a spin-phonon interaction 
applying the second-order perturbation theory to 
a model with the spin-orbit interaction. 
Here, the coefficient of the spin-phonon interaction contained 
the local strain on the spin site due to the phonon.\cite{nanomagnet} 
In these spin-phonon interactions, 
the structure of the systems 
was assumed to be a periodic structure (e.g., a bulk crystal), 
where 
a unit cell consisted of 
the magnetic ion and the surrounding ions.\cite{ex_negle} 
The phonon then had a continuous energy spectrum 
on the assumption that 
the phonon frequency was proportional to 
the magnitude of the wave vector. 


Such spin-phonon interactions, however, may not be 
suitable for the atomic spin on substrate (or surface), 
because the vibration of the adsorbate is not described 
in their interactions. 
The vibration of the adsorbate on surface is obviously different from 
the above-mentioned phonon. 
The local vibrational density of states at the adsorbate 
shows sharp peaks at certain energies;\cite{Liu} 
that is, 
this adsorbate has discrete vibrational energy levels. 
In contrast, the above phonon 
exhibits the broad vibrational density of states,\cite{Liu} 
i.e., continuous vibrational energy levels. 



\begin{figure}[ht]
\begin{center}
\includegraphics[width=0.5\linewidth]{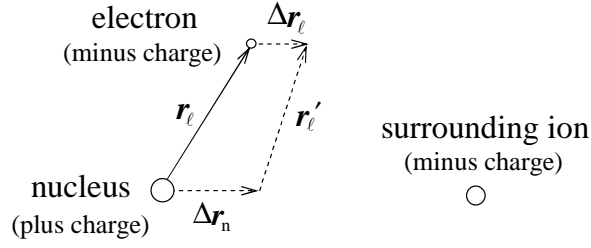}
\caption{
Positions and displacements of the effective nucleus (the large circle) 
and the $\ell$th electron (the small circle) in the single magnetic ion. 
The magnetic ion is surrounded by several ions. 
Here, one of the surrounding ions is shown by the medium circle. 
${\mbox{\boldmath $r$}}_\ell$: 
position vector of the $\ell$th electron 
measured from the nucleus at an equilibrium position. 
$\Delta {\mbox{\boldmath $r$}}_{\rm n}$: displacement of the nucleus. 
$\Delta {\mbox{\boldmath $r$}}_\ell$: displacement of the $\ell$th electron. 
${\mbox{\boldmath $r$}}_\ell'$: 
position vector of the displaced $\ell$th electron 
measured from the displaced nucleus. 
}
\label{dis1}
\end{center}
\end{figure}

Furthermore, in the spin-phonon interactions, 
the difference in vibration displacement 
between an effective nucleus and electrons 
has not been taken into account. 
Here, this nucleus consists of a nucleus and core electrons. 
As shown in Appendix \ref{ele_pol}, 
the difference in displacement can be actually obtained, 
when the nucleus (plus charge) and electrons (minus charge) 
have the displacement 
in the presence of a crystalline electric field 
due to the surrounding ions (minus charge) 
(see Fig. \ref{dis1}). 
The displacements of the nucleus and the electrons 
are written by 
$\Delta r_{\rm n}$ and $( 1 - \eta) \Delta r_{\rm n}$, respectively, 
where $\eta$ ($0 \le \eta < 1$) 
is a dimensionless quantity characterizing the difference. 
For instance, 
this $\eta$ is evaluated to be about 0.05 
for a single Fe ion on the CuN surface (see Appendix \ref{ele_pol}).

On the basis of 
the above-mentioned facts, 
we aim to derive 
a spin-atomic vibration interaction of 
a single atomic spin surrounded by several ions, 
where the ions are present in the substrate. 
Here, the spin-atomic vibration interaction represents 
an interaction between the spin of the single magnetic ion 
and the vibration of this ion. 
This interaction also takes into account the above $\eta$. 

In this paper, we derived the spin-atomic vibration interaction 
and a spin-flip Hamiltonian of a single atomic spin 
in a crystal field, 
applying a perturbation theory to a model with 
the spin-orbit interaction and 
the kinetic and potential energies of electrons. 
The spin-flip Hamiltonian 
contains spin-flip operators such as $S_XS_Z$ and $S_YS_Z$, 
but no atomic vibration operators, 
where $S_I$ is the $I$ component of the spin 
${\mbox{\boldmath $S$}}$=$(S_X,S_Y,S_Z)$. 
From the coefficients of the spin-atomic vibration interaction 
and the spin-flip Hamiltonian, 
we found that 
the former appears for $\eta$$\ne$0, 
while the latter is present independently of $\eta$. 
We also obtained their coefficients 
in the case of the Fe ion in a crystal field of tetragonal symmetry. 
The magnitude of the coefficient of the spin-atomic vibration interaction 
was estimated to be larger than that of 
the conventional spin-phonon interaction 
in a specific region of the vibration frequency. 
In addition, 
we discussed 
transition probabilities 
per unit time due to the spin-atomic vibration interaction 
and spin-flip Hamiltonian of 
the Fe ion.

The present paper is organized as follows: 
In \S \ref{Model}, 
we propose a model with 
the spin-orbit interaction and the kinetic and potential energies 
of electrons. 
This model also takes into account the vibration displacement. 
In \S \ref{SA_SF}, 
we derive the spin-atomic vibration interaction and spin-flip Hamiltonian. 
In \S \ref{Application}, 
we investigate the vibration frequency dependences of their coefficients 
of the Fe ion in the crystal field of tetragonal symmetry. 
In addition, 
their coefficients 
are evaluated 
by assigning appropriate values to the spin-orbit coupling constant. 
In \S \ref{Dis_comp}, 
we compare the spin-atomic vibration interaction 
with the conventional spin-phonon interactions. 
In \S \ref{Dis_tran}, 
we discuss 
the transition probabilities per unit time 
due to the spin-atomic vibration interaction and the spin-flip Hamiltonian. 
The conclusion is presented in \S \ref{Conc}. 
In Appendix \ref{ele_pol}, 
we evaluate 
the difference in vibration displacement 
between the effective nucleus and the electron. 
The vibration displacement of 
the nucleus is evaluated in Appendix \ref{D_rn_ap}. 
In Appendix \ref{Appendix}, 
we describe 
the expressions of the coefficients of the spin-atomic vibration interaction 
and spin-flip Hamiltonian of the Fe ion. 

\section{Model}
\label{Model}

Toward the calculation of perturbation energies in \S \ref{SA_SF} 
(i.e., the spin-atomic vibration interaction and 
the spin-flip Hamiltonian), 
we here derive the unperturbed and perturbed Hamiltonians 
for a model with 
the spin-orbit interaction and kinetic and potential energies 
of electrons. 
This model also 
takes into account the difference in vibration displacement 
between the effective nucleus and electrons. 
In addition, we introduce orbital and spin states 
for the unperturbed Hamiltonian. 

\subsection{
Hamiltonian in an equilibrium state 
}
\label{H_equili}

We first propose a Hamiltonian in an equilibrium state 
(with no vibration displacement) 
for 3d electrons 
of an iron-group ion (i.e., a magnetic ion) in a crystal field 
due to the surrounding ions. 
Here, the 3d shell of the magnetic ion is incompletely filled, 
and the surrounding ions have a minus charge. 
In particular, we treat the case of a weak crystal field. 
Namely, the electrons occupy energy levels from the bottom 
keeping their spins parallel, according to Hund's first rule.\cite{UJ} 
The system is here assumed 
to have the total spin ${\mbox{\boldmath $S$}}$.

We now focus on electrons existing 
in the partially filled 
$\sigma$-spin shell of the magnetic ion, with $\sigma$ being up or down. 
When 
the number of electrons of 
the ground state 
is less than half (more than half), 
the $\sigma$-spin shell 
corresponds to the up-spin shell (down-spin shell). 
The excited states are then assumed to consist of orbitals 
in the $\sigma$-spin shell. 
In addition, 
the electrons in the $\sigma$-spin shell have the spin-orbit interaction 
$V_{\mbox{\footnotesize so}}$.\cite{Yosida} 
Here, $|V_{\mbox{\footnotesize so}}|$ 
is considered to be 
smaller than the magnitude of 
the crystal field potential energy.\cite{Mattuck} 
When the number of electrons 
in the $\sigma$-spin shell is $N$, 
we have the following Hamiltonian for the electrons:
\begin{eqnarray}
\label{Hamil_start}
&&\hspace*{-0.5cm}{\cal H} = 
{\cal H}_{\mbox{\footnotesize e}} 
+ {V}_{\mbox{\footnotesize so}}, \\
\label{H_e}
&&\hspace*{-0.5cm}{\cal H}_{\mbox{\footnotesize e}} = 
\sum_{\ell=1}^{N} \left[ \frac{{\mbox{\boldmath $p$}}_\ell^2}{2m} 
+ {V}_{\mbox{\footnotesize en}} ({\mbox{\boldmath $r$}}_\ell) 
+ {V}_{\mbox{\footnotesize c}} ({\mbox{\boldmath $r$}}_\ell)
\right], \\
\label{so}
&&\hspace*{-0.5cm}{V}_{\mbox{\footnotesize so}}
= \lambda {\mbox{\boldmath $L$}} \cdot {\mbox{\boldmath $S$}}, 
\end{eqnarray}
with
\begin{eqnarray}
&&{\mbox{\boldmath $p$}}_\ell=(p_{\ell,x},~p_{\ell,y}, p_{\ell,z}) =
m \dot{\mbox{\boldmath $r$}}_\ell, \\
&&
\label{V_en}
{V}_{\mbox{\footnotesize en}} ({\mbox{\boldmath $r$}}_\ell)
= - \frac{Z_{\rm eff} e^2}{4 \pi \epsilon_0 r_\ell},\\
&&
\label{V_c}
{V}_{\mbox{\footnotesize c}} ({\mbox{\boldmath $r$}}_\ell)
= \sum_{\ell'} \frac{Z_{\ell'} e^2}
{4 \pi \epsilon_0 | 
{\mbox{\boldmath $r$}}_\ell -{\mbox{\boldmath $R$}}_{\ell'}|}, \\
\label{so_lam}
&&\lambda = C_\pm \left\langle \frac{1}{r^3} \right\rangle, \\ 
\label{distance}
&&\left\langle \frac{1}{r^3} \right\rangle
=
\displaystyle{\sum_{\ell=1}^N} 
\left( \Psi_1 \left| r_\ell^{-3} \right| \Psi_1 \right)
\left/ \displaystyle{\sum_{\ell=1}^N} 1 \right., \\
&&{\mbox{\boldmath $L$}}= (L_x, L_y, L_z)
=\frac{1}{\hbar}\sum_{\ell=1}^N \left( {\mbox{\boldmath $r$}}_\ell 
\times {\mbox{\boldmath $p$}}_\ell \right), 
\end{eqnarray}
with 
${\mbox{\boldmath $r$}}_\ell$=$(x_\ell,~y_\ell,~z_\ell)$, 
$r_\ell$=$|{\mbox{\boldmath $r$}}_\ell|$
=$\sqrt{ x_\ell^2 + y_\ell^2 + z_\ell^2}$, 
${\mbox{\boldmath $S$}}$=$(S_x, S_y, S_z)$, 
$C_\pm$=
$\pm (\mu_0/4 \pi) g \mu_{\mbox{\tiny B}}^2 Z_{\rm eff}/2S$, 
and 
$\mu_{\mbox{\tiny B}}$=$e\hbar/2m$. 
Here, 
${\cal H}_{\mbox{\footnotesize e}}$ denotes the Hamiltonian 
consisting of 
the kinetic energy $\sum_{\ell=1}^N {\mbox{\boldmath $p$}}_\ell^2/(2m)$ 
and the potential energy 
$\sum_{\ell=1}^N 
[ {V}_{\mbox{\footnotesize en}} ({\mbox{\boldmath $r$}}_\ell) 
+ {V}_{\mbox{\footnotesize c}} ({\mbox{\boldmath $r$}}_\ell)]$. 
The quantity $m$ is the electron mass, 
${\mbox{\boldmath $r$}}_\ell$ is 
the position of the $\ell$th electron 
measured from the nucleus (see Fig. \ref{dis1}), 
and ${\mbox{\boldmath $p$}}_\ell$ 
is the momentum of the $\ell$th electron. 
The operator 
${V}_{\mbox{\footnotesize en}} ({\mbox{\boldmath $r$}}_\ell)$ is 
a spherically symmetric potential energy of the $\ell$th electron 
created by a nucleus and the other electrons in the magnetic ion. 
The quantity 
$-e$ ($<0$) is the electronic charge, and 
$\epsilon_0$ is the electric constant. 
In addition, $Z_{\rm eff}$ represents the effective nuclear charge 
for the 3d electron,\cite{Z_eff} 
which can be evaluated from Slater's rule.\cite{Slater} 
The operator 
${V}_{\mbox{\footnotesize c}} ({\mbox{\boldmath $r$}}_\ell)$ 
is the crystal field potential energy, 
where 
${\mbox{\boldmath $R$}}_{\ell'}$ and $-Z_{\ell'} e$ ($<$0) are 
the position vector and the electric charge 
of the $\ell'$th surrounding ion, respectively. 
As to $\lambda$ of eq. (\ref{so_lam}), 
$C_+$ ($C_-$) is the coefficient in the case of 
an electron number of less than half (more than half) 
in an incompletely filled 3d shell.\cite{Yosida} 
The mean $\langle r^{-3} \rangle$ 
is defined by eq. (\ref{distance}), 
where $|\Psi_1 )$ is the orbital state 
of the electrons in the $\sigma$-spin shell 
of the ground state (see eq. (\ref{Psi_j})). 
The quantity 
$\mu_{\mbox{\tiny B}}$ is the Bohr magneton, 
$g$ is the g value, 
$\mu_0$ is the permeability, 
and $\hbar$ is the Planck constant $h$ divided by 2$\pi$. 
Also, 
$S$ is the magnitude of ${\mbox{\boldmath $S$}}$, 
and $\hbar {\mbox{\boldmath $L$}}$ is 
the total orbital angular momentum.

\subsection{Vibration displacement} 
\label{Hamil}

For the Hamiltonian in the equilibrium state of \S \ref{H_equili}, 
we take the effect of the vibration displacement into consideration. 
Regarding the vibration, 
we consider a simple model, in which 
only a single magnetic ion vibrates, 
while the surrounding ions are rigid. 
This model may be relevant to, for example, 
a single magnetic ion surrounded by several ions 
in the substrate. 
Here, the surrounding ions are assumed to behave as an oscillator with 
the magnitude of vibration displacement, 
which is much smaller than that of the magnetic ion. 
This assumption is justified 
under the following conditions: 
(i) The coupling within the surrounding ions is stronger than 
that between the magnetic ion and the surrounding ions. 
(ii) 
The total mass of the surrounding ions is sufficiently larger than 
the mass of the magnetic ion.\cite{disp_ref,Desjonqueres} 
For example, in the case of the Fe ion on the CuN surface,\cite{Hirjibehedin} 
the Fe ion is surrounded by two N ions and five Cu ions. 
The total mass of the surrounding ions 
is more than six times as large as the mass of the Fe ion. 
(iii) 
A force constant 
between the surrounding ions 
and the other ions adjacent to them in the substrate 
is much larger than that 
between the surrounding ions and the magnetic ion.\cite{disp_ref}


We now focus on the effective nucleus of the magnetic ion. 
Here, 
the nucleus consists of a nucleus and core electrons, 
where the core electrons correspond to 
the electrons other than those in the $\sigma$-spin shell. 
The effective nucleus is first assumed to have 
a slight vibration displacement, $\Delta {\mbox{\boldmath $r$}}_{\rm n}$ 
(see Fig. \ref{dis1} and Appendix \ref{D_rn_ap}), i.e.,  
\begin{eqnarray}
\label{Delta_rn_xyz}
\Delta {\mbox{\boldmath $r$}}_{\rm n}=
(\Delta x_{\rm n}, \Delta y_{\rm n}, \Delta z_{\rm n}).
\end{eqnarray}
The displacement of the $\ell$th electron in the $\sigma$-spin shell, 
$\Delta {\mbox{\boldmath $r$}}_\ell$, 
is then given by
\begin{eqnarray}
\label{del_r_ell}
\Delta {\mbox{\boldmath $r$}}_\ell=
(1-\eta) \Delta {\mbox{\boldmath $r$}}_{\rm n}, 
\end{eqnarray}
where $-\eta \Delta {\mbox{\boldmath $r$}}_{\rm n}$ represents 
the difference in displacement 
between the nucleus and the electron. 
This $-\eta \Delta {\mbox{\boldmath $r$}}_{\rm n}$ appears 
in the presence of the crystalline electric field 
due to the surrounding ions (see Appendix \ref{ele_pol}). 
Here, the dimensionless quantity characterizing the difference 
$\eta$ ($0 \le \eta < 1$) has been assumed 
to be independent of $\ell$ and $\Delta {\mbox{\boldmath $r$}}_{\rm n}$ 
for simplicity. 
For example, this $\eta$ may be written by eq. (\ref{eta_ex}). 
Regarding eq. (\ref{del_r_ell}), 
we have considered the following situation: 
First, the effective nucleus of the magnetic ion 
has the displacement of 
$\Delta {\mbox{\boldmath $r$}}_{\rm n}$. 
Second, each electron moves in a circular orbit 
around ``the nucleus fixed at $\Delta {\mbox{\boldmath $r$}}_{\rm n}$'' 
at a high speed, and simultaneously 
experiences the crystalline electric field due to the surrounding ions. 
This electric field shifts 
the center position of the circular orbit 
from $\Delta {\mbox{\boldmath $r$}}_{\rm n}$ 
to $(1-\eta) \Delta {\mbox{\boldmath $r$}}_{\rm n}$ 
(see Appendix \ref{ele_pol}). 

Thus, when 
the nucleus has the displacement of $\Delta {\mbox{\boldmath $r$}}_{\rm n}$, 
the position of the $\ell$th electron 
measured from the nucleus, ${\mbox{\boldmath $r$}}_\ell'$, 
is represented by 
\begin{eqnarray}
\label{r_ell'}
{\mbox{\boldmath $r$}}_\ell'= 
{\mbox{\boldmath $r$}}_\ell 
+ \Delta {\mbox{\boldmath $r$}}_\ell
- \Delta {\mbox{\boldmath $r$}}_{\rm n} 
= {\mbox{\boldmath $r$}}_\ell - \eta \Delta {\mbox{\boldmath $r$}}_{\rm n}. 
\end{eqnarray}
From the time differential of eq. (\ref{r_ell'}), 
we obtain the following momentum: 
\begin{eqnarray}
\label{p_ell'}
{\mbox{\boldmath $p$}}_\ell'= 
{\mbox{\boldmath $p$}}_\ell - \eta \Delta {\mbox{\boldmath $p$}}_{\rm e}, 
\end{eqnarray}
with 
\begin{eqnarray}
&&{\mbox{\boldmath $p$}}_\ell'=m \dot{\mbox{\boldmath $r$}}_\ell', \\
\label{delta_p}
&&\Delta {\mbox{\boldmath $p$}}_{\rm e}
=(\Delta p_{{\rm e},x},\Delta p_{{\rm e},y},\Delta p_{{\rm e},z})
=m \Delta \dot{\mbox{\boldmath $r$}}_{\rm n}. 
\end{eqnarray}

\subsection{Unperturbed and perturbed Hamiltonians}

We next derive the unperturbed and perturbed Hamiltonians 
from a Hamiltonian, ${\cal H}^{\rm disp}$, 
which contains 
the effect of the above-mentioned vibration displacement. 
This ${\cal H}^{\rm disp}$ can be obtained 
on the basis of ${\cal H}$ of eq. (\ref{Hamil_start}). 
Namely, 
when ${\mbox{\boldmath $r$}}_\ell'$ of eq. (\ref{r_ell'}) 
and ${\mbox{\boldmath $p$}}_\ell'$ of eq. (\ref{p_ell'}) 
are taken into account, ${\cal H}$ is rewritten as
\begin{eqnarray}
\label{H^d}
&&\hspace*{-0.5cm}{\cal H}^{\rm disp} = 
{\cal H}_{\mbox{\footnotesize e}}^{\rm disp} 
+ {V}_{\mbox{\footnotesize so}}^{\rm disp}, \\
\label{H_e^d}
&&\hspace*{-0.5cm}{\cal H}_{\mbox{\footnotesize e}}^{\rm disp} = 
\sum_{\ell=1}^N \left[ \frac{{{\mbox{\boldmath $p$}}_\ell'}^2}{2m} 
+{V}_{\mbox{\footnotesize en}} ({\mbox{\boldmath $r$}}_\ell') 
+ {V}_{\mbox{\footnotesize c}} ({\mbox{\boldmath $r$}}_\ell' 
+\Delta {\mbox{\boldmath $r$}}_{\rm n}) \right], \\
\label{V_so^d}
&&\hspace*{-0.5cm}{V}_{\mbox{\footnotesize so}}^{\rm disp}=
C_\pm 
\left\langle \frac{1}{{r'}^3} \right\rangle 
\sum_{\ell=1}^N 
\left( \frac{{\mbox{\boldmath $r$}}_\ell' \times 
{\mbox{\boldmath $p$}}_\ell'}{\hbar} 
\right)
\cdot 
{\mbox{\boldmath $S$}}, 
\end{eqnarray}
where
\begin{eqnarray}
&&\left\langle \frac{1}{{r'}^3} \right\rangle
=
\displaystyle{\sum_{\ell=1}^N} 
\left( \Psi_1 \left| {r'_\ell}^{-3} \right| \Psi_1 \right)
\left/ \displaystyle{\sum_{\ell=1}^N} 1 \right., \\
&&r'_\ell
=\sqrt{( x_\ell - \eta \Delta x_{\rm n} )^2 
+ ( y_\ell - \eta \Delta y_{\rm n} )^2 
+ ( z_\ell - \eta \Delta z_{\rm n} )^2 }, \\
\label{r_l'+Drn}
&&\mbox{\boldmath $r$}'_\ell 
+\Delta {\mbox{\boldmath $r$}}_{\rm n} = 
{\mbox{\boldmath $r$}}_\ell +(1-\eta) \Delta {\mbox{\boldmath $r$}}_{\rm n}. 
\end{eqnarray}

For eq. (\ref{H^d}), we expand 
${V}_{\mbox{\footnotesize en}} ({\mbox{\boldmath $r$}}_\ell')$
=${V}_{\mbox{\footnotesize en}} 
({\mbox{\boldmath $r$}}_\ell - \eta \Delta {\mbox{\boldmath $r$}}_{\rm n})$ 
and 
${V}_{\mbox{\footnotesize c}}({\mbox{\boldmath $r$}}_\ell'
+\Delta {\mbox{\boldmath $r$}}_{\rm n})$
=${V}_{\mbox{\footnotesize c}} 
({\mbox{\boldmath $r$}}_\ell +(1-\eta) \Delta {\mbox{\boldmath $r$}}_{\rm n})$ 
in power series in 
$\eta \Delta \xi_{\rm n}$ 
and $(1-\eta)\Delta \xi_{\rm n}$, respectively, 
with $\xi$=$x$, $y$, $z$. 
In addition, 
$\langle {r'}^{-3} \rangle$ of ${V}_{\mbox{\footnotesize so}}^{\rm disp}$ 
is expanded in power series in $\eta \Delta \xi_{\rm n}$. 
Since $|\eta \Delta \xi_{\rm n}|$, 
$|(1-\eta)\Delta \xi_{\rm n}|$, and 
$|\eta \Delta p_{{\rm e},\xi}|$ 
are considered to be sufficiently small, 
we neglect 
terms higher than the first order of 
$\eta \Delta \xi_{\rm n}$, 
$(1-\eta)\Delta \xi_{\rm n}$, and 
$\eta \Delta p_{{\rm e},\xi}$. 

For such ${\cal H}^{\rm disp}$, 
we adopt 
Van Vleck's approach;\cite{Van1} 
that is, 
the crystal field potential energy in the equilibrium state 
is contained in the unperturbed Hamiltonian, 
while the spin-orbit interaction and 
the modulation of crystal field potential energy 
due to the vibration displacement are treated as perturbations. 
We therefore present the following effective Hamiltonian consisting of 
the unperturbed term ${\cal H}_0$ 
and the perturbed term ${V}_{\mbox{\footnotesize p}}$: 
\begin{eqnarray}
&&{\cal H}_{\rm eff} = 
{\cal H}_0 + {V}_{\mbox{\footnotesize p}}, \\
\label{Vp}
&&{V}_{\mbox{\footnotesize p}}= {V}_{\mbox{\footnotesize k}}'
+ {V}_{\mbox{\footnotesize en}}'
+ {V}_{\mbox{\footnotesize c}}'
+ {V}_{\mbox{\footnotesize so}}, \\
\label{Vp_so}
&&
{V}_{\mbox{\footnotesize so}}=
{V}_{\mbox{\footnotesize so0}}+
{V}_{\mbox{\footnotesize so1}}+
{V}_{\mbox{\footnotesize so2}}, 
\end{eqnarray}
with
\begin{eqnarray}
\label{H_0}
&&\hspace*{-0.3cm}
{\cal H}_0 = 
\sum_{\ell=1}^N \left[ \frac{{\mbox{\boldmath $p$}}_\ell^2}{2m} 
+ {V}_{\mbox{\footnotesize en}}({\mbox{\boldmath $r$}}_\ell)
+ {V}_{\mbox{\footnotesize c}}({\mbox{\boldmath $r$}}_\ell)
\right], \\
\label{Vp_k}
&&\hspace*{-0.3cm}{V}_{\mbox{\footnotesize k}}'=
- \frac{\eta}{m} {\mbox{\boldmath $p$}}_t \cdot 
\Delta {\mbox{\boldmath $p$}}_{\rm e}, \\
\label{Vp_eff}
&&\hspace*{-0.3cm}{V}_{\mbox{\footnotesize en}}'= 
 -\eta \sum_{\ell=1}^N 
\left( \Delta x_{\rm n} \frac{\partial}{\partial x_\ell} 
+ \Delta y_{\rm n} \frac{\partial}{\partial y_\ell} 
+ \Delta z_{\rm n} \frac{\partial}{\partial z_\ell} \right) 
V_{\mbox{\footnotesize en}} ({\mbox{\boldmath $r$}}_\ell), \\
\label{Vp_c}
&&\hspace*{-0.3cm}{V}_{\mbox{\footnotesize c}}'= 
(1-\eta) \sum_{\ell=1}^N 
\left( \Delta x_{\rm n} \frac{\partial}{\partial x_\ell} 
+  \Delta y_{\rm n} \frac{\partial}{\partial y_\ell} 
+  \Delta z_{\rm n} \frac{\partial}{\partial z_\ell} \right) 
V_{\mbox{\footnotesize c}} ({\mbox{\boldmath $r$}}_\ell), \\
&&
\label{V_so0}
\hspace*{-0.3cm}
{V}_{\mbox{\footnotesize so0}}=
\lambda 
{\mbox{\boldmath $L$}} \cdot {\mbox{\boldmath $S$}}, \\
\label{V_so1}
&&\hspace*{-0.3cm}
{V}_{\mbox{\footnotesize so1}}=
\eta C_\pm 
\left\langle \frac{3x}{r^5} \Delta x_{\rm n} 
+ \frac{3y}{r^5} \Delta y_{\rm n} 
+ \frac{3z}{r^5} \Delta z_{\rm n} \right\rangle 
{\mbox{\boldmath $L$}} \cdot {\mbox{\boldmath $S$}},
\\
&&
\label{V_so2}
\hspace*{-0.3cm}
{V}_{\mbox{\footnotesize so2}}=- \eta 
\lambda 
\frac{1}{\hbar}
\left(
{\mbox{\boldmath $r$}}_t \times \Delta {\mbox{\boldmath $p$}}_{\rm e}
+ \Delta {\mbox{\boldmath $r$}}_{\rm n} \times {\mbox{\boldmath $p$}}_t
\right) \cdot 
{\mbox{\boldmath $S$}},
\end{eqnarray}
where 
\begin{eqnarray}
\label{r_t}
&&\hspace*{-0.5cm}
{\mbox{\boldmath $r$}}_t=(x_t,y_t,z_t)=
\sum_{\ell=1}^N {\mbox{\boldmath $r$}}_\ell, \\
\label{p_t}
&&\hspace*{-0.5cm}
{\mbox{\boldmath $p$}}_t=(p_{t,x}, p_{t,y}, p_{t,z})
=\sum_{\ell=1}^N {\mbox{\boldmath $p$}}_\ell, \\
&&\hspace*{-0.5cm}\left\langle 
\frac{3\xi}{ r^5} \Delta \xi_{\rm n} 
\right\rangle = \displaystyle{\sum_{\ell=1}^N} 
\left( \Psi_1 \left| 
3 \xi_\ell r_\ell^{-5} \Delta \xi_{\rm n} \right| \Psi_1 \right)
\left/ \displaystyle{\sum_{\ell=1}^N} 1 \right., 
\end{eqnarray}
for $\xi$=$x$, $y$, $z$. 
Here, eq. (\ref{H_0}) is 
the zeroth-order term of 
$\eta \Delta p_{{\rm e},\xi}$, 
$\eta \Delta \xi_{\rm n}$, 
and $(1-\eta) \Delta \xi_{\rm n}$ in eq. (\ref{H_e^d}). 
Equation (\ref{Vp_k}) is the first-order term of 
$\eta \Delta {\mbox{\boldmath $p$}}_{\rm e}$ in 
$\sum_{\ell=1}^N {{\mbox{\boldmath $p$}}_\ell'}^2 /2m$, 
while 
eq. (\ref{Vp_eff}) is that 
of $\eta \Delta \xi_{\rm n}$ 
in $\sum_{\ell=1}^N {V}_{\mbox{\footnotesize en}}
({\mbox{\boldmath $r$}}_\ell')$. 
Equation (\ref{Vp_c}) is the first-order term of 
$(1-\eta) \Delta \xi_{\rm n}$ in 
$\sum_{\ell=1}^N {V}_{\mbox{\footnotesize c}} 
({\mbox{\boldmath $r$}}_\ell 
+(1-\eta) \Delta {\mbox{\boldmath $r$}}_{\rm n})$. 
In addition, 
eq. (\ref{V_so0}) is the zeroth-order term of 
$\eta \Delta \xi_{\rm n}$ 
and $\eta \Delta p_{{\rm e},\xi}$ in eq. (\ref{V_so^d}), 
while 
eqs. (\ref{V_so1}) and (\ref{V_so2}) are 
the first-order terms of them in eq. (\ref{V_so^d}). 
Here, eq. (\ref{V_so1}) (eq. (\ref{V_so2})) 
corresponds to a term with a change of $\lambda$ due to the vibration 
(a term with a change of ${\mbox{\boldmath $L$}}$ due to the vibration). 
Note that 
$\Delta x_{\rm n}$, $\Delta y_{\rm n}$, and $\Delta z_{\rm n}$ 
in $\langle \hspace{0.2cm} \rangle$ 
of eq. (\ref{V_so1}) can be put outside of $\langle \hspace{0.2cm} \rangle$, 
which is related to electron systems. 
Also, the suffix $t$ in eqs. (\ref{r_t}) and (\ref{p_t}) 
represents the total sum.

\subsection{Orbital and spin states}

For the calculation of the perturbation energies, 
we here propose a specific orbital and spin state 
for ${\cal H}_0$ of eq. (\ref{H_0}). 
Simultaneously, we introduce a 3d orbital state 
that can be utilized in the principal axis transformation 
of \S \ref{transform}.

On the basis of eq. (\ref{H_0}), 
we first have the 3d orbital and spin state with the largest $S$, i.e.,  
\begin{eqnarray}
\label{PhijSz}
|\Phi_j \rangle |S_z \rangle, 
\end{eqnarray}
where 
$|S_z \rangle$ is a spin state with a $(2S+1)$-fold degeneracy 
of $S_z$=$-S$ - $S$, 
and 
$|\Phi_j \rangle$ is 
an orbital state of the electrons 
in the partially filled $\sigma$-spin shell. 
When the number of the electrons is $N$, 
the wave function for $|\Phi_j \rangle$, $\Phi_j$, 
is given by the following Slater determinant:
\begin{eqnarray}
\label{Phi_j}
\Phi_j= 
\frac{1}{\sqrt{N!}}
\left|
  \begin{array}{cccc}
   d_\alpha(1) & d_\alpha(2) & \cdots & d_\alpha(N) \\
   d_\beta(1) & d_\beta(2) & \cdots & d_\beta(N)  \\
   \vdots &  & \ddots & \vdots  \\
   d_\gamma(1) & d_\gamma(2) & \cdots & d_\gamma(N) 
  \end{array}
\right|. 
\end{eqnarray}
The suffix $j$ of $\Phi_j$ specifies 
the electron configuration, 
in which the occupied orbitals are represented by 
$\alpha$, $\beta$, $\cdots$, and $\gamma$. 
The function $d_\zeta (\ell)$ represents the 3d orbital 
of the $\sigma$-spin, 
where $\ell$=1, 2, $\cdots$, $N$ 
and $\zeta$=$\alpha$, $\beta$, $\cdots$, and $\gamma$. 
We also have the following eigenvalue equation: 
\begin{eqnarray}
\label{H0_Phi}
&&\hspace{-1cm}{\cal H}_{0}|\Phi_j \rangle = e_j|\Phi_j \rangle, 
\end{eqnarray}
with
\begin{eqnarray}
\label{e_d_j}
&&\hspace{-1cm}e_j = 
\varepsilon_{d_\alpha} + \varepsilon_{d_\beta} 
+ \cdots + \varepsilon_{d_\gamma} 
= \sum_{\zeta \in j} \varepsilon_{d_\zeta}, \\
\label{Sch}
&&\hspace{-1cm}\left[ - \frac{\hbar^2}{2m} \nabla_\ell^2 
+ {V}_{\mbox{\footnotesize en}} ({\mbox{\boldmath $r$}}_\ell) 
+ {V}_{\mbox{\footnotesize c}} ({\mbox{\boldmath $r$}}_\ell)
\right] d_\zeta (\ell) = \varepsilon_{d_\zeta} d_\zeta (\ell), 
\end{eqnarray}
where $e_j$ is the eigenenergy for $|\Phi_j \rangle$, 
and $\varepsilon_{d_\zeta}$ is that for $d_\zeta (\ell)$. 
In addition, $|\Phi_1 \rangle$ is assumed to be 
a nondegenerate ground state with $e_1$. 
Since ${\cal H}_{0}$ 
is a real function, 
$\Phi_1$ can be a real one.\cite{Yosida1} 
This $|\Phi_j \rangle$ 
will be used to perform the principal axis transformation 
as described in \S \ref{transform}.

Second, we propose the orbital and spin state 
to obtain the perturbation energy 
(i.e., the spin-atomic vibration interaction 
and the spin-flip Hamiltonian); that is,
\begin{eqnarray}
\label{Psi_j_Sz}
|\Psi_j ) |S_z \rangle, 
\end{eqnarray}
where $|S_z \rangle$ is the above-mentioned spin state, 
and 
$|\Psi_j )$ is an orbital state of the electrons in the $\sigma$-spin shell. 
The wave function for $|\Psi_j )$, $\Psi_j$, is assumed to be
\begin{eqnarray}
\label{Psi_j}
\Psi_j= 
\frac{1}{\sqrt{N!}}
\left|
  \begin{array}{cccc}
   \psi_\alpha (1) & \psi_\alpha (2) & \cdots & \psi_\alpha (N) \\
   \psi_\beta (1) & \psi_\beta (2) & \cdots & \psi_\beta (N)  \\
   \vdots &  & \ddots & \vdots  \\
   \psi_\gamma (1) & \psi_\gamma (2) & \cdots & \psi_\gamma (N) 
  \end{array}
\right|, 
\end{eqnarray}
with
\begin{eqnarray}
\label{phi_a(l)}
&&\hspace*{-0.4cm} \psi_\zeta (\ell) = 
C_\zeta \left( d_\zeta (\ell) 
+ \sum_{m=1 (\ne \zeta)}^5 c_{d_m}^{(\zeta)} \overline{d_m}(\ell) 
+ \sum_{n=1}^3 c_{p_n}^{(\zeta)} \overline{p_n}(\ell) \right), \\
&&\hspace*{-0.4cm} C_\zeta=
\left( 1 + \sum_{m=1 (\ne \zeta)}^5 |c_{d_m}^{(\zeta)}|^2 
+ \sum_{n=1}^3 |c_{p_n}^{(\zeta)}|^2 \right)^{-1/2}, 
\end{eqnarray}
where $N$ is the number of the electrons. 
The suffix $j$ of $\Psi_j$ specifies 
the electron configuration, 
in which the occupied orbitals are represented by 
$\alpha$, $\beta$, $\cdots$, and $\gamma$. 
The function $d_\zeta (\ell)$ is the dominant 3d orbital of the $\sigma$-spin, 
while $\overline{d_m} (\ell)$ and 
$\overline{p_n} (\ell)$ are the other 3d orbital of the $\sigma$-spin 
and the 4p orbital of the $\sigma$-spin 
in the same magnetic ion, respectively. 
Owing to the d-d and d-p hybridizations in the magnetic ion, 
$\overline{d_m}(\ell)$ and $\overline{p_n}(\ell)$ 
are included in eq. (\ref{phi_a(l)}). 
The hybridization originate from, for example, 
the mixing of atomic orbitals via the surrounding ions. 
The bars of $\overline{d_m}(\ell)$ and 
$\overline{p_n}(\ell)$ 
are a mark to distinguish the components hybridized to $d_\zeta (\ell)$. 
The quantities $c_{d_m}^{(\zeta)}$ and $c_{p_n}^{(\zeta)}$ 
are the coefficients for 
$\overline{d_m}(\ell)$ and $\overline{p_n}(\ell)$, respectively, 
and $C_\zeta$ is the normalization factor, 
where $|c_{d_m}^{(\zeta)}|^2\ll$1 and $|c_{p_n}^{(\zeta)}|^2\ll$1 are assumed. 
Note that $\overline{d_m}(\ell)$ and $\overline{p_n}(\ell)$ 
play important roles in 
the presence of 
the spin-atomic vibration interaction, as will be described later.

We will also use the energy of $|\Psi_j)$ 
in the perturbation calculation. 
For simplicity, the energy is assumed to be 
\begin{eqnarray}
\label{E_j_diagonal}
E_j = (\Psi_j| {\cal H}_0 |\Psi_j), 
\end{eqnarray}
where 
it is noted that 
${\cal H}_0$ of eq. (\ref{H_0}) (or eq. (\ref{H_e})) 
is the Hamiltonian in the subspace of the 3d orbitals. 
We here write $E_j$ as
\begin{eqnarray}
\label{E_j}
E_j = e_j  + \Delta e_j, 
\end{eqnarray}
where $e_j$ is eq. (\ref{e_d_j}), 
while $\Delta e_j$ is given by 
$E_j - e_j$. 
This $\Delta e_j$ 
consists of 
terms higher than the first order of $c_{d_m}^{(\zeta)}$ 
and $c_{p_n}^{(\zeta)}$, 
where 
$| \Delta e_i - \Delta e_j | \ll | e_i - e_j |$ for $i\ne j$.

In this paper, 
eq. (\ref{E_j_diagonal}) is simultaneously regarded as 
the matrix element of 
the effective Hamiltonian of ${\cal H}_0$. 
Namely, we neglect the off-diagonal matrix element 
$(\Psi_{j'}| {\cal H}_0 |\Psi_j)$ for $j \ne j'$, 
which appears due to the hybridization, 
because the magnitudes of the off-diagonal elements are much smaller than 
$|E_j - E_{j'}|$. 
The effective Hamiltonian 
will be used in eqs. (\ref{jpi}) and (\ref{2nd-order}).

Furthermore, 
$|\Psi_1)$ is assumed to be a nondegenerate ground state with $E_1$. 
The function $\Psi_1$ 
is regarded as a real function 
because ${\cal H}_0$ is the real Hamiltonian.\cite{Yosida1}

\section{Spin-Atomic Vibration Interaction and 
Spin-Flip Hamiltonian}
\label{SA_SF}

We obtain the first- and second-order perturbation energies 
of the model in \S \ref{Model}. 
By means of the principal axis transformation, 
we finally derive 
the spin-atomic vibration interaction and the spin-flip Hamiltonian 
as well as the conventional anisotropy spin Hamiltonian. 

\subsection{First-order perturbation energy}
\label{First}

Using ${V}_{\mbox{\footnotesize p}}$ of eq. (\ref{Vp}) 
and $|\Psi_j ) |S_z \rangle$ of eq. (\ref{Psi_j_Sz}), 
we obtain the first-order perturbation energy ${V}^{(1)}$, i.e., 
\begin{eqnarray}
{V}^{(1)}=\langle {S_z}'| ( \Psi_1 | 
{V}_{\mbox{\footnotesize p}} |\Psi_1)| S_z \rangle. 
\end{eqnarray}
We here have 
$(\Psi_1 |{V}^{(1)}_{\mbox{\footnotesize so0}}|\Psi_1 )$=
$(\Psi_1 |{V}^{(1)}_{\mbox{\footnotesize so1}}|\Psi_1 )$=0, 
because ${\mbox{\boldmath $L$}}$ 
is quenched for the nondegenerate ground state 
whose wave function is the real function, 
i.e., 
$(\Psi_1 |{\mbox{\boldmath $L$}} |\Psi_1 )$=0.\cite{Yosida1,Pryce,LLLLL} 
The first-order energy ${V}^{(1)}$ 
is then expressed as follows: 
\begin{eqnarray}
\label{V^(1)}
&&\hspace*{-0.5cm}
{V}^{(1)}= {V}^{(1)}_{\mbox{\footnotesize so2}} 
+ {V}^{(1)}_{\mbox{\footnotesize en}} 
+ {V}^{(1)}_{\mbox{\footnotesize c}}
+ {V}^{(1)}_{\mbox{\footnotesize k}}, \\
\label{so2^1}
&&\hspace*{-0.5cm}
{V}^{(1)}_{\mbox{\footnotesize so2}} =  
\langle {S_z}'| ( \Psi_1 | 
V_{\mbox{\footnotesize so2}}
|\Psi_1)| S_z \rangle 
=
\sum_{\mu=x,y,z} \Pi_\mu S_\mu, \\
&&\hspace*{-0.5cm}
\label{V_en^1}
{V}^{(1)}_{\mbox{\footnotesize en}} = 
\langle {S_z}'| ( \Psi_1 | 
{V}_{\mbox{\footnotesize en}}'
|\Psi_1)| S_z \rangle \nonumber \\
&&\hspace*{0.25cm}=
-\eta \sum_{\xi=x,y,z} 
\left( \Psi_1 \left| \sum_{\ell=1}^N \frac{\partial}{\partial \xi_\ell} 
V_{\mbox{\footnotesize en}} ({\mbox{\boldmath $r$}}_\ell) 
\right| \Psi_1 \right) 
\sqrt{\frac{\hbar}{2M \omega_\xi}} \left(a_\xi + a_\xi^\dag \right) 
\delta_{{S_z}',S_z}, 
\\
&&\hspace*{-0.5cm}
\label{V_c^1}
{V}^{(1)}_{\mbox{\footnotesize c}} = 
\langle {S_z}'| ( \Psi_1 | 
{V}_{\mbox{\footnotesize c}}'
|\Psi_1)| S_z \rangle \nonumber \\
&&\hspace*{0.25cm}=
(1-\eta) \sum_{\xi=x,y,z} 
\left( \Psi_1 \left| \sum_{\ell=1}^N \frac{\partial}{\partial \xi_\ell} 
V_{\mbox{\footnotesize c}} ({\mbox{\boldmath $r$}}_\ell) 
\right| \Psi_1 \right) 
\sqrt{\frac{\hbar}{2M \omega_\xi}} \left(a_\xi + a_\xi^\dag \right)
\delta_{{S_z}',S_z}, 
\\
\label{zero}
&&\hspace*{-0.5cm}
{V}^{(1)}_{\mbox{\footnotesize k}} = 
\langle {S_z}'| ( \Psi_1 | 
{V}_{\mbox{\footnotesize k}}'
|\Psi_1)| S_z \rangle  \nonumber \\
&&\hspace*{0.25cm}=
- \frac{\eta}{M} 
\sum_{\xi=x,y,z}
\left( \Psi_1 \left| 
p_{t,\xi} \right| \Psi_1 \right)
{\rm i} \sqrt{\frac{M \hbar\omega_\xi}{2}} 
\left(- a_\xi + a_\xi^\dag \right)\delta_{{S_z}',S_z}
\nonumber \\
&&\hspace*{0.25cm}=0,
\end{eqnarray}
with 
\begin{eqnarray}
\label{G_x}
&& \Pi_x = - \eta \lambda \frac{1}{\hbar} \left[ 
\left( \Psi_1 \left| y_t \right| \Psi_1 \right) \Delta p_{{\rm e},z} 
-\left( \Psi_1 \left| z_t \right| \Psi_1 \right) \Delta p_{{\rm e},y} \right]   \nonumber \\
&&\hspace*{0.55cm} =
\Lambda_{y,z}^{(1)} \left(-a_z + a_z^\dag \right)
-\Lambda_{z,y}^{(1)}  \left( -a_y + a_y^\dag \right), \\
\label{G_y}
&& \Pi_y 
=- \eta \lambda \frac{1}{\hbar} \left[ 
\left( \Psi_1 \left| z_t \right| \Psi_1 \right) \Delta p_{{\rm e},x} 
-\left( \Psi_1 \left| x_t \right| \Psi_1 \right) \Delta p_{{\rm e},z} \right]   \nonumber \\
&&\hspace*{0.55cm}
= \Lambda_{z,x}^{(1)} \left( -a_x + a_x^\dag \right)
-\Lambda_{x,z}^{(1)} \left( -a_z + a_z^\dag \right), \\
\label{G_z}
&& \Pi_z 
=- \eta \lambda \frac{1}{\hbar} \left[ 
\left( \Psi_1 \left| x_t \right| \Psi_1 \right) \Delta p_{{\rm e},y} 
-\left( \Psi_1 \left| y_t \right| \Psi_1 \right) \Delta p_{{\rm e},x} \right]   \nonumber \\
&&\hspace*{0.55cm}
=\Lambda_{x,y}^{(1)} \left(- a_y +  a_y^\dag \right)
- \Lambda_{y,x}^{(1)} \left( -a_x + a_x^\dag \right), 
\end{eqnarray}
where
\begin{eqnarray}
\label{L_mxi^(1)}
\Lambda_{\mu,\xi}^{(1)} = 
-\eta
\lambda 
\frac{1}{\hbar}
{\rm i} \frac{m}{M} \sqrt{\frac{M \hbar \omega_\xi}{2}}
( \Psi_1 | \mu_t  | \Psi_1). 
\end{eqnarray}
In the last result of eq. (\ref{so2^1}), 
$\langle S_z'|$ and $|S_z \rangle$ are omitted 
to use the operator representation. 
The vibration displacement of the effective nucleus, 
$\Delta {\mbox{\boldmath $r$}}_{\rm n}$ of eq. (\ref{Delta_rn_xyz}), 
has been replaced here with
\begin{eqnarray}
\label{Delta_rn}
&&\Delta {\mbox{\boldmath $r$}}_{\rm n}
=(\Delta x_{\rm n}, \Delta y_{\rm n}, \Delta z_{\rm n}) \nonumber \\
&&\hspace*{0.8cm}
=\left( 
\sqrt{\frac{\hbar}{2M \omega_x}} \left(a_x + a_x^\dag \right), 
\sqrt{\frac{\hbar}{2M \omega_y}} \left(a_y + a_y^\dag \right), 
\sqrt{\frac{\hbar}{2M \omega_z}} \left(a_z + a_z^\dag \right)
\right). 
\end{eqnarray}
The operator $a_\xi$ ($a_\xi^\dag$) denotes 
the annihilation operator (creation operator) of 
the atomic vibration in the $\xi$ direction,\cite{Hartmann} 
where the atomic vibration represents 
the vibration of the magnetic ion. 
The quantity 
$M$ is the mass of the magnetic ion, and 
$\omega_\xi$ is the angular frequency in the $\xi$ direction 
of the magnetic ion. 
On the basis of eq. (\ref{delta_p}), 
$\Delta {\mbox{\boldmath $p$}}_{\rm e}$ has been written as 
\begin{eqnarray}
\label{Delta_p}
\Delta {\mbox{\boldmath $p$}}_{\rm e}
=m \Delta \dot{\mbox{\boldmath $r$}}_{\rm n}=
m \frac{\Delta {\mbox{\boldmath $p$}}_{\rm n}}{M}, 
\end{eqnarray}
where 
$\Delta {\mbox{\boldmath $p$}}_{\rm n}$ is 
the momentum of the effective nucleus. 
This $\Delta {\mbox{\boldmath $p$}}_{\rm n}$ also has been replaced with
\begin{eqnarray}
\label{Delta_pn}
&&\hspace*{-0.5cm}\Delta {\mbox{\boldmath $p$}}_{\rm n} 
=(\Delta p_{{\rm n},x}, \Delta p_{{\rm n},y}, \Delta p_{{\rm n},z}) 
\nonumber \\
&&\hspace*{0.3cm} =\left( {\rm i} \sqrt{\frac{M \hbar\omega_x}{2}} 
\left(- a_x + a_x^\dag \right), 
{\rm i} 
\sqrt{\frac{M \hbar\omega_y}{2}} \left( - a_y + a_y^\dag \right),~
{\rm i} 
\sqrt{\frac{M \hbar\omega_z}{2}} \left( - a_z + a_z^\dag \right) \right). 
\end{eqnarray}
Note here that 
$\Delta {\mbox{\boldmath $r$}}_{\rm n}$ 
(i.e., the displacement of the effective nucleus) 
may correspond to 
the position of the gravity point of the magnetic ion, 
because the mass of the nucleus is considerably larger than 
that of the electron. 
In addition, 
$\Delta {\mbox{\boldmath $r$}}_{\rm n}$ 
($\Delta {\mbox{\boldmath $p$}}_{\rm n}$) 
has been assumed to be equal to 
the vibration displacement 
(momentum) of a mass point 
acting as a harmonic oscillator, 
in which a restoring force is linear in the displacement 
(see eq. (\ref{eq_mo}) and Appendix \ref{D_rn_ap}). 
The mass of the mass point is $M$, 
and its angular frequency in the $\xi$ direction is $\omega_\xi$.

In the derivation of eqs. (\ref{zero}) - (\ref{G_z}), 
we have used $( \Psi_1 | p_{t,\mu} | \Psi_1)$=0 with $\mu$=$x$, $y$, $z$. 
Here, $( \Psi_{j'} | p_{t,\mu} | \Psi_j)$ is represented by
\begin{eqnarray}
\label{jpi}
&&\hspace*{-2cm}(\Psi_{j'} | p_{t,\mu} | \Psi_j )
=\frac{m}{{\rm i}\hbar}
\left( \Psi_{j'} \left| \left[ \mu_t, {\cal H}_0 \right] \right| 
\Psi_j \right) 
\nonumber \\
&&=\frac{m(E_j - E_{j'})}{{\rm i}\hbar}(\Psi_{j'} | \mu_t | \Psi_j ), 
\end{eqnarray}
where 
$\left[ \mu_t, {\cal H}_0 \right]$
=$\sum_{\ell=1}^N \left[ \mu_\ell, p_{\ell,\mu}^2/(2m) \right]$ 
=${\rm i}\hbar p_{t,\mu}/m$ and 
eq. (\ref{E_j_diagonal}) have been utilized. 
We also mention that 
$\Pi_\mu$ of eqs. (\ref{G_x}) - (\ref{G_z}) (or eq. (\ref{so2^1})) 
is the $\mu$ component of 
$( \Psi_1 |- \eta \lambda \frac{1}{\hbar}\left(
{\mbox{\boldmath $r$}}_t \times 
\Delta {\mbox{\boldmath $p$}}_{\rm e} \right) | \Psi_1)$, 
where 
$( \Psi_1 | p_{t,\mu} | \Psi_1)$=0 
has been used in ${V}^{(1)}_{\mbox{\footnotesize so2}}$ 
of eq. (\ref{so2^1}) with eq. (\ref{V_so2}). 

From now on, 
we will make no mention of eq. (\ref{V_en^1}) or (\ref{V_c^1}), 
which either contains only $a_\xi$ and $a_\xi^\dag$ 
but no spin operators. 

\subsection{Second-order perturbation energy}
\label{Second}

Using ${V}_{\mbox{\footnotesize p}}$ of eq. (\ref{Vp}), 
$|\Psi_j ) |S_z \rangle$ of eq. (\ref{Psi_j_Sz}), 
and $E_j$ of eq. (\ref{E_j_diagonal}), 
we obtain the second-order perturbation energy ${V}^{(2)}$, i.e., 
\begin{eqnarray}
\label{2nd-order}
&&
\hspace*{-0.3cm} 
{V}^{(2)}=
\displaystyle{\sum_{j (\ne 1)}}
\displaystyle{\sum_{{S_z}''}}
\frac{  \langle {S_z}' | (\Psi_1 |{V}_{\mbox{\footnotesize p}} | \Psi_j )| {S_z}'' \rangle
\langle {S_z}'' |( \Psi_j | {V}_{\mbox{\footnotesize p}} 
| \Psi_1)|S_z \rangle }
{ E_1 - E_j}. 
\end{eqnarray}
When terms higher than the first order of 
$\eta\Delta \xi_{\rm n}$ 
and $\eta\Delta p_{{\rm e},\xi}$ with $\xi$=$x$, $y$, $z$ 
are neglected, 
${V}^{(2)}$ is expressed as follows:
\begin{eqnarray}
\label{pertub}
&&\hspace*{-0.7cm}
{V}^{(2)}
={V}_{\mbox{\footnotesize so0,so0}}^{(2)} 
+ {V}_{\mbox{\footnotesize so0,so1}}^{(2)}
+ {V}_{\mbox{\footnotesize so0,so2}}^{(2)} 
+ {V}_{\mbox{\footnotesize so0,k}}^{(2)}, \\
\label{so0so0^2}
&&
\hspace*{-0.7cm}
{V}_{\mbox{\footnotesize so0,so0}}^{(2)}= 
f({V}_{\mbox{\footnotesize so0}},{V}_{\mbox{\footnotesize so0}})
=\displaystyle{\sum_{\mu,\nu=x,y,z}} \Lambda_{\mu,\nu} 
S_\mu S_\nu, \\
\label{so0so1^2}
&&
\hspace*{-0.7cm}
{V}_{\mbox{\footnotesize so0,so1}}^{(2)}=
f({V}_{\mbox{\footnotesize so0}},{V}_{\mbox{\footnotesize so1}})
+f({V}_{\mbox{\footnotesize so1}},{V}_{\mbox{\footnotesize so0}}) \nonumber \\
&&\hspace*{0.54cm}=\sum_{\mu,\nu,\xi=x,y,z}
\Lambda_{\mu,\nu;\xi}S_\mu S_\nu (a_\xi + a_\xi^\dag ),  \\
\label{so0so2^2}
&&
\hspace*{-0.7cm}
{V}_{\mbox{\footnotesize so0,so2}}^{(2)}= 
f({V}_{\mbox{\footnotesize so0}},{V}_{\mbox{\footnotesize so2}})
+f({V}_{\mbox{\footnotesize so2}},{V}_{\mbox{\footnotesize so0}})\nonumber \\
&&\hspace*{0.54cm}
=\sum_{\mu,\nu=x,y,z}
\Gamma_{\mu,\nu} S_\mu S_\nu, \\
\label{so0k^2}
&&\hspace*{-0.7cm}
{V}_{\mbox{\footnotesize so0,k}}^{(2)} = 
f({V}_{\mbox{\footnotesize so0}},{V}_{\mbox{\footnotesize k}}')
+f({V}_{\mbox{\footnotesize k}}',{V}_{\mbox{\footnotesize so0}})
=\sum_{\mu=x,y,z} k_\mu S_\mu, 
\end{eqnarray}
where 
\begin{eqnarray}
\label{f_XY}
&&\hspace*{-1cm}f({X},{Y})= 
\sum_{j (\ne 1)}\sum_{{S_z}''}
\frac{ \langle {S_z}' |(\Psi_1 | {X} | \Psi_j ) | {S_z}'' \rangle
 \langle {S_z}'' |( \Psi_j | {Y} | \Psi_1) |S_z \rangle}{E_1 - E_j}. 
\end{eqnarray}
In the last results of eqs. (\ref{so0so0^2}) - (\ref{so0k^2}), 
$\langle S_z'|$ and $|S_z \rangle$ are omitted 
to use the operator representation. 
The quantities 
$\Lambda_{\mu,\nu}$, $\Lambda_{\mu,\nu;\xi}$, $\Gamma_{\mu,\nu}$, 
and $k_\mu$ are given by 
\begin{eqnarray}
\label{lam_1}
&&\hspace*{-0.5cm} 
\Lambda_{\mu,\nu} =\lambda^2 
{\cal L}_{\mu,\nu}, \\
\label{lam_2}
&&\hspace*{-0.5cm} 
\Lambda_{\mu,\nu;\xi} =2 \eta 
\lambda^2 
\frac{\left\langle 3 \xi r^{-5} \right\rangle }
{\left\langle r^{-3} \right\rangle }
\sqrt{\frac{\hbar}{2M \omega_\xi}} {\cal L}_{\mu,\nu}, \\
&&\hspace*{-0.5cm} 
\label{G_xx}
\Gamma_{x,x}=
(-\Lambda_{x;z,y}^{+} + \Lambda_{x;z,y}^{-})(a_y + a_y^\dag)
+ (\Lambda_{x;y,z}^{+} - \Lambda_{x;y,z}^{-})(a_z + a_z^\dag), \\
&&\hspace*{-0.5cm} 
\Gamma_{x,y}=
\Lambda_{x;z,x}^{+}a_x - \Lambda_{x;z,x}^{-}a_x^\dag 
+ \Lambda_{y;z,y}^{-}a_y - \Lambda_{y;z,y}^{+}a_y^\dag  
+ (-\Lambda_{x;x,z}^{+} - \Lambda_{y;y,z}^{-})a_z 
+ (\Lambda_{y;y,z}^{+}+\Lambda_{x;x,z}^{-})a_z^\dag, \nonumber \\\\
&&\hspace*{-0.5cm} 
\Gamma_{x,z} =
- \Lambda_{x;y,x}^{+} a_x + \Lambda_{x;y,x}^{-} a_x^\dag 
+ (\Lambda_{x;x,y}^{+} + \Lambda_{z;z,y}^{-} ) a_y  
+ ( - \Lambda_{z;z,y}^{+}-\Lambda_{x;x,y}^{-} )a_y^\dag 
- \Lambda_{z;y,z}^{-} a_z + \Lambda_{z;y,z}^{+} a_z^\dag, \nonumber \\\\
&&\hspace*{-0.5cm} 
\Gamma_{y,x}=
-\Lambda_{x;z,x}^{-}a_x + \Lambda_{x;z,x}^{+}a_x^\dag 
- \Lambda_{y;z,y}^{+}a_y + \Lambda_{y;z,y}^{-}a_y^\dag  
+ (\Lambda_{y;y,z}^{+} + \Lambda_{x;x,z}^{-})a_z 
+ ( - \Lambda_{x;x,z}^{+}-\Lambda_{y;y,z}^{-})a_z^\dag, \nonumber \\\\
&&\hspace*{-0.5cm} 
\Gamma_{y,y}=
(\Lambda_{y;z,x}^{+} - \Lambda_{y;z,x}^{-})(a_x + a_x^\dag) 
+ (- \Lambda_{y;x,z}^{+} + \Lambda_{y;x,z}^{-})(a_z + a_z^\dag), \\
&&\hspace*{-0.5cm} 
\Gamma_{y,z}=
(- \Lambda_{y;y,x}^{+} - \Lambda_{z;z,x}^{-} )a_x 
+ ( \Lambda_{z;z,x}^{+} + \Lambda_{y;y,x}^{-} ) a_x^\dag 
+ \Lambda_{y;x,y}^{+} a_y - \Lambda_{y;x,y}^{-} a_y^\dag 
+ \Lambda_{z;x,z}^{-} a_z - \Lambda_{z;x,z}^{+} a_z^\dag, \nonumber \\\\
&&\hspace*{-0.5cm} 
\Gamma_{z,x} =
\Lambda_{x;y,x}^{-} a_x - \Lambda_{x;y,x}^{+} a_x^\dag 
+ (  - \Lambda_{z;z,y}^{+}- \Lambda_{x;x,y}^{-}) a_y 
+(\Lambda_{x;x,y}^{+} + \Lambda_{z;z,y}^{-} )a_y^\dag 
+ \Lambda_{z;y,z}^{+} a_z - \Lambda_{z;y,z}^{-} a_z^\dag, \nonumber \\\\
&&\hspace*{-0.5cm} 
\Gamma_{z,y}=
(\Lambda_{z;z,x}^{+} + \Lambda_{y;y,x}^{-}) a_x 
+ (- \Lambda_{y;y,x}^{+}- \Lambda_{z;z,x}^{-} ) a_x^\dag 
- \Lambda_{y;x,y}^{-}  a_y + \Lambda_{y;x,y}^{+}  a_y^\dag 
- \Lambda_{z;x,z}^{+}  a_z + \Lambda_{z;x,z}^{-} a_z^\dag, \nonumber \\\\
&&\hspace*{-0.5cm} 
\label{G_zz}
\Gamma_{z,z}=
(-\Lambda_{z;y,x}^{+} + \Lambda_{z;y,x}^{-} ) (a_x + a_x^\dag) 
+ (\Lambda_{z;x,y}^{+} - \Lambda_{z;x,y}^{-} ) (a_y + a_y^\dag), \\
\label{k_mu}
&&\hspace*{-0.5cm} k_\mu = \sum_{\xi=x,y,z} 
\Lambda_{\mu,\xi}^k (-a_\xi + a_\xi^\dag), 
\end{eqnarray}
respectively, where
\begin{eqnarray}
\label{cal L}
&&\hspace*{-0.4cm} 
{\cal L}_{\mu,\nu} =
\sum_{j (\ne 1)} 
\frac{( \Psi_1 |L_\mu | \Psi_j )(\Psi_j | L_\nu | \Psi_1) }
{ E_1 - E_j}, \\
\label{Lam^k}
&&\hspace*{-0.4cm} \Lambda_{\mu,\xi}^k = 
-2 \eta \lambda \frac{m}{M} \frac{1}{\hbar}
\sqrt{\frac{M\hbar \omega_\xi}{2}}
\sum_{j (\ne 1)} (\Psi_1 |L_\mu|\Psi_j )(\Psi_j |\xi_t|\Psi_1), \\
&&\hspace*{-0.4cm} 
\label{lam_d}
\Lambda_{\mu;\nu,\xi}^{\pm}=\eta
\lambda^2 
\frac{1}{\hbar}
\sum_{j (\ne 1)}  \frac{
(\Psi_1 | L_\mu | \Psi_j) 
(\Psi_j | F_{\nu,\xi}^{\pm} | \Psi_1)}{ E_1 - E_j}, \\
\label{jFi}
&& \hspace*{-0.4cm} ( \Psi_{j'} | F_{\nu,\xi}^{\pm} |\Psi_j )=
{\rm i} \left[ 
\frac{m}{M} \sqrt{\frac{M \hbar\omega_\xi}{2}} 
\pm  \sqrt{\frac{\hbar}{2M \omega_\xi}} 
\frac{m}{\hbar} ( E_{j'}-E_j) 
\right] (\Psi_{j'} | \nu_t |\Psi_j ),  \\
&&\hspace*{-0.4cm} 
\label{F_mu_nu}
F_{\nu,\xi}^{\pm}=
{\rm i} \frac{m}{M} \sqrt{\frac{M \hbar \omega_\xi}{2}}\nu_t 
\pm \sqrt{\frac{\hbar}{2M \omega_\xi}} p_{t,\nu}. 
\end{eqnarray}
In the derivation of eq. (\ref{Lam^k}), 
we have used eq. (\ref{jpi}), 
$(\Psi_{j'} | {\mbox{\boldmath $L$}} | \Psi_j) 
= -(\Psi_j | {\mbox{\boldmath $L$}} | \Psi_{j'})$,\cite{LLLLL} 
and 
$(\Psi_{j'} | \xi_t | \Psi_j) = (\Psi_j | \xi_t | \Psi_{j'})$. 
In addition, eq. (\ref{jFi}) has been obtained by using eqs. (\ref{F_mu_nu}) 
and (\ref{jpi}). 
Here, we mention the relations of 
$(\Psi_{j'} | F_{\nu,\xi}^{\pm} | \Psi_j)^*$ 
= 
$(\Psi_{j} | ( F_{\nu,\xi}^{\pm} )^\dag | \Psi_{j'})$ 
=
$-(\Psi_{j} | F_{\nu,\xi}^{\mp} | \Psi_{j'})$ and 
$(\Psi_{j'} | F_{\nu,\xi}^{\pm} | \Psi_j)$ 
= 
$(\Psi_j | F_{\nu,\xi}^{\mp} | \Psi_{j'})$, 
where 
$(\Psi_{j'} | \nu_t | \Psi_j)$ 
= 
$(\Psi_j | \nu_t | \Psi_{j'})$. 
We also show the relations of 
${\cal L}_{\mu,\nu}$ = ${\cal L}_{\nu,\mu}$,\cite{LL} 
$\Lambda_{\mu,\nu}$=$\Lambda_{\nu,\mu}$, 
$\Lambda_{\mu,\nu;\xi}$=$\Lambda_{\nu,\mu;\xi}$, 
$\Lambda_{\mu;\nu,\xi}^{\pm}$=$-\Lambda_{\nu,\xi;\mu}^{\mp}$
=$(\Lambda_{\mu;\nu,\xi}^{\pm})^*$,\cite{Lambda} 
and $(\Gamma_{\mu,\nu})^\dag$=$\Gamma_{\nu,\mu}$, 
where 
\begin{eqnarray}
\label{lam_dd}
\Lambda_{\nu,\xi;\mu}^{\pm}=\eta
\lambda^2 
\frac{1}{\hbar}
\sum_{j (\ne 1)}  \frac{
(\Psi_1 | F_{\nu,\xi}^{\pm} | \Psi_j)
(\Psi_j| L_\mu | \Psi_1) }{E_1 - E_j}. 
\end{eqnarray}

Note that 
the two terms consisting of 
${V}_{\mbox{\footnotesize so0}}$ and 
${V}_{\mbox{\footnotesize en}}' + {V}_{\mbox{\footnotesize c}}'$ 
in eq. (\ref{2nd-order}) cancel each other, i.e.,
\begin{eqnarray}
\label{vanish}
&& \hspace*{-0.5cm}
f({V}_{\mbox{\footnotesize so0}},{V}_{\mbox{\footnotesize en}}' 
+ {V}_{\mbox{\footnotesize c}}')
+f({V}_{\mbox{\footnotesize en}}' + {V}_{\mbox{\footnotesize c}}',
{V}_{\mbox{\footnotesize so0}})=0,
\end{eqnarray}
where $f({X},{Y})$ is given by eq. (\ref{f_XY}). 
Here, we use $(\Psi_j |{\mbox{\boldmath $L$}}|\Psi_1)$=
$-(\Psi_1 |{\mbox{\boldmath $L$}}|\Psi_j)$ 
(see ref. \citen{LLLLL}) 
and 
$(\Psi_j | 
{V}_{\mbox{\footnotesize en}}'
+
{V}_{\mbox{\footnotesize c}}'
|\Psi_1)$=
$(\Psi_1 | 
{V}_{\mbox{\footnotesize en}}'
+
{V}_{\mbox{\footnotesize c}}'
|\Psi_j)$. 
In particular, 
$f({V}_{\mbox{\footnotesize so0}},
{V}_{\mbox{\footnotesize c}}')
+f({V}_{\mbox{\footnotesize c}}',
{V}_{\mbox{\footnotesize so0}})$=0 
corresponds to the so-called 
Van Vleck cancellation.\cite{Van1,Kondo}

\subsection{Principal axis transformation}
\label{transform}

To obtain 
expressions of 
${V}^{(1)}_{\mbox{\footnotesize so2}}$ of eq. (\ref{so2^1}) 
and ${V}^{(2)}$ of eq. (\ref{pertub}) 
in the principal axis coordinate system, 
we first introduce the rotational transformation matrix 
for the principal axis transformation. 
This transformation is based on the conventional theory 
for the anisotropy spin Hamiltonian.\cite{Yosida1,nanomagnet1} 
Here, $x$, $y$, and $z$ are regarded as the initial axes. 

In accordance with 
refs. \citen{Yosida1} and \citen{nanomagnet1}, 
we focus on $\sum_{\mu,\nu=x,y,z} {\cal L}_{\mu,\nu} S_\mu S_\nu$ 
included in 
${V}_{\mbox{\footnotesize so0,so0}}^{(2)}$ of eq. (\ref{so0so0^2}) 
and ${V}_{\mbox{\footnotesize so0,so1}}^{(2)}$ of eq. (\ref{so0so1^2}). 
The expression in the matrix representation is written as
\begin{eqnarray}
\label{Vs_mat}
&&\hspace*{-1.5cm}
\displaystyle{\sum_{\mu,\nu=x,y,z}} {\cal L}_{\mu,\nu} 
S_\mu S_\nu 
=
(S_x, S_y, S_z) 
\left(
  \begin{array}{ccc}
   {\cal L}_{x,x} & {\cal L}_{x,y} & {\cal L}_{x,z}  \\
   {\cal L}_{y,x} & {\cal L}_{y,y} & {\cal L}_{y,z}  \\
   {\cal L}_{z,x} & {\cal L}_{z,y} & {\cal L}_{z,z} 
  \end{array}
\right)
\left(
  \begin{array}{c}
   S_x  \\
   S_y  \\
   S_z
  \end{array}
\right) \nonumber \\
&&\hspace*{1.4cm}\equiv 
{\mbox{\boldmath $S$}} {\mbox{\boldmath ${\cal L}$}} {\mbox{\boldmath $S$}}^T, 
\end{eqnarray}
with
\begin{eqnarray}
&&
{\mbox{\boldmath ${\cal L}$}} =
\left(
  \begin{array}{ccc}
   {\cal L}_{x,x} & {\cal L}_{x,y} & {\cal L}_{x,z}  \\
   {\cal L}_{y,x} & {\cal L}_{y,y} & {\cal L}_{y,z}  \\
   {\cal L}_{z,x} & {\cal L}_{z,y} & {\cal L}_{z,z} 
  \end{array}
\right) 
\nonumber \\
&&\hspace*{0.4cm} = {\mbox{\boldmath ${\cal L}$}}^{(0)} 
+ {\mbox{\boldmath ${\cal L}$}}', \\
\label{lam0}
&&
{\mbox{\boldmath ${\cal L}$}}^{(0)} 
=
\left(
  \begin{array}{ccc}
   {\cal L}_{x,x}^{(0)} & {\cal L}_{x,y}^{(0)} & {\cal L}_{x,z}^{(0)}  \\
   {\cal L}_{y,x}^{(0)} & {\cal L}_{y,y}^{(0)} & {\cal L}_{y,z}^{(0)}  \\
   {\cal L}_{z,x}^{(0)} & {\cal L}_{z,y}^{(0)} & {\cal L}_{z,z}^{(0)}  
  \end{array}
\right), \\
&&
{\mbox{\boldmath ${\cal L}$}}'=
\left(
  \begin{array}{ccc}
   {\cal L}_{x,x}' & {\cal L}_{x,y}' & {\cal L}_{x,z}'  \\
   {\cal L}_{y,x}' & {\cal L}_{y,y}' & {\cal L}_{y,z}'  \\
   {\cal L}_{z,x}' & {\cal L}_{z,y}' & {\cal L}_{z,z}'  
  \end{array}
\right)  \\
\label{lam_0}
&&
{\cal L}_{\mu,\nu}^{(0)}=
\sum_{j (\ne 1)} \frac{ \langle \Phi_1 |L_\mu | \Phi_j \rangle 
\langle \Phi_j | L_\nu | \Phi_1 \rangle }{e_1 - e_j}, \\
&&
\label{L_mu,nu'}
{\cal L}_{\mu,\nu}'=
{\cal L}_{\mu,\nu} - {\cal L}_{\mu,\nu}^{(0)},
\end{eqnarray}
where 
${\mbox{\boldmath $S$}}^T$ is the transposed vector of 
${\mbox{\boldmath $S$}}$=$(S_x,S_y,S_z)$. 
Here, ${\cal L}_{\mu,\nu}^{(0)}$ is described by 
$\Phi_j$ consisting of only $d_\zeta (\ell)$, 
whereas 
${\cal L}_{\mu,\nu}'$ is related to $\overline{d_m}(\ell)$, 
$\overline{p_n}(\ell)$, and $d_\zeta (\ell)$. 
We also have 
${\cal L}_{\mu,\nu}^{(0)}$=${\cal L}_{\nu,\mu}^{(0)}$ and 
${\cal L}_{\mu,\nu}'$=${\cal L}_{\nu,\mu}'$.\cite{LL}

The principal axis of the crystal, named $X$, $Y$, and $Z$, are 
considered to be 
the coordinate system that diagonalizes 
${\mbox{\boldmath ${\cal L}$}}^{(0)}$.\cite{Yosida1,nanomagnet1}
We then introduce the matrices ${\mbox{\boldmath $U$}}$ and 
${\mbox{\boldmath $U$}}^T$ 
that diagonalize ${\mbox{\boldmath ${\cal L}$}}^{(0)}$. 
Here, ${\mbox{\boldmath $U$}}$ is the rotational transformation matrix, while 
the transposed matrix of ${\mbox{\boldmath $U$}}$, ${\mbox{\boldmath $U$}}^T$, 
satisfies ${\mbox{\boldmath $U$}}^T$=${\mbox{\boldmath $U$}}^{-1}$. 
We thus have
\begin{eqnarray}
\label{lam0_diago}
&&{\mbox{\boldmath $U$}} {\mbox{\boldmath ${\cal L}$}}^{(0)} 
{\mbox{\boldmath $U$}}^T= 
\left(
  \begin{array}{ccc}
   {\cal L}^{(0)}_{X,X} & 0 & 0  \\
   0 & {\cal L}^{(0)}_{Y,Y} & 0  \\
   0 & 0 & {\cal L}^{(0)}_{Z,Z}  
  \end{array}
\right), \\
\label{lam0_diagod}
&&{\mbox{\boldmath $U$}} {\mbox{\boldmath ${\cal L}$}}' 
{\mbox{\boldmath $U$}}^T=
\left(
  \begin{array}{ccc}
   {\cal L}_{X,X}' & {\cal L}_{X,Y}' & {\cal L}_{X,Z}'  \\
   {\cal L}_{Y,X}' & {\cal L}_{Y,Y}' & {\cal L}_{Y,Z}'  \\
   {\cal L}_{Z,X}' & {\cal L}_{Z,Y}' & {\cal L}_{Z,Z}'  
  \end{array}
\right), 
\end{eqnarray}
where
\begin{eqnarray}
\label{Unitary}
&&\hspace*{-1cm} {\mbox{\boldmath $U$}}=
\left(
  \begin{array}{ccc}
   \cos \theta_{X,x} & \cos \theta_{X,y} & \cos \theta_{X,z}  \\
   \cos \theta_{Y,x} & \cos \theta_{Y,y} & \cos \theta_{Y,z}  \\
   \cos \theta_{Z,x} & \cos \theta_{Z,y} & \cos \theta_{Z,z} 
  \end{array}
\right), \\
\label{Unitary^T}
&&\hspace*{-1cm} {\mbox{\boldmath $U$}}^T=
\left(
  \begin{array}{ccc}
   \cos \theta_{X,x} & \cos \theta_{Y,x} & \cos \theta_{Z,x}  \\
   \cos \theta_{X,y} & \cos \theta_{Y,y} & \cos \theta_{Z,y}  \\
   \cos \theta_{X,z} & \cos \theta_{Y,z} & \cos \theta_{Z,z} 
  \end{array}
\right), \\
&&\hspace*{-1cm} 
{\cal L}^{(0)}_{I,I}=
\sum_{\mu,\nu=x,y,z} 
\cos \theta_{I,\mu} \cos \theta_{I,\nu} 
{\cal L}^{(0)}_{\mu,\nu}, \\
&&\hspace*{-1cm} 
\label{L_I,J'}
{\cal L}_{I,J}'=
\sum_{\mu,\nu=x,y,z} 
\cos \theta_{I,\mu} \cos \theta_{J,\nu} {\cal L}_{\mu,\nu}',
\end{eqnarray}
with $I$, $J$=$X$, $Y$, $Z$. 
The angle $\theta_{I,\mu}$ is the relative angle 
between the $I$ and $\mu$ axes, 
with $I$=$X$, $Y$, $Z$ and $\mu$=$x$, $y$, $z$. 
By utilizing 
${\cal L}_{\mu,\nu}'$=${\cal L}_{\nu,\mu}'$,\cite{LL} 
${\cal L}_{I,J}'$=
$\sum_{\mu,\nu=x,y,z} 
\cos \theta_{J,\nu} \cos \theta_{I,\mu} {\cal L}_{\nu,\mu}'$
=${\cal L}_{J,I}'$ is obtained.

Using ${\mbox{\boldmath $U$}}$ and ${\mbox{\boldmath $U$}}^T$, 
we write eq. (\ref{Vs_mat}) as 
\begin{eqnarray}
\label{Vs_UU}
&&
\displaystyle{\sum_{\mu,\nu=x,y,z}} {\cal L}_{\mu,\nu} 
S_\mu S_\nu 
=(S_x, S_y, S_z) {\mbox{\boldmath $U$}}^T {\mbox{\boldmath $U$}}
\left(
  \begin{array}{ccc}
   {\cal L}_{x,x} & {\cal L}_{x,y} & {\cal L}_{x,z}  \\
   {\cal L}_{y,x} & {\cal L}_{y,y} & {\cal L}_{y,z}  \\
   {\cal L}_{z,x} & {\cal L}_{z,y} & {\cal L}_{z,z} 
  \end{array}
\right) {\mbox{\boldmath $U$}}^T {\mbox{\boldmath $U$}}
\left(
  \begin{array}{c}
   S_x  \\
   S_y  \\
   S_z
  \end{array}
\right) \nonumber \\
&&\hspace*{3cm}=
(S_X, S_Y, S_Z) 
\left(
  \begin{array}{ccc}
   {\cal L}^{(0)}_{X,X} & 0 & 0  \\
   0 & {\cal L}^{(0)}_{Y,Y} & 0  \\
   0 & 0 & {\cal L}^{(0)}_{Z,Z}  
  \end{array}
\right) 
\left(
  \begin{array}{c}
   S_X  \\
   S_Y  \\
   S_Z
  \end{array}
\right) \nonumber \\
&&\hspace*{3.5cm}
+(S_X, S_Y, S_Z) 
\left(
  \begin{array}{ccc}
   {\cal L}_{X,X}' & {\cal L}_{X,Y}' & {\cal L}_{X,Z}'  \\
   {\cal L}_{Y,X}' & {\cal L}_{Y,Y}' & {\cal L}_{Y,Z}'  \\
   {\cal L}_{Z,X}' & {\cal L}_{Z,Y}' & {\cal L}_{Z,Z}'  
  \end{array}
\right)
\left(
  \begin{array}{c}
   S_X  \\
   S_Y  \\
   S_Z
  \end{array}
\right) \nonumber \\
&&\hspace*{3cm}=
\sum_{I=X,Y,Z} {\cal L}_{I,I}^{(0)} S_I^2
+ \sum_{I,J=X,Y,Z} {\cal L}_{I,J}' S_I S_J, 
\end{eqnarray}
where the relation between ($S_X$, $S_Y$, $S_Z$) in the principal axis 
($X$, $Y$, $Z$) and 
($S_x$, $S_y$, $S_z$) in the initial axes ($x$, $y$, $z$) has been defined by
\begin{eqnarray}
\label{unitary1}
\left(
  \begin{array}{c}
   S_X  \\
   S_Y  \\
   S_Z
  \end{array}
\right) ={\mbox{\boldmath $U$}}
\left(
  \begin{array}{c}
   S_x  \\
   S_y  \\
   S_z
  \end{array}
\right). 
\end{eqnarray}

\subsection{Expressions of spin-atomic vibration interaction 
and spin-flip Hamiltonian}
\label{final_exp}

Using the above-mentioned ${\mbox{\boldmath $U$}}$ 
and ${\mbox{\boldmath $U$}}^T$, 
we can obtain the expressions of 
${V}^{(1)}_{\mbox{\footnotesize so2}}$ of eq. (\ref{so2^1}) 
and ${V}^{(2)}$ 
(=${V}_{\mbox{\footnotesize so0,so0}}^{(2)} 
+ {V}_{\mbox{\footnotesize so0,so1}}^{(2)}
+ {V}_{\mbox{\footnotesize so0,so2}}^{(2)} 
+ {V}_{\mbox{\footnotesize so0,k}}^{(2)}$) 
of eq. (\ref{pertub}) 
in the principal axis coordinate system. 
Each term is named 
any one of the spin-atomic vibration interaction, 
spin-flip Hamiltonian, 
or anisotropy spin Hamiltonian. 
In addition, we consider the feature of the respective terms. 
Their coefficients and operators are listed in Table \ref{tab1}.

\subsubsection{\rm ${v}_{\mbox{\footnotesize so0,so0,1}}^{(2)}$ 
and ${v}_{\mbox{\footnotesize so0,so0,2}}^{(2)}$}
\label{v_so0so0}
Using eq. (\ref{Vs_UU}), 
we first rewrite 
${V}_{\mbox{\footnotesize so0,so0}}^{(2)}$ of 
eq. (\ref{so0so0^2}) as
\begin{eqnarray}
\label{V_eff}
&&
\hspace*{-1cm} 
{V}_{\mbox{\footnotesize so0,so0}}^{(2)}=
{v}_{{\mbox{\footnotesize so0,so0}},1}^{(2)} 
+ {v}_{{\mbox{\footnotesize so0,so0}},2}^{(2)}, \\
\label{v_S^(1)}
&&
\hspace*{-1cm} 
{v}_{{\mbox{\footnotesize so0,so0}},1}^{(2)}=
D S_Z^2 + E (S_X^2 - S_Y^2) + FS(S+1), \\
&&
\hspace*{-1cm} 
\label{v_S^(2)}
{v}_{{\mbox{\footnotesize so0,so0}},2}^{(2)}=
\sum_{I,J=X,Y,Z} 
\Lambda_{I,J}^{\rm SF} S_I S_J, 
\end{eqnarray}
with
\begin{eqnarray}
&&\hspace*{-1cm} 
\label{DDDDD}
D= 
\lambda^2 
\left[ {\cal L}_{Z,Z}^{(0)} - 
( {\cal L}_{X,X}^{(0)} + {\cal L}_{Y,Y}^{(0)} )/2 \right], \\
&&\hspace*{-1cm} 
\label{EEEEE}
E= 
\lambda^2 
( {\cal L}_{X,X}^{(0)} - {\cal L}_{Y,Y}^{(0)} )/2, \\
&&\hspace*{-1cm} 
\label{FFFFF}
F=
\lambda^2 
({\cal L}_{X,X}^{(0)} + {\cal L}_{Y,Y}^{(0)})/2, \\
&&\hspace*{-1cm} 
\label{Lam^SF}
\Lambda_{I,J}^{\rm SF} =
\lambda^2 
{\cal L}_{I,J}', 
\end{eqnarray}
where we have $\Lambda_{I,J}^{\rm SF}$=$\Lambda_{J,I}^{\rm SF}$ 
using ${\cal L}_{I,J}'$=${\cal L}_{J,I}'$ (see \S \ref{transform}). 
The Hamiltonian ${v}_{{\mbox{\footnotesize so0,so0}},1}^{(2)}$ is 
the so-called anisotropy spin Hamiltonian,\cite{Yosida1} 
where $D$ is an anisotropy constant. 
The Hamiltonian ${v}_{{\mbox{\footnotesize so0,so0}},2}^{(2)}$, 
which contains 
the spin-flip operators such as 
$S_X S_Z$ and $S_Y S_Z$ but no atomic vibration operators, 
is named the spin-flip Hamiltonian. 

From now on, $FS(S+1)$ in eq. (\ref{v_S^(1)}) 
will be neglected because of a constant term. 

\subsubsection{\rm 
${v}_{{\mbox{\footnotesize so0,so1}},1}^{(2)}$ 
and ${v}_{{\mbox{\footnotesize so0,so1}},2}^{(2)}$
}
\label{v_so1o1}
Using eq. (\ref{Vs_UU}), we express 
${V}_{\mbox{\footnotesize so0,so1}}^{(2)}$ of eq. (\ref{so0so1^2}) as
\begin{eqnarray}
\label{V_so0so1^2}
&&
\hspace*{-1cm} 
{V}_{\mbox{\footnotesize so0,so1}}^{(2)}=
{v}_{{\mbox{\footnotesize so0,so1}},1}^{(2)} 
+ {v}_{{\mbox{\footnotesize so0,so1}},2}^{(2)}, \\
&&
\hspace*{-1cm} 
\label{v_Sv^(1)}
{v}_{{\mbox{\footnotesize so0,so1}},1}^{(2)}=
\left[ D S_Z^2 + E (S_X^2 - S_Y^2) + FS(S+1)
\right]  \sum_{\xi=x,y,z} G_\xi (a_\xi + a_\xi^\dag),  \\
&&
\hspace*{-1cm} 
\label{v_Sv^(2)}
{v}_{{\mbox{\footnotesize so0,so1}},2}^{(2)}=
\sum_{\xi=x,y,z} \sum_{I,J=X,Y,X}
\Lambda_{I,J;\xi}' S_I S_J(a_\xi + a_\xi^\dag ), 
\end{eqnarray}
with
\begin{eqnarray}
&&\hspace*{-1cm} 
\label{G_xi}
G_\xi=2 \eta
\frac{ \left\langle 3 \xi r^{-5} \right\rangle}
{\left\langle r^{-3} \right\rangle } 
\sqrt{\frac{ \hbar}{2 M \omega_\xi}}, \\
&&\hspace*{-1cm} 
\label{Lam_IJxi}
\Lambda_{I,J;\xi}' =
2 \eta 
\lambda^2 
\frac{\left\langle 3 \xi r^{-5} \right\rangle }
{ \left\langle r^{-3} \right\rangle }
\sqrt{\frac{\hbar}{2M \omega_\xi}}{\cal L}_{I,J}', 
\end{eqnarray}
where we have 
$\Lambda_{I,J;\xi}'$=$\Lambda_{J,I;\xi}'$ 
using ${\cal L}_{I,J}'$=${\cal L}_{J,I}'$ (see \S \ref{transform}). 
The interactions 
${v}_{{\mbox{\footnotesize so0,so1}},1}^{(2)}$ and 
${v}_{{\mbox{\footnotesize so0,so1}},2}^{(2)}$, 
which contain the spin and vibration operators, 
are named the spin-atomic vibration interactions.

\subsubsection{\rm 
${V}_{\mbox{\footnotesize so0,so2}}^{(2)}$
}
Using ${\mbox{\boldmath $U$}}$ and ${\mbox{\boldmath $U$}}^T$, 
we rewrite ${V}_{\mbox{\footnotesize so0,so2}}^{(2)}$ of 
eq. (\ref{so0so2^2}) as 
\begin{eqnarray}
\label{v_22}
&& {V}_{\mbox{\footnotesize so0,so2}}^{(2)}= 
\displaystyle{\sum_{\mu,\nu=x,y,z}} \Gamma_{\mu,\nu} 
S_\mu S_\nu \nonumber \\
&&=(S_x, S_y, S_z) {\mbox{\boldmath $U$}}^T {\mbox{\boldmath $U$}}
\left(
  \begin{array}{ccc}
   \Gamma_{x,x} & \Gamma_{x,y} & \Gamma_{x,z}  \\
   \Gamma_{y,x} & \Gamma_{y,y} & \Gamma_{y,z}  \\
   \Gamma_{z,x} & \Gamma_{z,y} & \Gamma_{z,z} 
  \end{array}
\right) {\mbox{\boldmath $U$}}^T {\mbox{\boldmath $U$}}
\left(
  \begin{array}{c}
   S_x  \\
   S_y  \\
   S_z
  \end{array}
\right) \nonumber \\
&&=
(S_X, S_Y, S_Z) 
\left(
  \begin{array}{ccc}
   \Gamma_{X,X} & \Gamma_{X,Y} & \Gamma_{X,Z}  \\
   \Gamma_{Y,X} & \Gamma_{Y,Y} & \Gamma_{Y,Z}  \\
   \Gamma_{Z,X} & \Gamma_{Z,Y} & \Gamma_{Z,Z}  
  \end{array}
\right)
\left(
  \begin{array}{c}
   S_X  \\
   S_Y  \\
   S_Z
  \end{array}
\right) \nonumber \\
\label{vv_so0so2}
&&= \sum_{I,J=X,Y,Z} \Gamma_{I,J} S_I S_J, 
\end{eqnarray}
where 
\begin{eqnarray}
\Gamma_{I,J}=\sum_{\mu,\nu=x,y,z} \cos \theta_{I,\mu} \cos \theta_{J,\nu} 
\Gamma_{\mu,\nu}, 
\end{eqnarray}
with $\Gamma_{\mu,\nu}$ being eqs. (\ref{G_xx}) - (\ref{G_zz}). 
We here mention a relation of 
$(\Gamma_{I,J})^\dag=
\sum_{\mu,\nu=x,y,z}  \cos \theta_{J,\nu} \cos \theta_{I,\mu}
\Gamma_{\nu,\mu}$=$\Gamma_{J,I}$ 
using $(\Gamma_{\mu,\nu})^\dag$=$\Gamma_{\nu,\mu}$ (see \S \ref{Second}). 
The interaction 
${V}_{\mbox{\footnotesize so0,so2}}^{(2)}$, 
which contains the spin and vibration operators, 
is the spin-atomic vibration interaction.

\subsubsection{\rm 
${V}_{\mbox{\footnotesize so0,k}}^{(2)}$
}
We write ${V}_{\mbox{\footnotesize so0,k}}^{(2)}$ 
of eq. (\ref{so0k^2}) by
\begin{eqnarray}
&& \hspace*{-0.97cm}
{V}_{\mbox{\footnotesize so0,k}}^{(2)}
= \sum_{\mu =x,y,z} k_\mu S_\mu  \nonumber \\
&&=(k_x, k_y, k_z) {\mbox{\boldmath $U$}}^T {\mbox{\boldmath $U$}}
\left(
  \begin{array}{c}
   S_x \\
   S_y \\
   S_z
  \end{array}
\right) \nonumber \\
\label{vv_so0k}
&&= \sum_{I=X,Y,Z} k_I S_I, 
\end{eqnarray}
with 
\begin{eqnarray}
&&k_I = 
\sum_{\mu=x,y,z} k_\mu \cos \theta_{I,\mu},
\end{eqnarray}
where $k_\mu$ is eq. (\ref{k_mu}). 
The interaction 
${V}_{\mbox{\footnotesize so0,k}}^{(2)}$, 
which contains the spin and vibration operators, 
is the spin-atomic vibration interaction.

\subsubsection{\rm 
${V}_{\mbox{\footnotesize so2}}^{(1)}$
}
\label{V_so2^1}
We express ${V}_{\mbox{\footnotesize so2}}^{(1)}$ 
of eq. (\ref{so2^1}) by 
\begin{eqnarray}
&& \hspace*{-0.75cm}
{V}_{\mbox{\footnotesize so2}}^{(1)}= 
\sum_{\mu =x,y,z} \Pi_\mu S_\mu 
\nonumber \\
&& =(\Pi_x, \Pi_y, \Pi_z) {\mbox{\boldmath $U$}}^T {\mbox{\boldmath $U$}}
\left(
  \begin{array}{c}
   S_x \\
   S_y \\
   S_z
  \end{array}
\right) \nonumber \\
&&= \Pi_X S_X + \Pi_Y S_Y + \Pi_Z S_Z \nonumber \\
\label{vv_so2_1}
&&= \sum_{I=X,Y,Z} \Pi_I S_I, 
\end{eqnarray}
with 
\begin{eqnarray}
&&\Pi_I = 
\sum_{\mu=x,y,z} \Pi_\mu \cos \theta_{I,\mu},
\end{eqnarray}
where $\Pi_x$, $\Pi_y$, and $\Pi_z$ are 
eqs. (\ref{G_x}) - (\ref{G_z}), respectively. 
The interaction 
${V}_{\mbox{\footnotesize so2}}^{(1)}$, 
which contains the spin and vibration operators, 
is the spin-atomic vibration interaction.

\subsubsection{Considerations}
\label{consider}
On the basis of \S \ref{v_so0so0} - \S \ref{V_so2^1}, 
we finally express the perturbation energy, 
${V}^{(1)}_{\mbox{\footnotesize so2}}+{V}^{(2)}$, 
as
\begin{eqnarray}
\label{VVVVV}
&&\hspace*{-0.55cm}{V} =
{V}_{\mbox{\footnotesize so2}}^{(1)}
+{v}_{\mbox{\footnotesize so0,so0,1}}^{(2)}
+{v}_{\mbox{\footnotesize so0,so0,2}}^{(2)}
+{v}_{\mbox{\footnotesize so0,so1,1}}^{(2)}
+{v}_{\mbox{\footnotesize so0,so1,2}}^{(2)}
+{V}_{\mbox{\footnotesize so0,so2}}^{(2)}
+{V}_{\mbox{\footnotesize so0,k}}^{(2)} \nonumber \\
&&\hspace*{-0.15cm}\equiv {V}_{\rm A} + {V}_{\rm SA} + {V}_{\rm SF}, 
\end{eqnarray}
where 
${V}_{\rm A}$, ${V}_{\rm SA}$, and ${V}_{\rm SF}$ 
are the anisotropy spin Hamiltonian, 
spin-atomic vibration interaction, 
and spin-flip Hamiltonian, respectively. 
These terms 
are written by the following simplified expressions: 
\begin{eqnarray}
\label{V_A}
&&\hspace*{-0.5cm}
{V}_{\rm A} \equiv 
{v}_{\mbox{\footnotesize so0,so0,1}}^{(2)}
=
D S_Z^2 + E (S_X^2 - S_Y^2), \\
\label{V_SA}
&&\hspace*{-0.5cm}
{V}_{\rm SA} \equiv 
{V}_{\mbox{\footnotesize so2}}^{(1)}
+{v}_{\mbox{\footnotesize so0,so1,1}}^{(2)}
+{v}_{\mbox{\footnotesize so0,so1,2}}^{(2)}
+{V}_{\mbox{\footnotesize so0,so2}}^{(2)}
+{V}_{\mbox{\footnotesize so0,k}}^{(2)} \nonumber \\
&&\hspace*{0.2cm}=
\sum_{I=X,Y,Z}
\sum_{\xi=x,y,z}
c_{I,\xi} S_I \left( -a_\xi + a_\xi^\dag \right) \nonumber \\
&&\hspace*{0.6cm}+ \sum_{I,J=X,Y,Z} \sum_{\xi=x,y,z} S_I S_J 
\left(c_{1,I,J,\xi} a_\xi 
+ c_{2,I,J,\xi} a_\xi^\dag \right) \nonumber
\\
&&\hspace*{0.6cm}
+ \left[ D S_Z^2 + E (S_X^2 - S_Y^2) + FS(S+1)\right] 
\sum_{\xi=x,y,z} G_\xi (a_\xi + a_\xi^\dag), \\
\label{V_SF}
&&\hspace*{-0.5cm}
{V}_{\rm SF} \equiv {v}_{\mbox{\footnotesize so0,so0,2}}^{(2)}
=\sum_{I,J=X,Y,Z} \Lambda_{I,J}^{\rm SF} S_I S_J, 
\end{eqnarray}
where
$c_{I,\xi}$ is 
the coefficient of the $S_I (- a_\xi + a_\xi^\dag)$ term, 
and 
$c_{1,I,J,\xi}$ ($c_{2,I,J,\xi}$) is 
that of the $S_I S_J a_\xi$ term (that of the $S_I S_J a_\xi^\dag$ term). 
Here, 
$c_{I,\xi}$ consists of 
$\Lambda_{\mu,\xi}^{(1)}$ of eq. (\ref{L_mxi^(1)}) 
and $\Lambda_{\mu,\xi}^k$ of eq. (\ref{Lam^k}), 
while $c_{n,I,J,\xi}$ ($n$=1, 2) 
is composed of 
$\Lambda_{I,J;\xi}'$ of eq. (\ref{Lam_IJxi}) 
and $\Lambda_{\mu;\nu,\xi}^{\pm}$ of eq. (\ref{lam_d}). 
In other words, 
${V}_{\mbox{\footnotesize so2}}^{(1)}$ and 
${V}_{\mbox{\footnotesize so0,k}}^{(2)}$ are 
terms with $S_I (-a_\xi + a_\xi^\dag)$, 
while 
${v}_{{\mbox{\footnotesize so0,so1}},2}^{(2)}$ 
and ${V}_{\mbox{\footnotesize so0,so2}}^{(2)}$ 
are those with $S_IS_J a_\xi$ and $S_IS_J a_\xi^\dag$.

We consider the features of 
${V}_{\mbox{\footnotesize so2}}^{(1)}$, 
${V}_{\mbox{\footnotesize so0,k}}^{(2)}$, 
${v}_{{\mbox{\footnotesize so0,so1}},1}^{(2)}$, 
${v}_{{\mbox{\footnotesize so0,so1}},2}^{(2)}$, 
${V}_{\mbox{\footnotesize so0,so2}}^{(2)}$ of ${V}_{\rm SA}$, 
and 
${v}_{{\mbox{\footnotesize so0,so0}},2}^{(2)}$ of ${V}_{\rm SF}$. 
We first find that 
${V}_{\mbox{\footnotesize so2}}^{(1)}$ and 
${V}_{\mbox{\footnotesize so0,k}}^{(2)}$ are proportional to $\lambda$, 
while 
${v}_{{\mbox{\footnotesize so0,so1}},1}^{(2)}$, 
${v}_{{\mbox{\footnotesize so0,so1}},2}^{(2)}$, 
${V}_{\mbox{\footnotesize so0,so2}}^{(2)}$, and 
${v}_{{\mbox{\footnotesize so0,so0}},2}^{(2)}$ 
are proportional to $\lambda^2$ (see Table \ref{tab1}). 
Second, the presence or absence of the respective terms 
is shown by examining the respective coefficients 
for each set of $\eta$, $c_p$, $c_d$, 
where 
$c_d$ and $c_p$ are defined as 
$c_d$=$c_{d_m}^{(\zeta)}$ and $c_p$=$c_{p_n}^{(\zeta)}$, 
respectively (see Table \ref{tab1}). 
Here, $\omega_\xi$ is put to be 
$\omega_\xi$=2$\pi f_\xi$ for $\xi$=$x$, $y$, $z$, 
where $f_\xi$ is the vibration frequency in the $\xi$ direction. 
In addition, 
$e_1-e_j$=$E_1-E_j$ is roughly set 
by taking into account $|\Delta e_1-\Delta e_j|$$\ll$$|e_1-e_j|$ 
for $j \ne 1$ (see eq. (\ref{E_j})). 
The details are described as follows: 
\begin{itemize}
\item 
Spin-atomic vibration interactions ${V}_{\rm SA}$\\
\vspace{0.01cm}\\
$S_I (- a_\xi + a_\xi^\dag)$ term 
\begin{itemize}
\item ${V}_{\mbox{\footnotesize so2}}^{(1)}$ with 
$\Lambda_{\mu,\xi}^{(1)} \propto \eta \sqrt{f_\xi}$~~
(see eq. (\ref{L_mxi^(1)}))\\
The interaction ${V}_{\mbox{\footnotesize so2}}^{(1)}$ 
appears for 
$\eta$$\ne$0 and $c_p$$\ne$0, 
because 
each term contains $\eta(\Psi_{j'} | \mu_t |\Psi_j )$, 
with $\mu$=$x$, $y$, $z$. 
Here, $(\Psi_{j'} | \mu_t |\Psi_j )$ 
is the same formula as 
a matrix element of the so-called electric dipole transition; 
it is nonzero for $c_p$$\ne$0. 
In addition, ${V}_{\mbox{\footnotesize so2}}^{(1)}$ is 
proportional to $\sqrt{f_\xi}$, reflecting that 
${V}_{\mbox{\footnotesize so2}}^{(1)}$ contains 
$\Delta p_{{\rm n},\xi}$ of eq. (\ref{Delta_pn}), with $\xi$=$x$, $y$, $z$. 
\item ${V}_{\mbox{\footnotesize so0,k}}^{(2)}$ 
with $\Lambda_{\mu,\xi}^k \propto \eta \sqrt{f_\xi}$~~
(see eq. (\ref{Lam^k}))\\
The interaction ${V}_{\mbox{\footnotesize so0,k}}^{(2)}$, 
in which each term contains $\eta (\Psi_{j} | \xi_t |\Psi_1 )$ 
with $\xi$=$x$, $y$, $z$, 
is present for 
$\eta$$\ne$0 and $c_p$$\ne$0. 
In addition, ${V}_{\mbox{\footnotesize so0,k}}^{(2)}$ is 
proportional to $\sqrt{f_\xi}$ 
because ${V}_{\mbox{\footnotesize so0,k}}^{(2)}$ 
(or ${V}_{\mbox{\footnotesize k}}'$) 
contains 
$\Delta p_{{\rm n},\xi}$. 
\end{itemize}
$S_IS_J a_\xi$ and $S_IS_J a_\xi^\dag$ terms 
\begin{itemize}
\item 
${V}_{\mbox{\footnotesize so0,so2}}^{(2)}$ 
with 
$\Lambda_{\mu;\nu,\xi}^{\pm} \propto \eta$~~
(see eq. (\ref{lam_d}))\\
The interaction ${V}_{\mbox{\footnotesize so0,so2}}^{(2)}$ appears for 
$\eta$$\ne$0 and $c_p$$\ne$0, 
because 
each term has $\eta (\Psi_{j'} | \nu_t |\Psi_1 )$, 
with $\nu$=$x$, $y$, $z$. 
In addition, ${V}_{\mbox{\footnotesize so0,so2}}^{(2)}$ 
contains both terms with $\sqrt{f_\xi}$ and those with $1/\sqrt{f_\xi}$, 
reflecting that 
${V}_{\mbox{\footnotesize so0,so2}}^{(2)}$ 
(or ${V}_{\mbox{\footnotesize so2}}$) 
has 
$\Delta p_{{\rm n},\xi}$ and 
$\Delta \xi_{\rm n}$ of eq. (\ref{Delta_rn}), with $\xi$=$x$, $y$, $z$. 
\item ${v}_{{\mbox{\footnotesize so0,so1}},1}^{(2)}$ 
with 
$G_\xi \propto \eta \langle 3 \xi r^{-5} \rangle /\sqrt{f_\xi}$~~
(see eq. (\ref{G_xi})) \\
The interaction ${v}_{{\mbox{\footnotesize so0,so1}},1}^{(2)}$ 
exists for $\eta$$\ne$0 and $c_p$$\ne$0, 
because it is proportional to 
$\eta \langle 3 \xi r^{-5} \rangle$, 
with $\xi$=$x$, $y$, $z$. 
Here, $\langle 3 \xi r^{-5} \rangle$ is nonzero for $c_p$$\ne$0. 
It is noted that the operator is proportional to $\xi$ 
as with 
the matrix element of the electric dipole transition. 
In addition, ${v}_{{\mbox{\footnotesize so0,so1}},1}^{(2)}$ is 
proportional to $1/\sqrt{f_\xi}$ 
because ${v}_{{\mbox{\footnotesize so0,so1}},1}^{(2)}$ 
(or ${V}_{\mbox{\footnotesize so1}}$) 
contains 
$\Delta \xi_{\rm n}$. 
\item ${v}_{{\mbox{\footnotesize so0,so1}},2}^{(2)}$ 
with 
$\Lambda_{I,J;\xi}' \propto 
\eta \langle 3 \xi r^{-5} \rangle {\cal L}_{I,J}' /\sqrt{f_\xi}$~~
(see eq. (\ref{Lam_IJxi})) \\ 
The interaction 
${v}_{{\mbox{\footnotesize so0,so1}},2}^{(2)}$ is present 
for $\eta$$\ne$0, $c_p$$\ne$0, and $c_d$$\ne$0, 
because it is proportional to 
$\eta \langle 3 \xi r^{-5} \rangle {\cal L}_{I,J}'$ with $\xi$=$x$, $y$, $z$. 
Here, ${\cal L}_{I,J}'$ is nonzero for $c_d$$\ne$0. 
In addition, ${v}_{{\mbox{\footnotesize so0,so1}},2}^{(2)}$ is 
proportional to $1/\sqrt{f_\xi}$, 
reflecting that 
${v}_{{\mbox{\footnotesize so0,so1}},2}^{(2)}$ 
(or ${V}_{\mbox{\footnotesize so1}}$) 
has 
$\Delta \xi_{\rm n}$. 
\end{itemize}
\end{itemize}
\begin{itemize}
\item Spin-flip Hamiltonian ${V}_{\rm SF}$
\begin{itemize}
\item ${v}_{{\mbox{\footnotesize so0,so0}},2}^{(2)}$ 
with 
$\Lambda_{I,J}^{\rm SF} \propto {\cal L}_{I,J}'$~~
(see eq. (\ref{Lam^SF}))\\
The Hamiltonian ${v}_{{\mbox{\footnotesize so0,so0}},2}^{(2)}$, 
which is proportional to ${\cal L}_{I,J}'$, 
exists for $c_d\ne$0. 
Note that 
${v}_{{\mbox{\footnotesize so0,so0}},2}^{(2)}$ 
is independent of the atomic vibration, 
and it does not contain $\eta$ and $f_\xi$. 
\end{itemize}
\end{itemize}


\begin{table}[ht]
\caption{
We show the presence or absence of 
the spin-atomic vibration interactions ${V}_{\mbox{\footnotesize so2}}^{(1)}$, 
${V}_{\mbox{\footnotesize so0,k}}^{(2)}$, 
${v}_{{\mbox{\footnotesize so0,so1}},1}^{(2)}$, 
${v}_{{\mbox{\footnotesize so0,so1}},2}^{(2)}$, and 
${V}_{\mbox{\footnotesize so0,so2}}^{(2)}$, 
and 
the spin-flip Hamiltonian ${v}_{{\mbox{\footnotesize so0,so0}},2}^{(2)}$ 
for each set of $\eta$, $c_p$, $c_d$. 
Here, 
$e_1-e_j$=$E_1-E_j$ is set 
by taking into account $|\Delta e_1-\Delta e_j|$$\ll$$|e_1-e_j|$ 
for $j \ne 1$. 
The spin-atomic vibration interaction 
(spin-flip Hamiltonian) is named SA (SF). 
The presence and absence are represented by 
$\surd$ and $\times$, respectively. 
The operators and coefficients are also written for reference. 
The coefficient 
$\Lambda_{\mu;\nu,\xi}^{\pm}$ 
contains both terms with $1/\sqrt{f_\xi}$ and those with $\sqrt{f_\xi}$. 
The coefficient 
$\Lambda_{I,J}^{\rm SF}$ does not contain $\eta$, $f_\xi$, and $M$. 
}
\begin{tabular}{lllllll} 
\hline \\[-0.6cm]
 Name  & SA  & SA & SA & SA & SA & SF \\
\cline{1-7}
\\[-0.3cm]
 Energy & 
${V}_{\mbox{\footnotesize so2}}^{(1)}$ 
& ${V}_{\mbox{\footnotesize so0,k}}^{(2)}$ &
${v}_{{\mbox{\footnotesize so0,so1}},1}^{(2)}$ 
& ${v}_{{\mbox{\footnotesize so0,so1}},2}^{(2)}$ 
& ${V}_{\mbox{\footnotesize so0,so2}}^{(2)}$ 
& ${v}_{{\mbox{\footnotesize so0,so0}},2}^{(2)}$\\
\\[-0.3cm]
& eq. (\ref{vv_so2_1}) 
& eq. (\ref{vv_so0k})
& eq. (\ref{v_Sv^(1)})
& eq. (\ref{v_Sv^(2)})
& eq. (\ref{vv_so0so2})
& eq. (\ref{v_S^(2)})
\\
\\[-0.5cm]
 Operator & 
$S_I (-a_\xi + a_\xi^\dag)$ 
& $S_I (-a_\xi + a_\xi^\dag)$  &
$S_I^2 (a_\xi + a_\xi^\dag)$  
& $S_I S_J (a_\xi + a_\xi^\dag)$ 
& $S_I S_J a_\xi$ 
& $S_I S_J$ \\
& & & &  &$S_I S_J a_\xi^\dag$ & \\
\\[-0.5cm]
\\[-0.5cm]
 Coefficient & 
$\Lambda_{\mu,\xi}^{(1)}$
& $\Lambda_{\mu,\xi}^k$ &
$G_\xi D$
& $\Lambda_{I,J;\xi}'$ 
& $\Lambda_{\mu;\nu,\xi}^{\pm}$ 
& $\Lambda_{I,J}^{\rm SF}$\\
 & &  &$G_\xi E$ & & & \\
 & &  &$G_\xi F$ & & & \\
\\[-0.5cm]
 & $\propto \eta \lambda \displaystyle{\sqrt{\frac{f_\xi}{M}}}$ 
& $\propto \eta \lambda \displaystyle{\sqrt{\frac{f_\xi}{M}}}$ & 
$\propto \displaystyle{\frac{\eta \lambda^2}{\sqrt{M f_\xi}}}$ & 
$\propto \displaystyle{\frac{\eta \lambda^2}{\sqrt{M f_\xi}}}$ & 
$\propto \displaystyle{\frac{\eta \lambda^2}{\sqrt{M}}}$ &  
$\propto \lambda^2$ \\
\\[-0.3cm]
& eq. (\ref{L_mxi^(1)}) 
& eq. (\ref{Lam^k})
& eq. (\ref{G_xi})
& eq. (\ref{Lam_IJxi})
& eq. (\ref{lam_d})
& eq. (\ref{Lam^SF})\\
\\[-0.4cm]
\hline \\[-0.4cm]
$\eta$=0  &  &  &  &  &  &  \\
$c_d$=0, $c_p$=0 & $\times$ & $\times$ & $\times$ & $\times$ & $\times$ & $\times$ \\
  $c_d$=0, $c_p$$\ne$0 & $\times$ & $\times$ & $\times$ & $\times$ & $\times$ & $\times$ \\
  $c_d$$\ne$0, $c_p$=0 & $\times$ & $\times$ & $\times$ & $\times$ & $\times$ & $\surd$ \\
  $c_d$$\ne$0, $c_p$$\ne$0 & $\times$ & $\times$ & $\times$ & $\times$ & $\times$ & $\surd$ \\ \\[-0.3cm] \hline \\[-0.4cm]
$\eta$$\ne$0  &  &  &  &  &  &  \\
$c_d$=0, $c_p$=0 & $\times$ & $\times$ & $\times$ & $\times$ & $\times$ & $\times$ \\
  $c_d$=0, $c_p$$\ne$0 & $\surd$ & $\surd$ & $\surd$ & $\times$ & $\surd$ & $\times$ \\
  $c_d$$\ne$0, $c_p$=0 & $\times$ & $\times$ & $\times$ & $\times$ & $\times$ & $\surd$ \\
  $c_d$$\ne$0, $c_p$$\ne$0 & $\surd$ & $\surd$ & $\surd$ & $\surd$ & $\surd$ & $\surd$ \\ \\[-0.3cm]\hline 
\end{tabular}
\label{tab1}
\end{table}


\begin{figure}[ht]
\begin{center}
\includegraphics[width=0.35\linewidth]{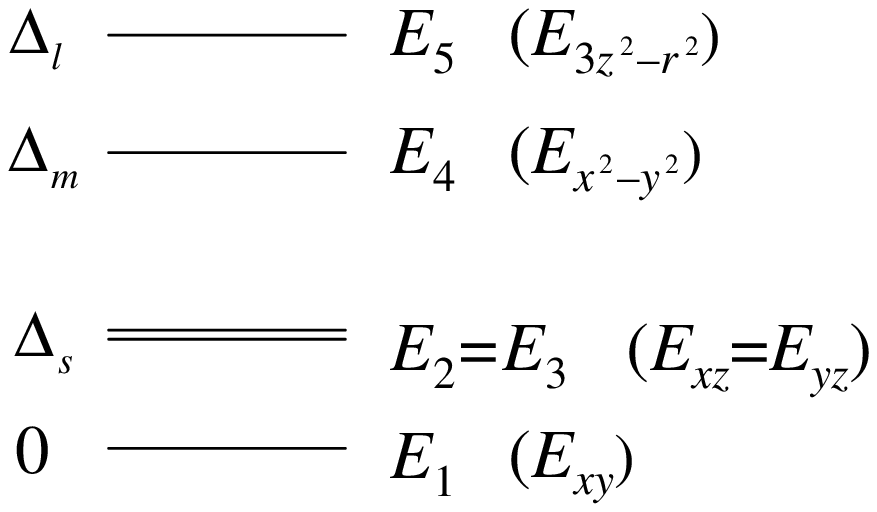}
\caption{
Energy level $E_j$ of 
the down-spin shell of the 3d orbitals of the Fe$^{2+}$ (3d$^6$) 
in the crystal field of tetragonal symmetry. 
We use the notation of 
$E_1$=$E_{xy}$, 
$E_2$=$E_{xz}$, 
$E_3$=$E_{yz}$, 
$E_4$=$E_{x^2 -y^2}$, and 
$E_5$=$E_{3z^2 -r^2}$, 
where $xy$, $xz$, $yz$, $x^2 -y^2$, and $3z^2-r^2$ correspond to 
$d_1$, $d_2$, $d_3$, $d_4$, and $d_5$, respectively 
(see eqs. (\ref{xyf32}) - (\ref{3z2f32})). 
Here, the relation of $E_1<E_2~(=E_3) < E_4 < E_5$ is considered, 
where the second excited states 
are doubly degenerate, i.e., $E_2$=$E_3$. 
In addition, 
$\Delta_s$, $\Delta_m$, and $\Delta_l$ represent 
$E_2 - E_1$, $E_4 - E_1$, and $E_5 - E_1$, respectively. 
}
\label{energy}
\end{center}
\end{figure}

\section{Application to F\lowercase{e} Ion} 	
\label{Application}
As an application, 
we consider the spin-atomic vibration interaction 
and the spin-flip Hamiltonian of the Fe ion (Fe$^{2+}$) 
in a crystal field of tetragonal symmetry, 
where Fe$^{2+}$ has six 3d electrons (i.e., 3d$^6$). 
We first investigate the $f$ dependence of their coefficients. 
Second, their coefficients are evaluated 
by assigning appropriate values to $\lambda$.

\subsection{Orbital and spin states}
As to the Fe ion (3d$^6$), 
the ground state consists of five up-spin electrons 
and one down-spin electron, 
according to Hund's first rule. 
The spin $S$ is therefore considered to be $S$=2. 
In addition, since the up-spin shell is filled, 
$\Psi_j$ of eq. (\ref{Psi_j}) is given by 
the orbital wave function 
of the down-spin electron, i.e., 
\begin{eqnarray}
\label{dj}
&&
\Psi_j = \psi_j 
= C_j \left( 
d_j 
+ \sum_{m=1 (\ne j)}^5 c_{d_m}^{(j)} \overline{d_m } 
+ \sum_{n=1}^3 c_{p_n}^{(j)} \overline{ p_n } 
\right), \\ 
\label{C_j}
&&
C_j=
\left( 
1 
+ \sum_{m=1 (\ne j)}^5 |c_{d_m}^{(j)}|^2 
+ \sum_{n=1}^3 |c_{p_n}^{(j)}|^2 
\right)^{-1/2}. 
\end{eqnarray}
Equation (\ref{Phi_j}) also becomes 
\begin{eqnarray}
\Phi_j = d_j.
\end{eqnarray}
The energy of $\Psi_j$ is given by 
$E_j$ = $e_j  + \Delta e_j$ (see eq. (\ref{E_j})), 
while that of $\Phi_j$ is $e_j$ (see eq. (\ref{e_d_j})). 
Here, $E_j$ is assumed to be $E_j$ in Fig. \ref{energy} 
by taking into account the crystal field of tetragonal symmetry. 
The wave functions $d_j$ and $p_j$ are then written as 
\begin{eqnarray}
\label{xyf32}
&&\hspace*{-0.5cm} d_1=xy f_{3,2}(r), \\
&&\hspace*{-0.5cm} d_2=xz f_{3,2}(r), \\
&&\hspace*{-0.5cm} d_3=yz f_{3,2}(r), \\
&&\hspace*{-0.5cm} d_4=\frac{1}{2} (x^2 -y^2) f_{3,2}(r), \\
\label{3z2f32}
&&\hspace*{-0.5cm} d_5=\frac{1}{2 \sqrt{3}}(3z^2 -r^2) f_{3,2}(r), 
\end{eqnarray}
and
\begin{eqnarray}
&&\hspace*{-0.5cm} p_1=x f_{4,1}(r), \\
&&\hspace*{-0.5cm} p_2=y f_{4,1}(r), \\
\label{zf31}
&&\hspace*{-0.5cm} p_3=z f_{4,1}(r), 
\end{eqnarray}
respectively. 
The function $f_{n,l}(r)$ with the principal quantum number $n$ 
and the azimuthal quantum number $l$ is given by
\begin{eqnarray}
&&\hspace*{-0.5cm}
f_{3,2}(r)= \frac{2}{81 \sqrt{2 \pi} a_{\mbox{\tiny B,d}}^{7/2}} 
\exp \left( - \frac{r}{3a_{\mbox{\tiny B,d}}} \right), \\
&&\hspace*{-0.5cm}
f_{4,1}(r)= 
\frac{1}{512 \sqrt{5 \pi} a_{\mbox{\tiny B,p}}^{5/2}}
\left[ 80 - \frac{20r}{a_{\mbox{\tiny B,p}}} +
\left( \frac{r}{a_{\mbox{\tiny B,p}}} \right)^2 \right]
\exp \left(- \frac{r}{4a_{\mbox{\tiny B,p}}} \right), \\
&&\hspace*{-0.5cm}
a_{\mbox{\tiny B,d}} = \frac{a_{\mbox{\tiny B}}}{Z_{\rm eff,d}}, \\
&&\hspace*{-0.5cm}
a_{\mbox{\tiny B,p}} = \frac{a_{\mbox{\tiny B}}}{Z_{\rm eff,p}}, 
\end{eqnarray}
with 
$a_{\mbox{\tiny B}}$=$\epsilon_0 h^2/(\pi m e^2)$, 
where $a_{\mbox{\tiny B}}$ is the Bohr radius 
and 
$Z_{\rm eff,d}$ ($Z_{\rm eff,p}$) is 
the effective nuclear charge for the 3d electron 
(4p electron).

\subsection{Notation and parameter setup}
\label{notation}
In this system, 
we first replace 
$\mbox{\boldmath $r$}_t$=$(x_t, y_t, z_t)$ 
and $\mbox{\boldmath $p$}_t$=$(p_{t,x}, p_{t,y}, p_{t,z})$ 
with 
$(x, y, z)$ and $(p_x, p_y, p_z)$ 
of the one-electron system, respectively. 
The angular frequency $\omega_\xi$ is set to be 
$\omega_x$=$\omega_y$=$\omega_z$=2$\pi f$, 
where $f$ is a vibration frequency. 
For eqs. (\ref{dj}) and (\ref{C_j}), 
we put 
$c_{d_m}^{(j)}$=$c_d$ for $m$=1 - 5 with $m \ne j$, 
$c_{p_n}^{(j)}$=$c_p$ 
for $n$=1 - 3, 
and $C_j$=$(1 + 4 c_d^2 + 3 c_p^2)^{-1/2}$$\equiv$ $C$. 
In addition, 
we use the notation of 
$e_1$=$e_{xy}$, 
$e_2$=$e_{xz}$, 
$e_3$=$e_{yz}$, 
$e_4$=$e_{x^2 -y^2}$, 
$e_5$=$e_{3z^2 -r^2}$, 
$E_1$=$E_{xy}$, 
$E_2$=$E_{xz}$, 
$E_3$=$E_{yz}$, 
$E_4$=$E_{x^2 -y^2}$, and 
$E_5$=$E_{3z^2 -r^2}$. 
The energy difference is then expressed as 
$E_{xz} - E_{xy}$=$E_{yz} - E_{xy}$$\equiv$$\Delta_s$ ($>0$), 
$E_{x^2 - y^2} - E_{xy}$$\equiv$$\Delta_m$ ($>0$), 
and $E_{3z^2 - r^2} - E_{xy}$$\equiv$$\Delta_l$ ($>0$) 
(see Fig. \ref{energy}). 
The energy difference 
$e_1-e_j$=$E_1-E_j$ is roughly set 
by taking into account $|\Delta e_1-\Delta e_j|$$\ll$$|e_1-e_j|$ 
for $j \ne 1$.

As a parameter set, we choose 
$c_d$=$c_p$=0.15,\cite{parameter} 
$\Delta_s$=$0.45\Delta_m$, $\Delta_l$=$1.45\Delta_m$, 
$\Delta_m$=0.5 eV, 
and 
$f$=0.01 THz - 100 THz, 
which correspond to typical atomic vibration frequencies.\cite{vibration1} 
The mass of Fe$^{2+}$, $M$, is 
$M$=(26+30)$\times$(1.67$\times$10$^{-27}$)+24$\times$(9.11$\times$10$^{-31}$) kg, 
where the number of protons (neutrons) is 26 (30), 
the mass of the proton (neutron) is 
1.67$\times$10$^{-27}$ kg (1.67$\times$10$^{-27}$ kg), 
the number of electrons is 24, 
and the electron mass is 9.11$\times$10$^{-31}$ kg. 
The effective nuclear charge $Z_{\rm eff,d}$ ($Z_{\rm eff,p}$) 
is evaluated to be 
$Z_{\rm eff,d}$=6.25 
($Z_{\rm eff,p}$=4.10) 
according to Slater's rules.\cite{Slater}

Note that from now on we will use 
the coordinate system $x$, $y$, and $z$ for some coefficients 
in accordance with the notation in \S \ref{Model}. 
Actually, however, 
$x$=$X$, $y$=$Y$, and $z$=$Z$ are realized in this system, 
as described in Appendix \ref{app_theta}. 

\begin{figure}[ht]
\begin{center}
\includegraphics[width=0.5\linewidth]{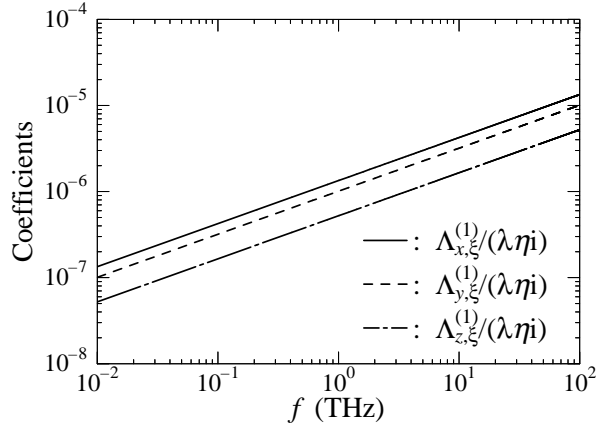}
\caption{
Vibration frequency $f$ dependence 
of $\Lambda_{\mu,\xi}^{(1)}/(\lambda \eta {\rm i})$ 
for $\mu$, $\xi$= $x$, $y$, $z$. 
The parameters are set to be 
$c_d$=$c_p$=0.15 and $\Delta_m$=0.5 eV. 
}
\label{a1_f}
\end{center}
\end{figure}

\begin{figure}[ht]
\begin{center}
\includegraphics[width=0.5\linewidth]{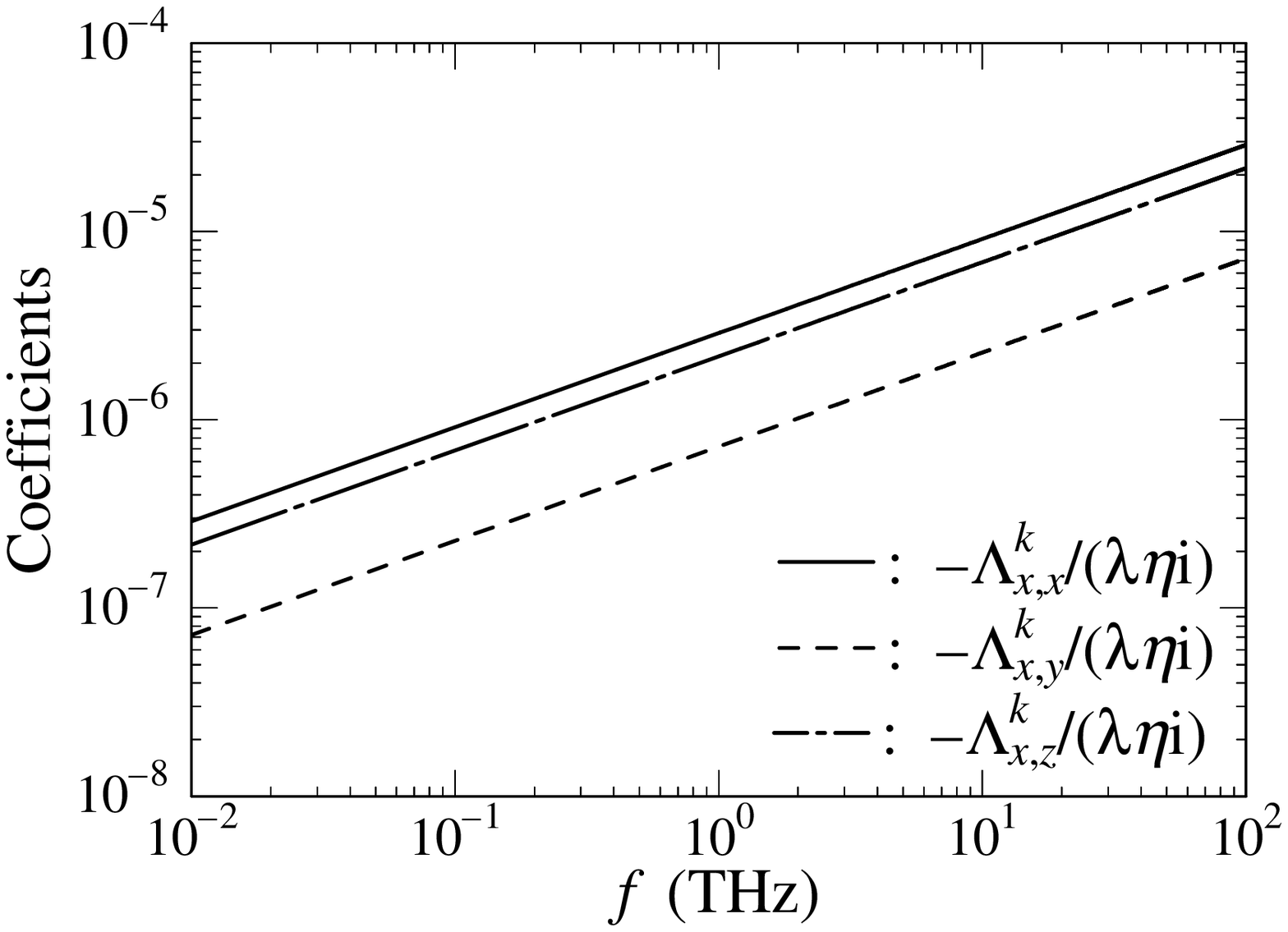}\\
\includegraphics[width=0.5\linewidth]{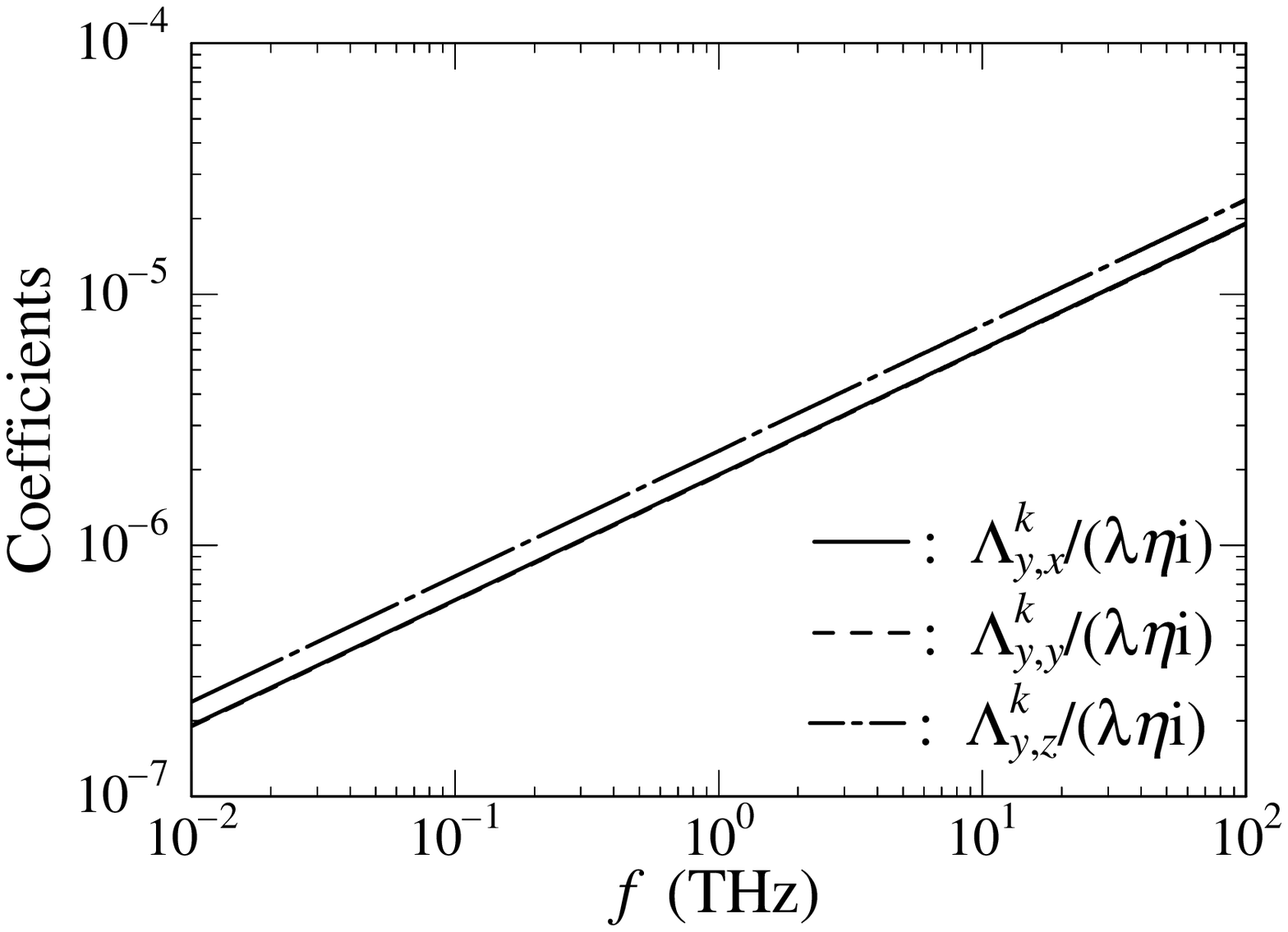}\\
\includegraphics[width=0.5\linewidth]{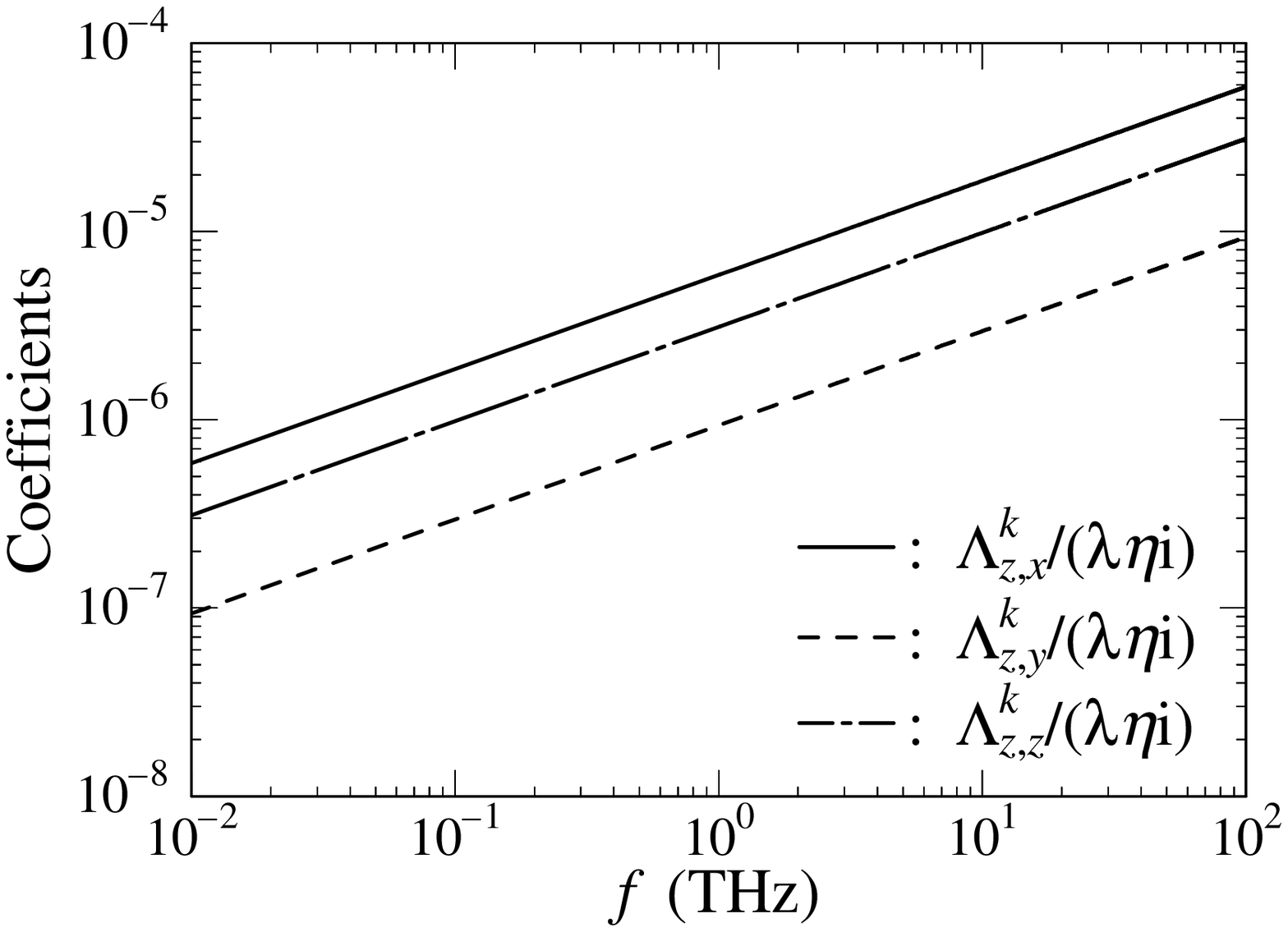}
\caption{
Vibration frequency $f$ dependences of the coefficients. 
Upper panel: $\Lambda_{x,\nu}^k/(\lambda \eta {\rm i})$ 
for $\nu$=$x$, $y$, $z$. 
Middle panel: $\Lambda_{y,\nu}^k/(\lambda \eta {\rm i})$ 
for $\nu$=$x$, $y$, $z$. 
Lower panel: $\Lambda_{z,\nu}^k/(\lambda \eta {\rm i})$ 
for $\nu$=$x$, $y$, $z$. 
Note that 
$\Lambda_{y,x}^k/(\lambda \eta {\rm i})$ takes almost the same value as 
$\Lambda_{y,y}^k/(\lambda \eta {\rm i})$ at each $f$. 
The parameters are set to be 
$c_d$=$c_p$=0.15 and $\Delta_m$=0.5 eV. 
}
\label{akx_f}
\end{center}
\end{figure}


\begin{figure}[ht]
\begin{center}
\includegraphics[width=0.5\linewidth]{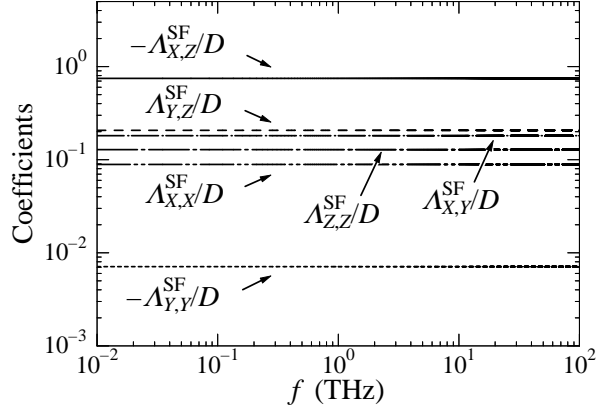}
\caption{
Vibration frequency $f$ dependence of $\Lambda_{I,J}^{\rm SF}/D$ 
for $I, J$= $X$, $Y$, $Z$, where 
$\Lambda_{I,J}^{\rm SF}$=$\Lambda_{J,I}^{\rm SF}$ 
(see \S \ref{v_so0so0}). 
The coefficients take the respective constant values. 
The parameters are set to be 
$c_d$=$c_p$=0.15 and $\Delta_m$=0.5 eV. 
}
\label{sf_f}
\end{center}
\end{figure}


\begin{figure}[ht]
\begin{center}
\includegraphics[width=0.5\linewidth]{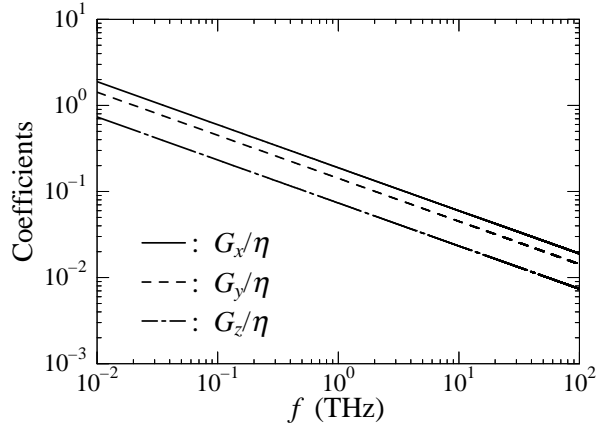}
\caption{
Vibration frequency $f$ dependence 
of $G_\xi/\eta$ for $\xi$=$x$, $y$, $z$. 
The parameters are set to be 
$c_d$=$c_p$=0.15 and $\Delta_m$=0.5 eV. 
}
\label{g_f}
\end{center}
\end{figure}


\begin{figure}[ht]
\begin{center}
\includegraphics[width=0.5\linewidth]{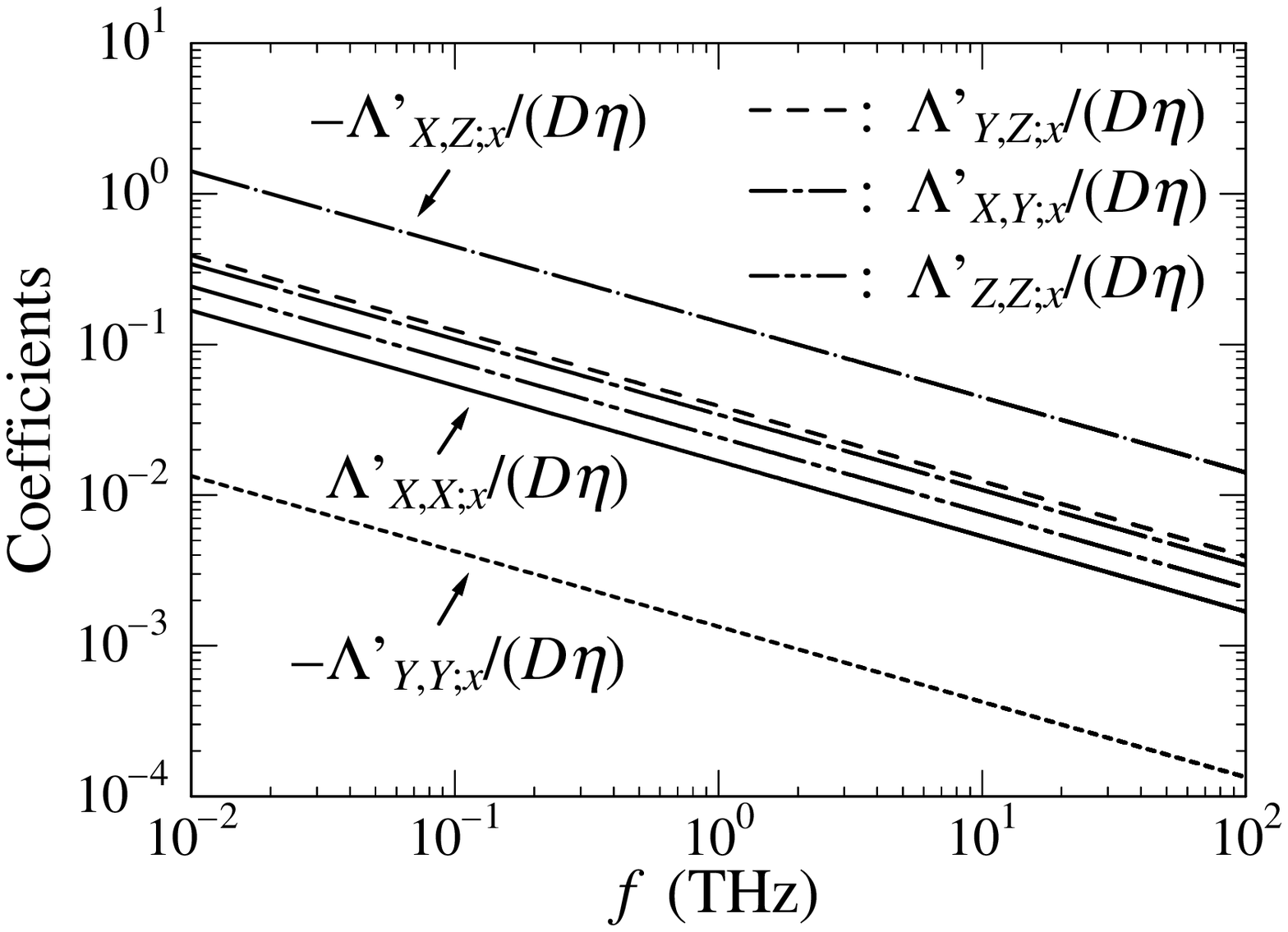}\\
\includegraphics[width=0.5\linewidth]{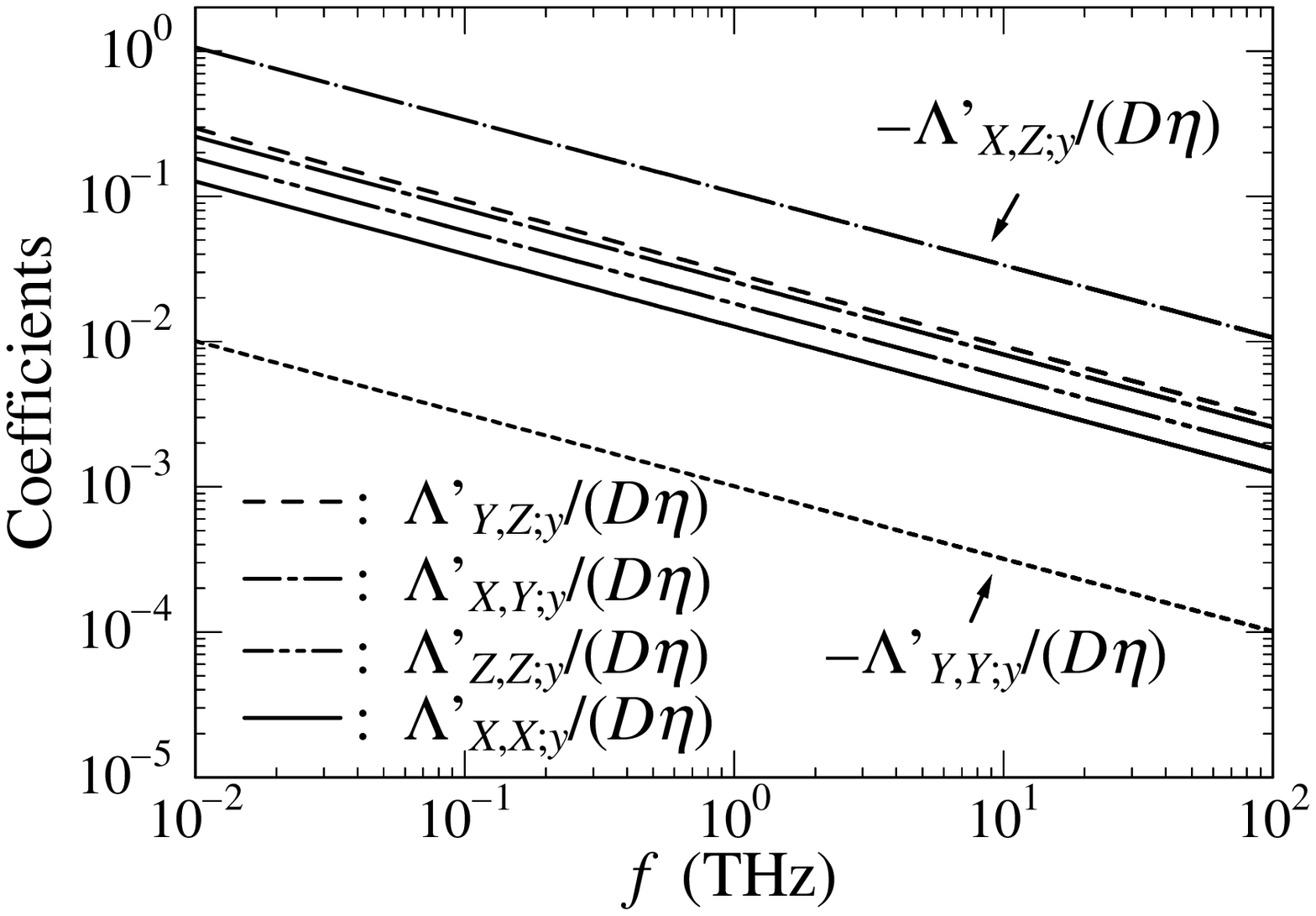}\\
\includegraphics[width=0.5\linewidth]{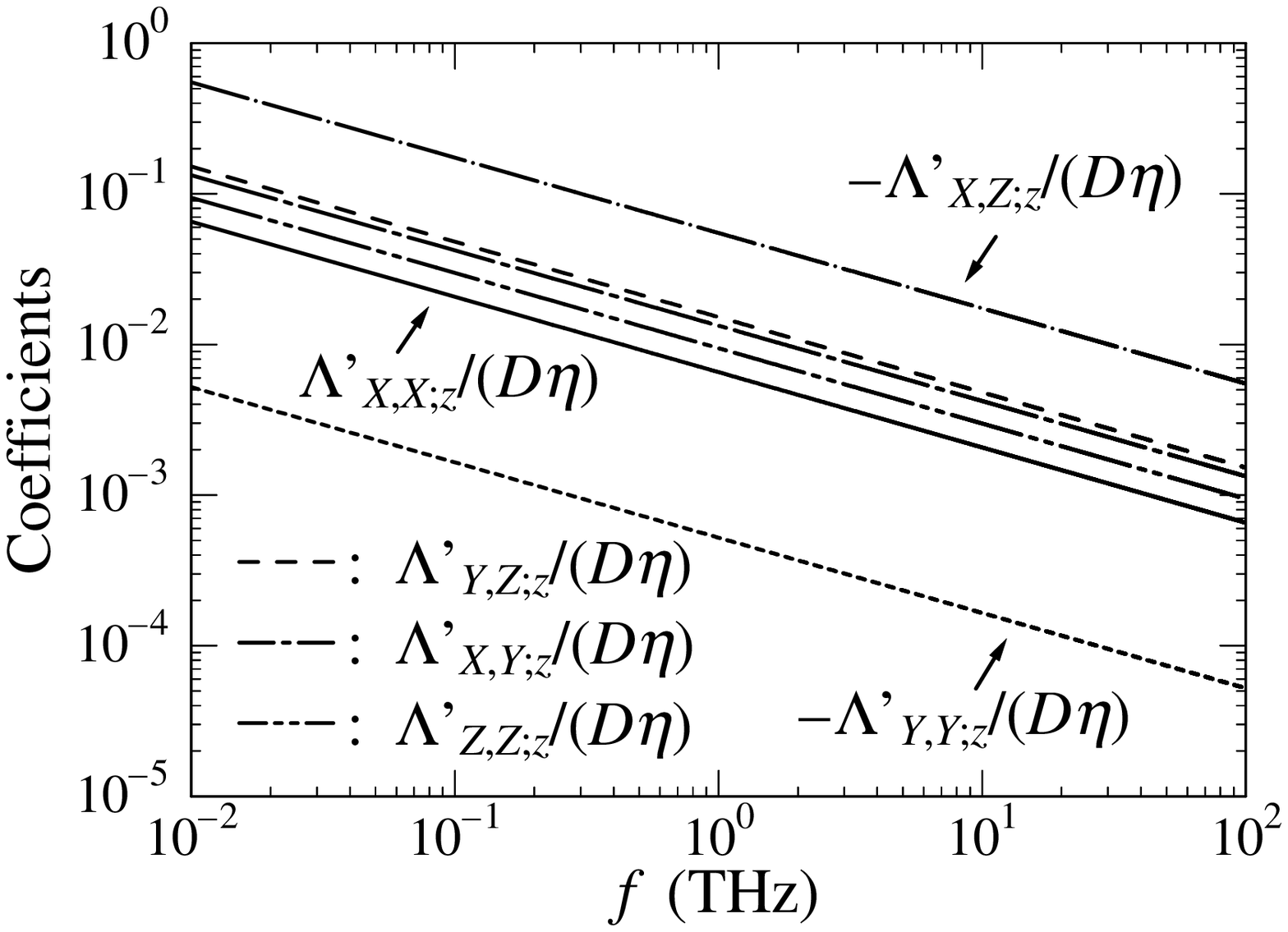}
\caption{
Vibration frequency $f$ dependences of the coefficients. 
Upper panel: $\Lambda_{I,J;x}'/(D \eta)$ for $I$, $J$=$X$, $Y$, $Z$. 
Middle panel: $\Lambda_{I,J;y}'/(D \eta)$ for $I$, $J$=$X$, $Y$, $Z$. 
Lower panel: $\Lambda_{I,J;z}'/(D \eta)$ for $I$, $J$=$X$, $Y$, $Z$. 
We have $\Lambda_{I,J;\xi}'$=$\Lambda_{J,I;\xi}'$ 
(see \S \ref{v_so1o1}). 
The parameters are set to be 
$c_d$=$c_p$=0.15 and $\Delta_m$=0.5 eV. 
}
\label{ad_x_f}
\end{center}
\end{figure}

\begin{figure}[ht]
\begin{center}
\includegraphics[width=0.5\linewidth]{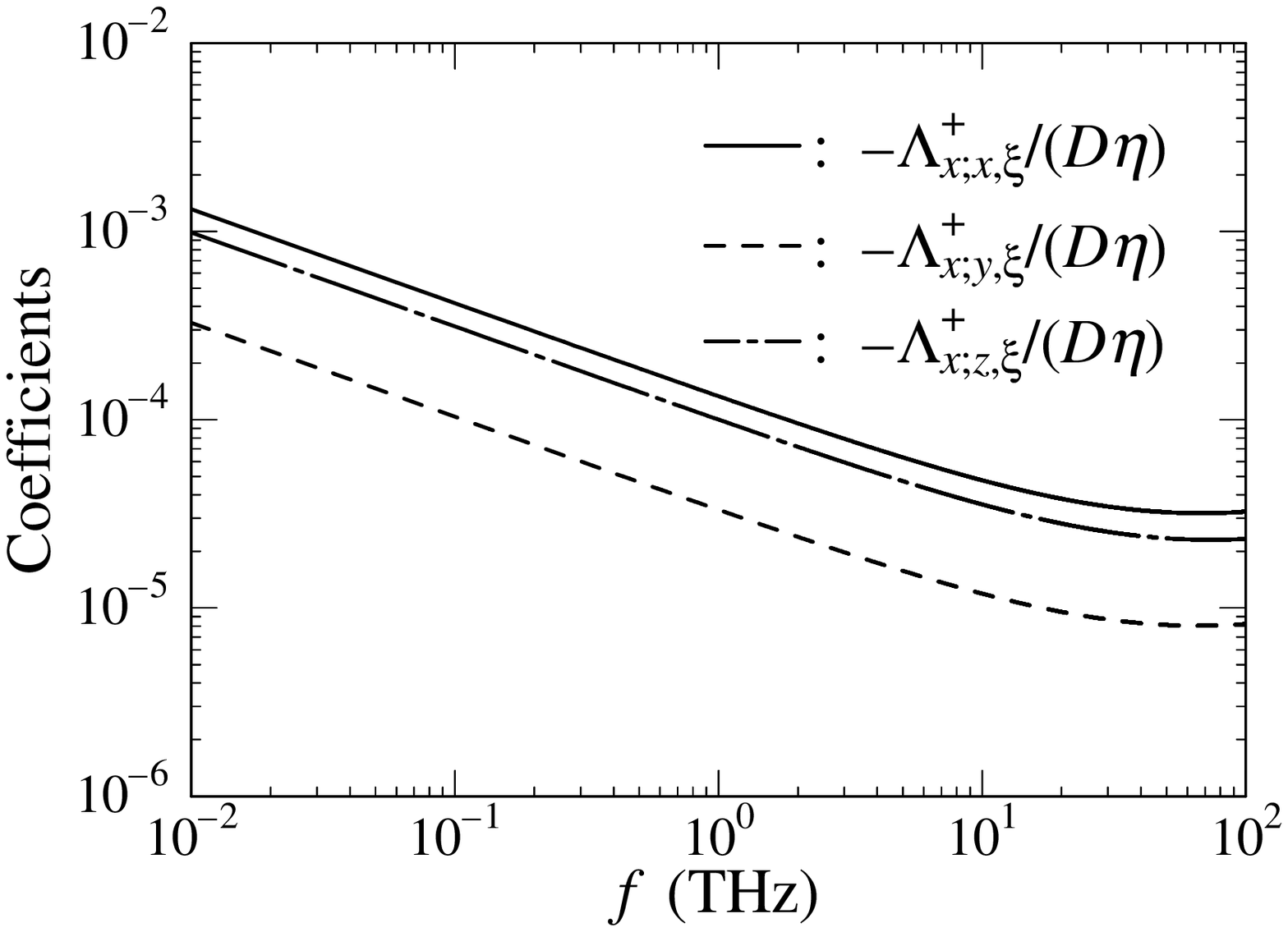}\\
\includegraphics[width=0.5\linewidth]{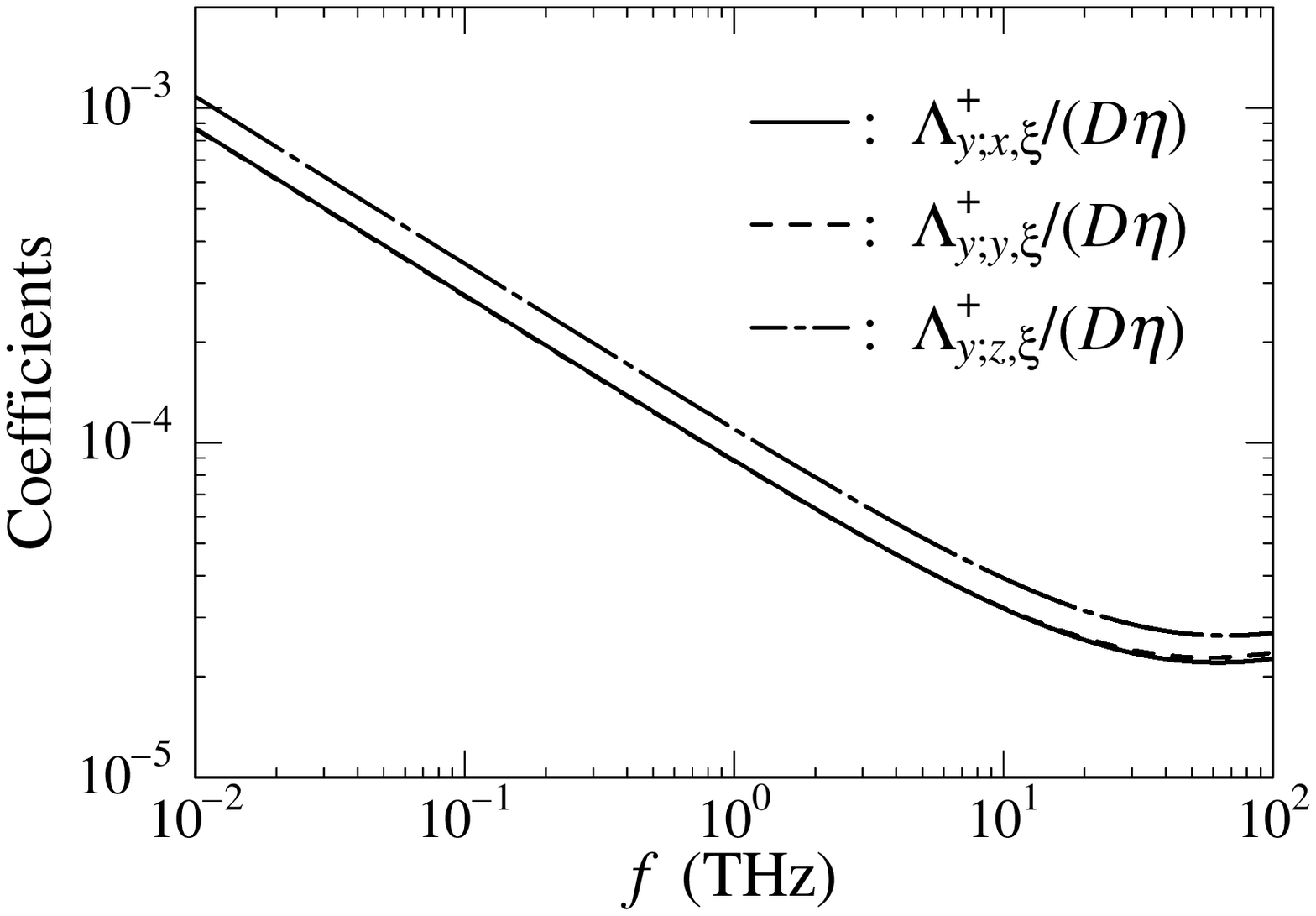}\\
\includegraphics[width=0.5\linewidth]{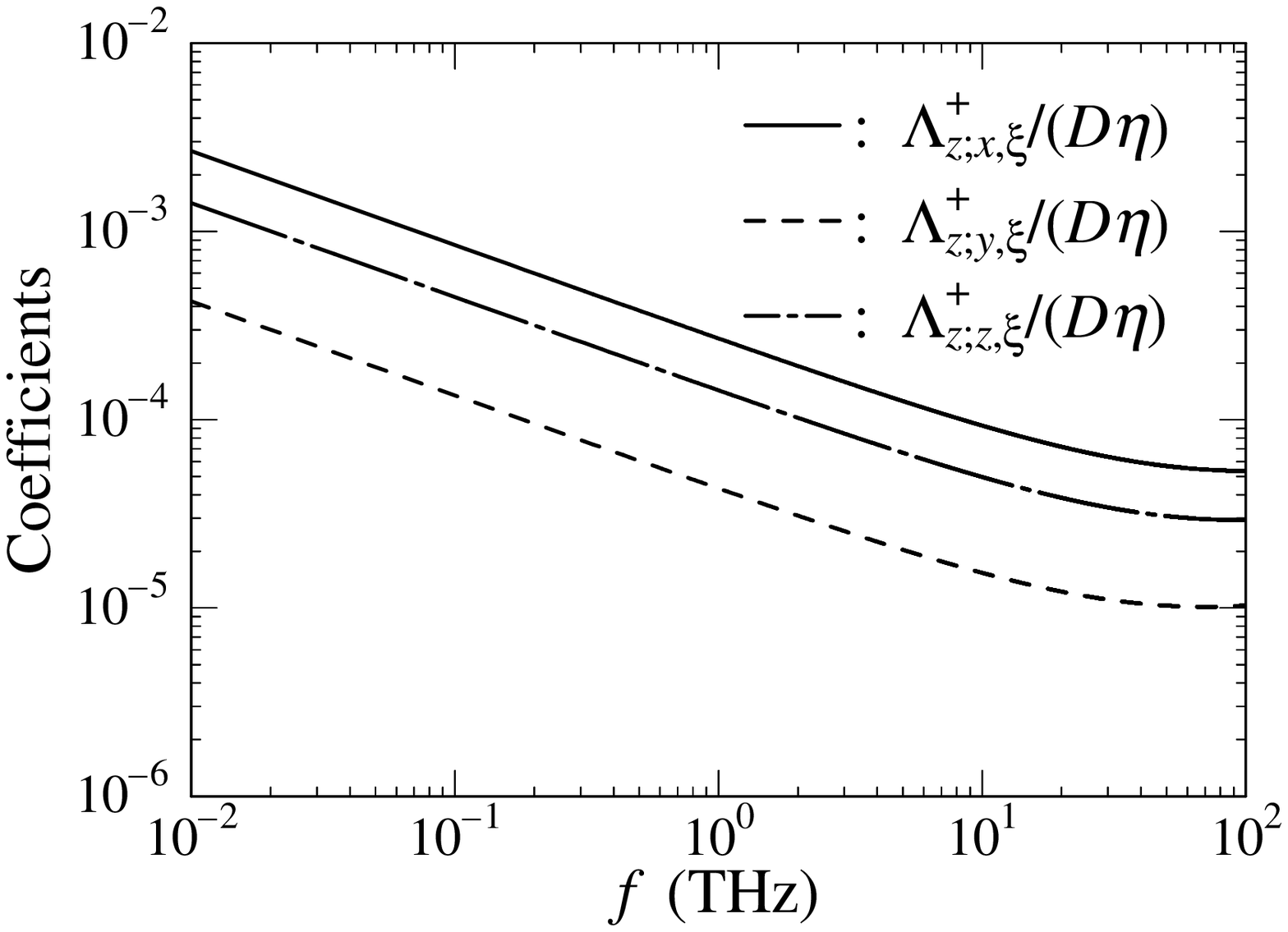}
\caption{
Vibration frequency $f$ dependences of the coefficients. 
Upper panel: $\Lambda_{x;\nu,\xi}^+/(D \eta)$ for $\nu$, $\xi$=$x$, $y$, $z$. 
Middle panel: $\Lambda_{y;\nu,\xi}^+/(D \eta)$ for $\nu$, $\xi$=$x$, $y$, $z$. 
Lower panel: $\Lambda_{z;\nu,\xi}^+/(D \eta)$ for $\nu$, $\xi$=$x$, $y$, $z$. 
We note that 
$\Lambda_{y;x,\xi}^+/(D \eta)$ takes almost the same value as 
$\Lambda_{y;y,\xi}^+/(D \eta)$ at each $f$. 
The parameters are set to be 
$c_d$=$c_p$=0.15 and $\Delta_m$=0.5 eV. 
}
\label{ap_x_f}
\end{center}
\end{figure}


\begin{figure}[ht]
\begin{center}
\includegraphics[width=0.5\linewidth]{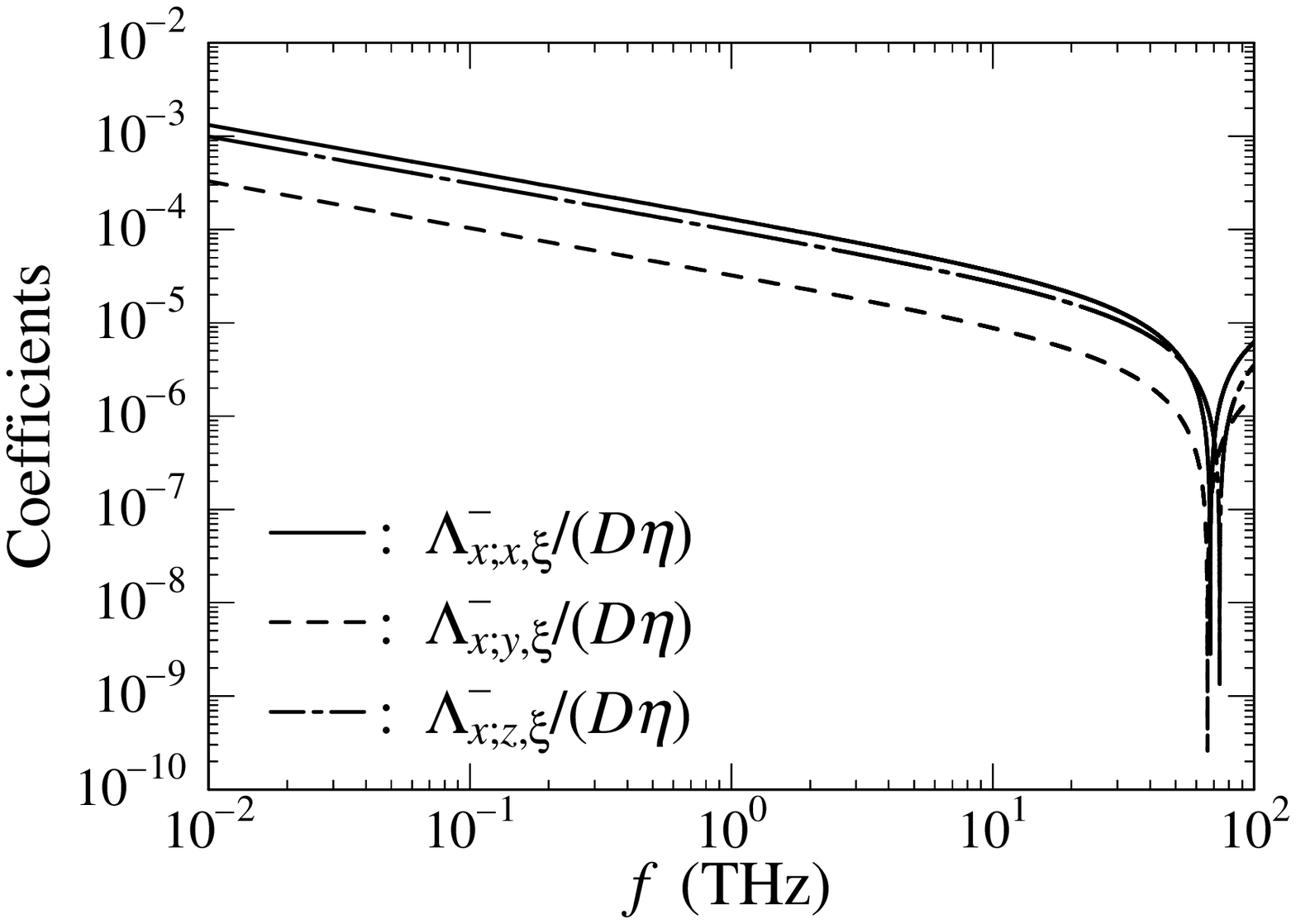}\\
\includegraphics[width=0.5\linewidth]{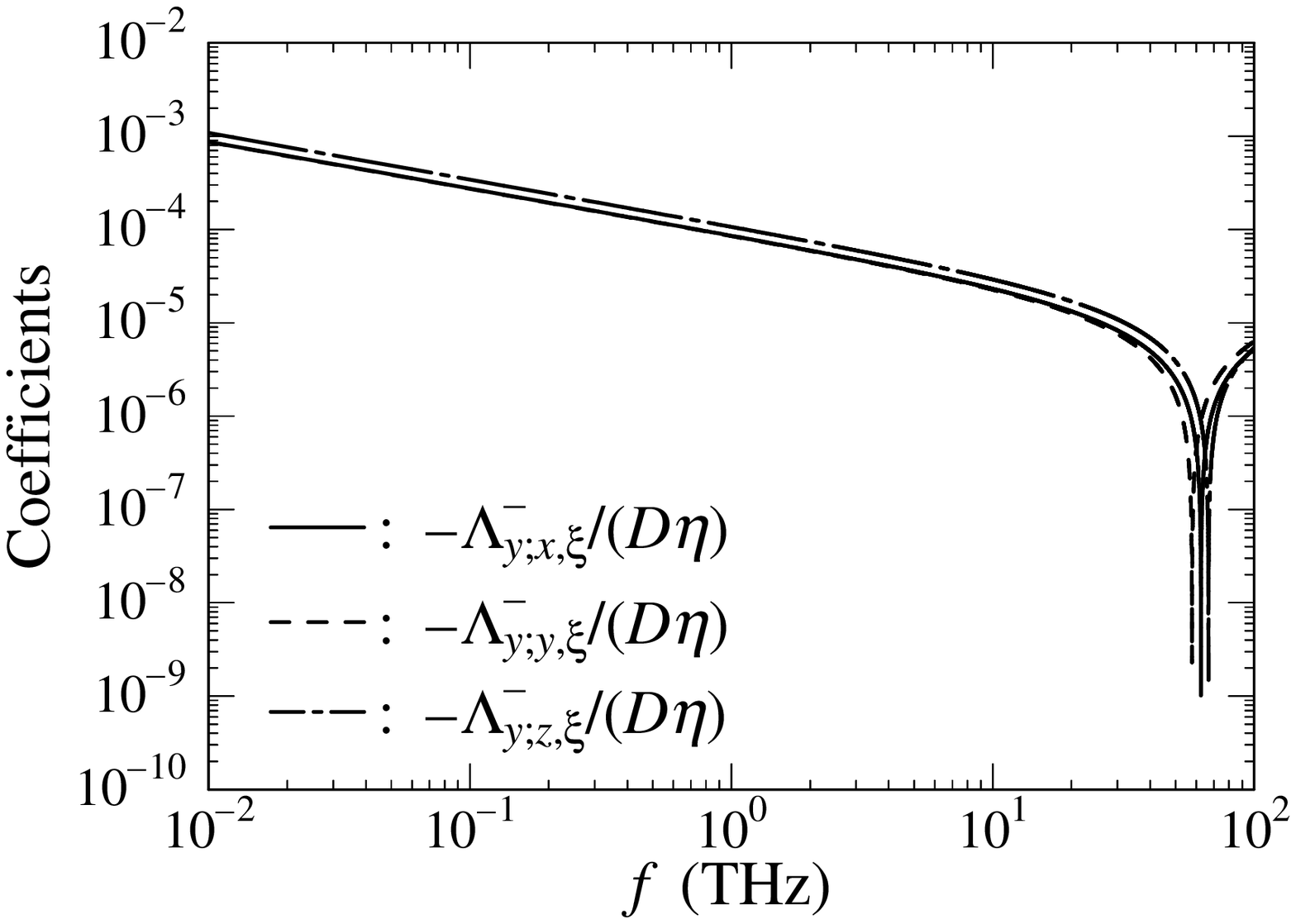}\\
\includegraphics[width=0.5\linewidth]{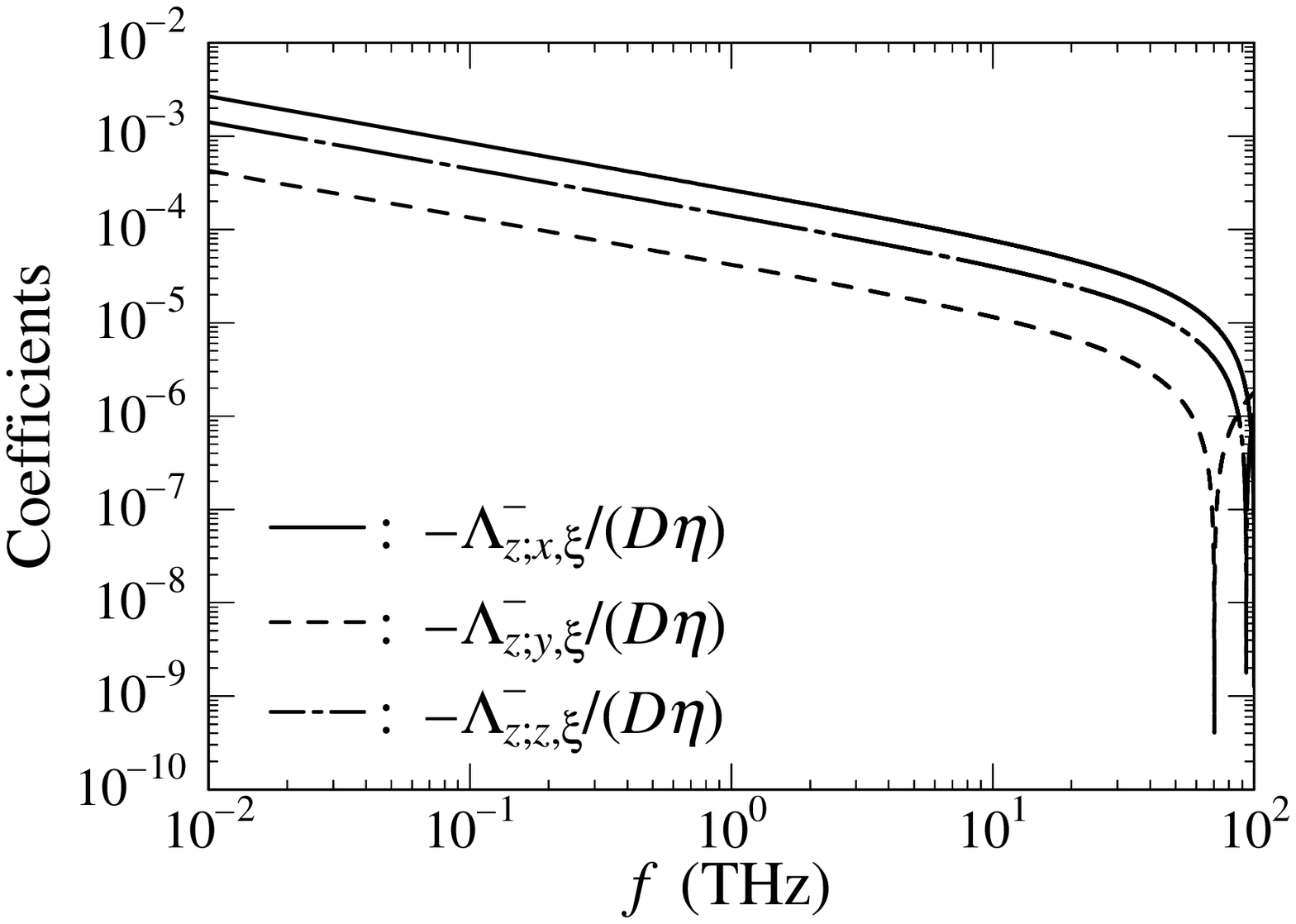}
\caption{
Vibration frequency $f$ dependences of the coefficients. 
Upper panel: $\Lambda_{x;\nu,\xi}^-/(D \eta)$ for $\nu$, $\xi$=$x$, $y$, $z$. 
Middle panel: $\Lambda_{y;\nu,\xi}^-/(D \eta)$ for $\nu$, $\xi$=$x$, $y$, $z$. 
Lower panel: $\Lambda_{z;\nu,\xi}^-/(D \eta)$ for $\nu$, $\xi$=$x$, $y$, $z$. 
We note that 
$\Lambda_{y;x,\xi}^-/(D \eta)$ takes almost the same value as 
$\Lambda_{y;y,\xi}^-/(D \eta)$ at each $f$. 
The parameters are set to be 
$c_d$=$c_p$=0.15 and $\Delta_m$=0.5 eV. 
}
\label{am_x_f}
\end{center}
\end{figure}

\subsection{Frequency dependence of coefficients}
\label{f-depen}

The coefficients of the spin-atomic vibration interaction are 
now written as 
$\Lambda_{\mu,\xi}^{(1)}/\lambda$, 
$\Lambda_{\mu,\xi}^k/\lambda$, 
$G_\xi$, $F/D$, 
$\Lambda_{I,J;\xi}'/D$, 
and $\Lambda_{\mu;\nu,\xi}^{\pm}/D$, 
while that of the spin-flip Hamiltonian is written as 
$\Lambda_{I,J}^{\rm SF}/D$. 
These expressions are dimensionless quantities. 
Note that 
$E$ of eq. (\ref{EEEEE}) becomes zero 
owing to the tetragonal symmetry (see eq. (\ref{E=0})). 
The expressions of all the coefficients are described 
in Appendix \ref{Appendix}.

In Figs. \ref{a1_f} and \ref{akx_f}, 
we show the $f$ dependences of 
$\Lambda_{\mu,\xi}^{(1)}/(\lambda \eta {\rm i})$ 
and $\Lambda_{\mu,\xi}^k/(\lambda \eta {\rm i})$, 
respectively. 
These expressions 
are the coefficients of the $S_I (-a_\xi + a_\xi^\dag)$ term. 
The magnitudes of the coefficients 
increase with increasing $f$, 
because they are proportional to $\sqrt{f}$. 

Figures \ref{sf_f} - \ref{am_x_f} 
show the $f$ dependences of 
$\Lambda_{I,J}^{\rm SF}/D$, 
$G_\xi/\eta$, 
$\Lambda_{I,J;\xi}'/(D \eta)$, 
$\Lambda_{\mu;\nu,\xi}^{+}/(D \eta)$, 
and 
$\Lambda_{\mu;\nu,\xi}^{-}/(D \eta)$, respectively. 
These expressions are the coefficients of 
the $S_I S_J a_\xi$ and $S_I S_J a_\xi^\dag$ terms. 
The coefficient $\Lambda_{I,J}^{\rm SF}/D$, 
which contains no $f$, 
certainly takes a constant value. 
The coefficients 
$G_\xi/\eta$ and $|\Lambda_{I,J;\xi}'/(D \eta)|$, 
which are proportional to $1/\sqrt{f}$, 
decrease with increasing $f$. 
This behavior is in contrast to that of 
$|\Lambda_{\mu,\xi}^{(1)}/(\lambda \eta {\rm i})|$ and 
$|\Lambda_{\mu,\xi}^k/(\lambda \eta {\rm i})|$. 
The coefficient 
$|\Lambda_{\mu;\nu,\xi}^{\pm}/(D \eta)|$ monotonically 
decreases with increasing $f$ in a region of $f \lesssim 20$ THz, 
because the dominant term of $\kappa_{u,\xi}^{\pm}$ in 
$\Lambda_{\mu;\nu,\xi}^{\pm}/D$ (see eq. (\ref{kappa})) is 
${\rm i} \eta(m /\hbar^2 ) \sqrt{\hbar/(2 M \omega_\xi)} \Delta_u $, 
where $\omega_\xi$=$2 \pi f$. 
In the vicinity of $f$=60 THz, 
$|\Lambda_{\mu;\nu,\xi}^{-}/D|$ drops, 
reflecting that some $\kappa_{u,\xi}^{-}$'s approach zero. 
Here, $\kappa_{u,\xi}^{-}$=0 reduces to $\hbar \omega_\xi$=$h f$=$\Delta_u$. 


\begin{figure}[ht]
\begin{center}
\includegraphics[width=0.5\linewidth]{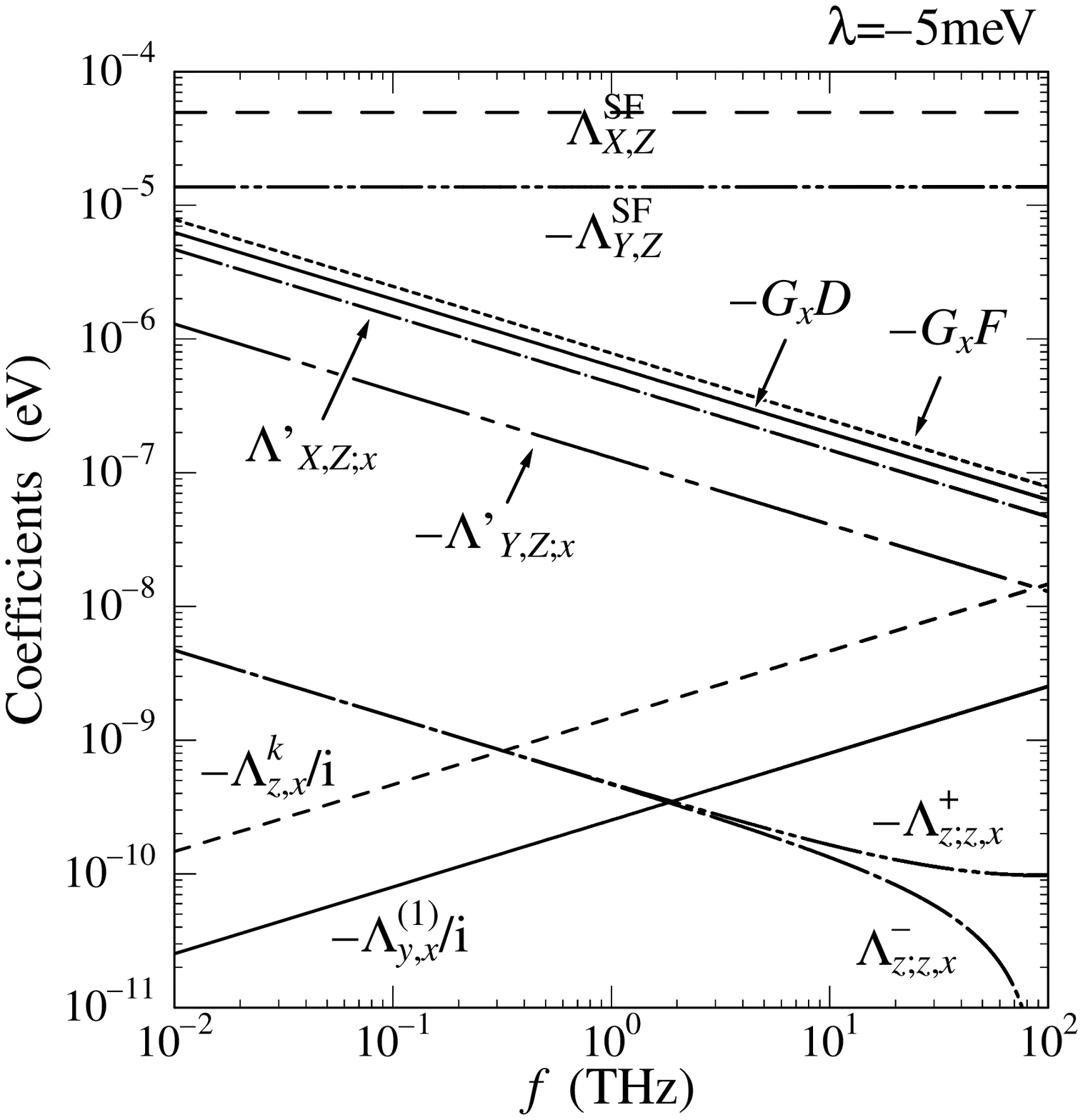}
\includegraphics[width=0.5\linewidth]{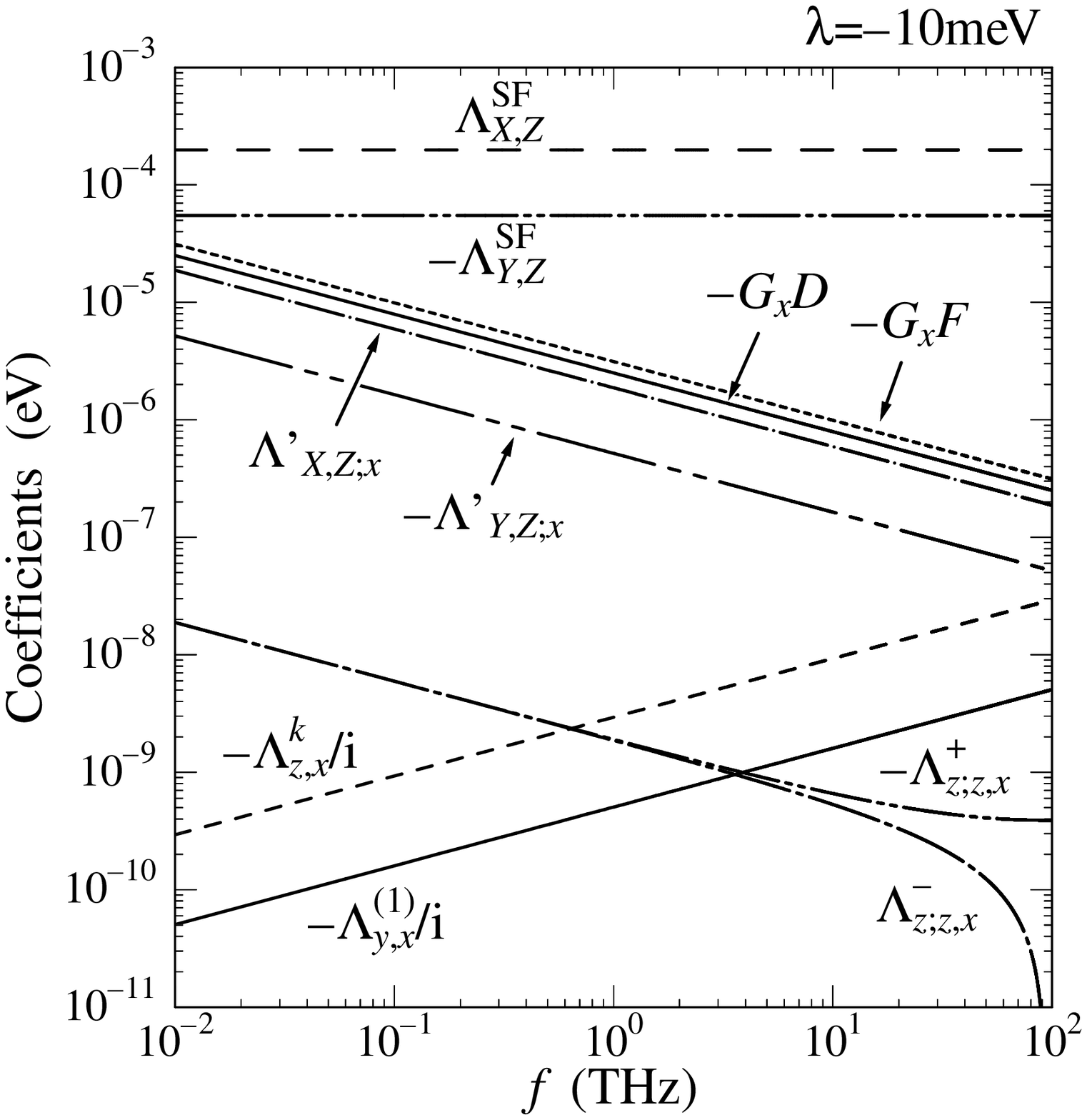}
\caption{
Vibration frequency $f$ dependences of 
$\Lambda_{y,x}^{(1)}$, 
$\Lambda_{z,x}^k$, 
$G_x D$, $G_x F$, 
$\Lambda_{X,Z;x}'$, $\Lambda_{Y,Z;x}'$, 
and $\Lambda_{z;z,x}^{\pm}$ in ${V}_{\rm SA}$, 
and 
$\Lambda_{X,Z}^{\rm SF}$ 
and $\Lambda_{Y,Z}^{\rm SF}$ in ${V}_{\rm SF}$. 
Upper panel: $\lambda$=$-$5 meV. 
Lower panel: $\lambda$=$-$10 meV. 
The cases of $\lambda$=$-$5 and $-$10 meV 
lead to 
$D$=$-$6.63$\times$10$^{-2}$ and 
$-$2.65$\times$10$^{-1}$ meV, respectively. 
The parameters are set to be 
$\eta$=0.05, 
$c_d$=$c_p$=0.15, and $\Delta_m$=0.5 eV. 
}
\label{comp}
\end{center}
\end{figure}


\begin{figure}[ht]
\begin{center}
\includegraphics[width=0.5\linewidth]{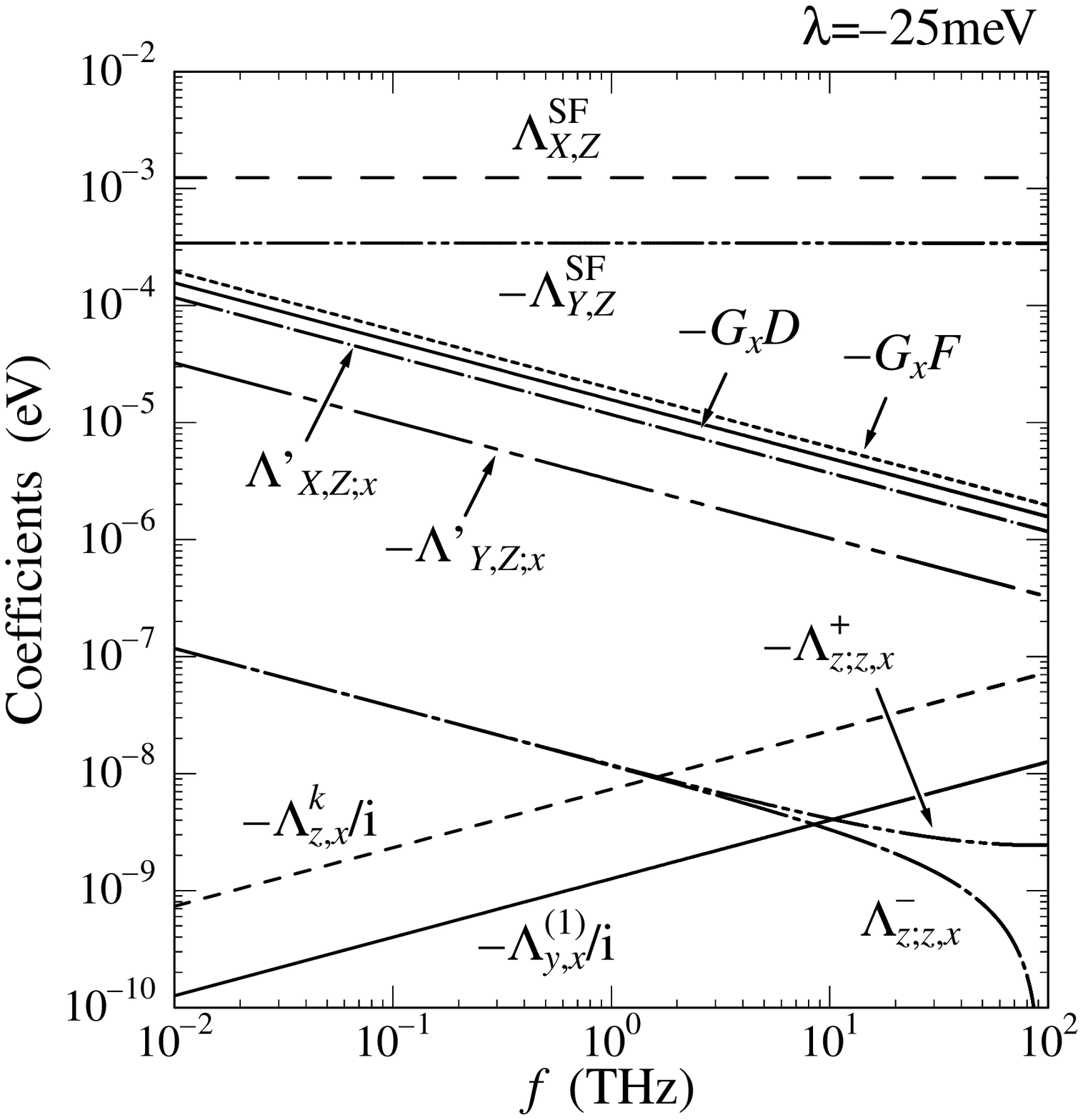}
\includegraphics[width=0.5\linewidth]{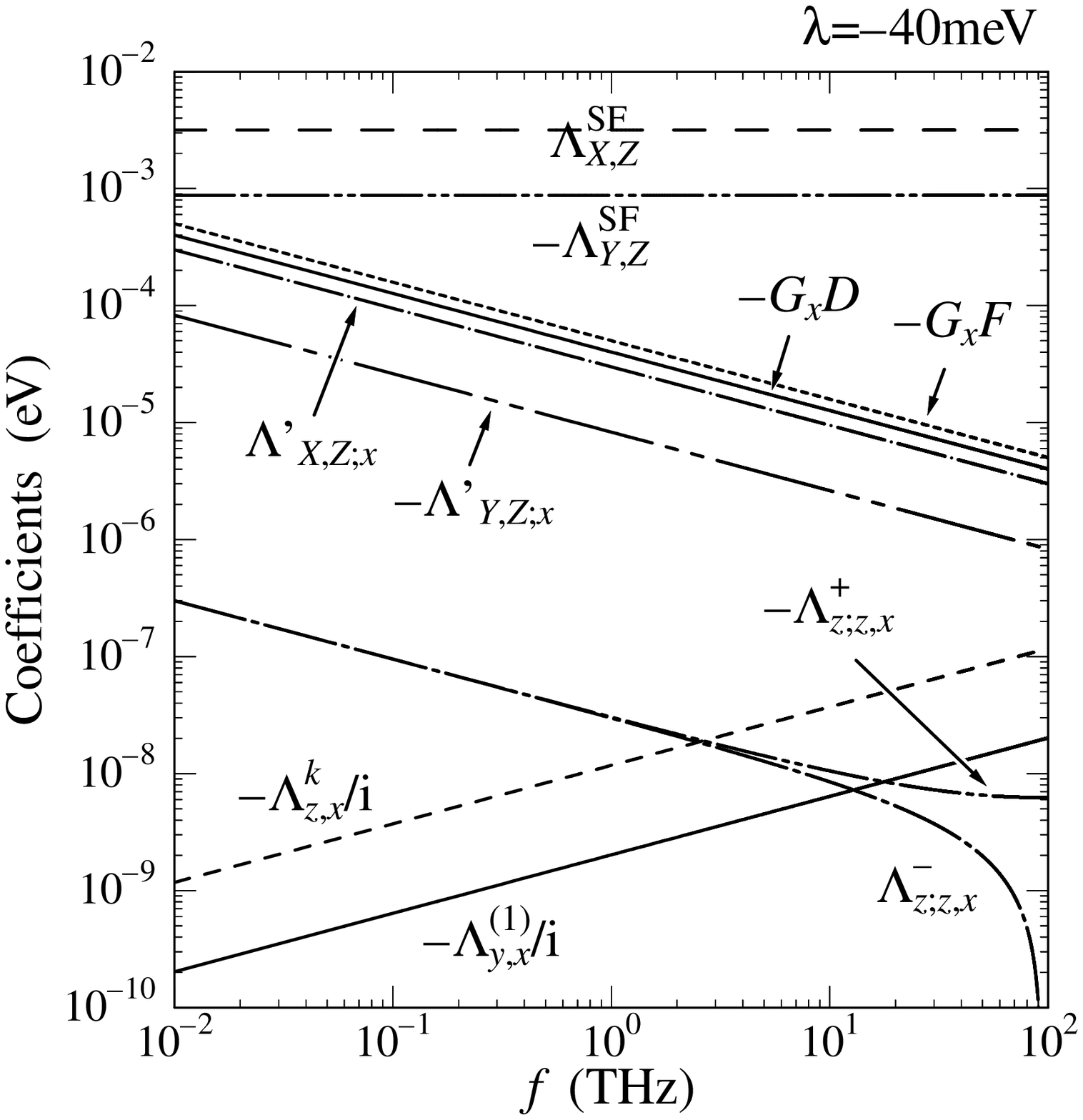}
\caption{
Vibration frequency $f$ dependences of 
$\Lambda_{y,x}^{(1)}$, 
$\Lambda_{z,x}^k$, 
$G_x D$, $G_x F$, 
$\Lambda_{X,Z;x}'$, $\Lambda_{Y,Z;x}'$, 
and $\Lambda_{z;z,x}^{\pm}$ in ${V}_{\rm SA}$, 
and 
$\Lambda_{X,Z}^{\rm SF}$ 
and $\Lambda_{Y,Z}^{\rm SF}$ in ${V}_{\rm SF}$. 
Upper panel: $\lambda$=$-$25 meV. 
Lower panel: $\lambda$=$-$40 meV. 
The cases of $\lambda$=$-$25 and $-$40 meV 
lead to 
$D$=$-$1.66 and $-$4.25 meV, respectively. 
The parameters are set to be 
$\eta$=0.05, 
$c_d$=$c_p$=0.15, and $\Delta_m$=0.5 eV. 
}
\label{comp1}
\end{center}
\end{figure}

\begin{figure}[ht]
\begin{center}
\includegraphics[width=0.5\linewidth]{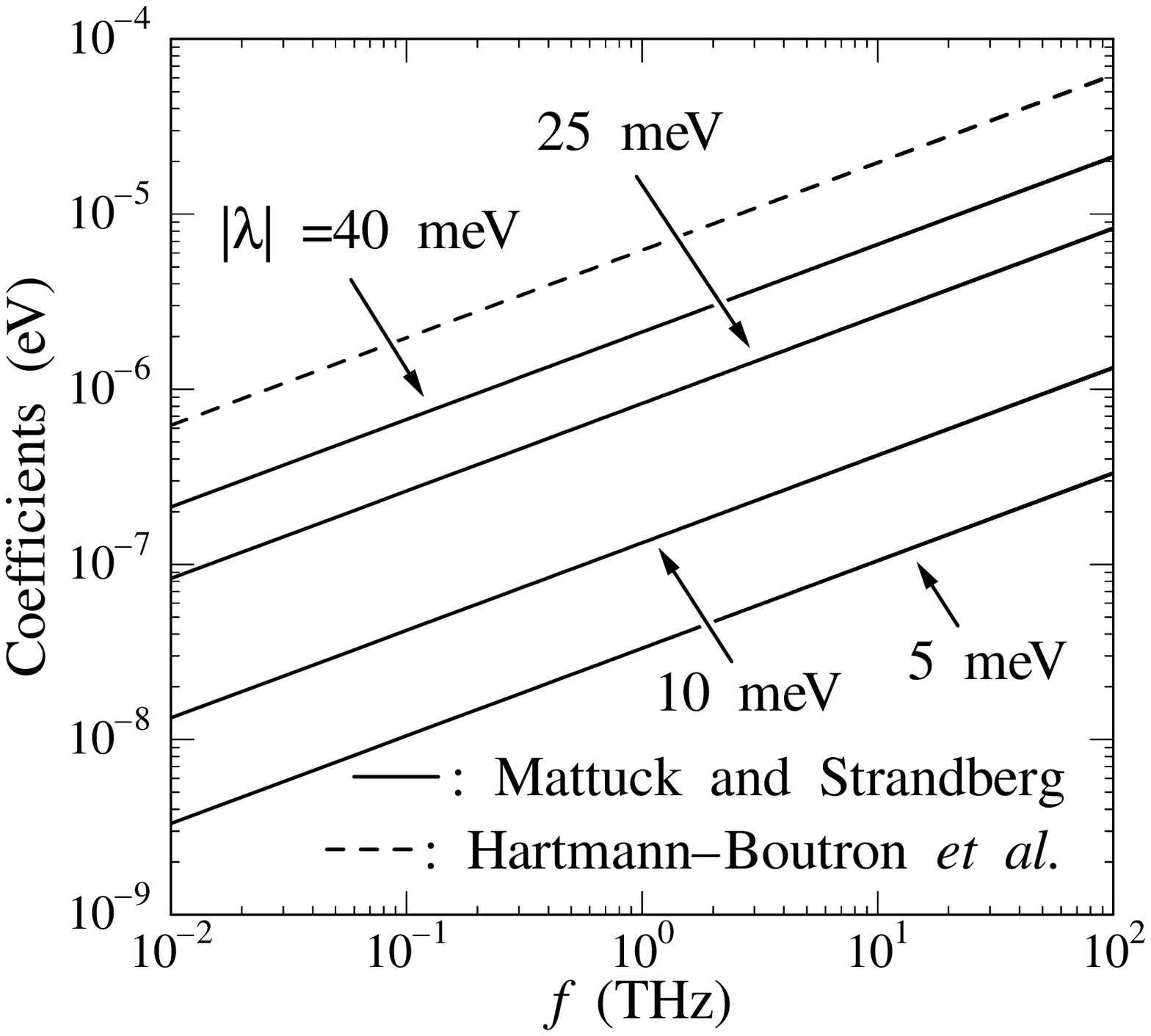}
\caption{
Vibration frequency $f$ dependence of 
the magnitude of coefficient of the spin-phonon interaction, $|c_q|$. 
Solid lines: 
The coefficients of Mattuck and Strandberg\cite{Mattuck} 
estimated for $|\lambda|$=5, 10, 25, and 40 meV. 
Dotted line: 
The coefficient of Hartmann-Boutron {\it et al}.\cite{Hartmann} 
In this coefficient, $\lambda$ is unspecified. 
}
\label{sp_coef}
\end{center}
\end{figure}

\subsection{Evaluation of coefficients}
\label{evaluation}

For the evaluation of the coefficients, 
we here consider a model in which 
the Fe ion (3d$^6$) in the crystal field of tetragonal symmetry 
vibrates in the $x$ direction. 
The spin-orbit coupling constant $\lambda$ 
is assumed to be 
$\lambda$=$-$5, $-$10, $-$25, and $-$40 meV, 
where $\lambda$ of Fe$^{2+}$ has been previously evaluated 
to be $-$12.6 meV.\cite{Yosida} 
The anisotropy constant $D$ can then be evaluated by 
substituting the above $\lambda$'s, 
$c_d$=$c_p$=0.15, $\Delta_m$=0.5 eV, and $\Delta_s$=0.45$\Delta_m$ 
into eq. (\ref{appen_D}). 
Namely, 
we have $D$=$-$6.63$\times$10$^{-2}$ meV for $\lambda$=$-$5 meV, 
$D$=$-$2.65$\times$10$^{-1}$ meV for $\lambda$=$-$10 meV, 
$D$=$-$1.66 meV for $\lambda$=$-$25 meV, 
and $D$=$-$4.25 meV for $\lambda$=$-$40 meV. 
We also use $\eta$=0.05, which is evaluated in 
Appendix \ref{ele_pol}.

Using the above $\lambda$'s and $D$'s and 
the results in Figs. \ref{a1_f} - \ref{am_x_f}, 
we can evaluate the coefficients of ${V}_{\rm SA}$ and ${V}_{\rm SF}$. 
In particular, 
we focus on large components in the respective coefficients, 
i.e., 
$\Lambda_{y,x}^{(1)}$, 
$\Lambda_{z,x}^k$, 
$G_x D$, $G_x F$, 
$\Lambda_{X,Z;x}'$, 
$\Lambda_{Y,Z;x}'$, 
and 
$\Lambda_{z;z,x}^{\pm}$ in ${V}_{\rm SA}$, 
and 
$\Lambda_{X,Z}^{\rm SF}$ 
and $\Lambda_{Y,Z}^{\rm SF}$ 
in ${V}_{\rm SF}$. 
In the upper and lower panels of Fig. \ref{comp}, 
we show the $f$ dependences of 
the coefficients with $\lambda$=$-$5 and $-$10 meV, respectively. 
In addition, 
the upper and lower panels of Fig. \ref{comp1} show those 
with $\lambda$=$-$25 and $-$40 meV, respectively. 

We first find that 
the magnitudes of the coefficients of ${V}_{\rm SF}$ are larger than 
those of the ${V}_{\rm SA}$. 
This relation mainly indicates that 
the coefficients of ${V}_{\rm SA}$ 
are proportional to $\eta$ 
(i.e., the degree of the difference in vibration displacement), 
whereas ${V}_{\rm SF}$ does not contain $\eta$. 
As found from eqs. (\ref{Lam_IJxi}) and (\ref{Lam^SF}), 
the major difference between 
$\Lambda_{I,J;\xi}'$ 
and 
$\Lambda_{I,J}^{\rm SF}$ 
may be the presence or absence of $\eta$.

In ${V}_{\rm SA}$, 
``$|G_x D|$, $|G_x F|$, $|\Lambda_{X,Z;x}'|$, and $|\Lambda_{Y,Z;x}'|$'' 
tend to be larger than 
$|\Lambda_{z,z;x}^{\pm}|$ and $|\Lambda_{z,x}^{k}|$, 
because of the relation of 
$|{V}_{\mbox{\footnotesize so1}}|
\gg |{V}_{\mbox{\footnotesize so2}}|$, 
$|{V}_{\mbox{\footnotesize k}}'|$ 
(see eqs. (\ref{so0so1^2}) - (\ref{so0k^2})). 
In particular, the relation of 
$|{V}_{\mbox{\footnotesize so1}}|
\gg |{V}_{\mbox{\footnotesize so2}}|$ 
indicates that 
the rate of the change in $\lambda$ 
due to the vibration, $|\Delta \lambda/\lambda|$, 
is larger than 
that in ${\mbox{\boldmath $L$}}$ 
due to the vibration, 
$|\Delta {\mbox{\boldmath $L$}}/{\mbox{\boldmath $L$}}|$, 
in a region with an atomic radius of $r \sim 0.1$ nm 
(see eqs. (\ref{V_so1}) and (\ref{V_so2})). 
Here, $\Delta \lambda$ and $\Delta {\mbox{\boldmath $L$}}$ 
are given by 
$\Delta \lambda$=
$\eta C_\pm \langle \partial r^{-3} /\partial x \rangle (-\Delta x_{\rm n})$
=$\eta C_\pm \langle 3 x r^{-5} \rangle \Delta x_{\rm n}$ 
and 
$\Delta {\mbox{\boldmath $L$}}$=
$-(\eta/\hbar)
\left(
{\mbox{\boldmath $r$}}_t \times \Delta {\mbox{\boldmath $p$}}_{\rm e}
+ \Delta {\mbox{\boldmath $r$}}_{\rm n} \times {\mbox{\boldmath $p$}}_t
\right)$, respectively, 
where $\Delta {\mbox{\boldmath $p$}}_{\rm e}$=$(\Delta p_{{\rm e},x},0,0)$ 
and $\Delta {\mbox{\boldmath $r$}}_{\rm n}$=$(\Delta x_{\rm n},0,0)$ 
are set in this section. 


\section{Comparison between ${V}_{\rm SA}$ and 
the Conventional Spin-Phonon Interaction}
\label{Dis_comp}

In this section, 
we compare ${V}_{\rm SA}$ with the conventional spin-phonon interactions 
derived by Mattuck and Strandberg\cite{Mattuck} and 
by Hartmann-Boutron {\it et al}.\cite{Hartmann}
In particular, we focus on 
the $f$ dependences and magnitudes of the coefficients 
in the case of the Fe ion. 
We also use $\eta$=0.05, which is evaluated in 
Appendix \ref{ele_pol}. 

\subsection{Modification of 
${V}_{\rm SA}$}
\label{V_SA_V_SF}

To compare ${V}_{\rm SA}$ with the conventional spin-phonon interaction 
(see \S \ref{subsub_SP}), 
we first modify ${V}_{\rm SA}$ to a novel spin-phonon interaction. 
We here consider the following one-dimensional lattice system: 
ions with the same mass $M$ are placed at even intervals along the $x$ axis, 
and 
vibrate in the $x$ direction. 
In addition, 
a specific ion located at the origin possesses ${\mbox{\boldmath $S$}}$, 
and also this ion is surrounded by several negative ions. 
The system may be just described by a simple model in which 
${\mbox{\boldmath $S$}}$ 
located at the origin 
is coupled to the phonon for the lattice system. 
Thus, $V_{\rm SA}$ in this model is obtained by replacing 
$\Delta x_{\rm n}$ in eq. (\ref{Delta_rn}) 
and $\Delta p_{{\rm n},x}$ in eq. (\ref{Delta_pn}) 
with $(1/\sqrt{N})
\sum_{q} \sqrt{\hbar/(2M \omega_{q})} 
\left(a_{q} + a_{q}^\dag \right)$ 
and 
$(1/\sqrt{N})
\sum_{q} 
{\rm i} \sqrt{M \hbar\omega_{q}/2} 
\left(- a_{q} + a_{q}^\dag \right)$, 
respectively. 
Here, $N$ is the number of unit cells, $q$ is the wave vector, 
$\omega_q$ is the angular frequency of the phonon at $q$, 
and $a_{q}$ ($a_{q}^\dag$) 
is the annihilation operator (creation operator) of 
the phonon with $q$. 
As a result, ${V}_{\rm SA}$ of eq. (\ref{V_SA}) is modified as follows: 
\begin{eqnarray}
&&\hspace*{-0.7cm}
{V}_{\rm SA}=
\sum_{I=X,Y,Z}S_I 
\frac{1}{\sqrt{N}}
\sum_{q} 
c_{I,x,q} 
\left( -a_{q} + a_{q}^\dag \right) \nonumber \\
&&
\hspace*{0.5cm}
+ \sum_{I,J=X,Y,Z} S_I S_J 
\frac{1}{\sqrt{N}}
\sum_{q} 
\left(c_{1,I,J,x,q} a_{q} 
+ c_{2,I,J,x,q} a_{q}^\dag \right) \nonumber
\\
&&
\hspace*{0.5cm}
+ \left[ D S_Z^2 + FS(S+1)\right] \frac{1}{\sqrt{N}}
\sum_{q} 
G_{x,q} (a_{q} + a_{q}^\dag). 
\end{eqnarray}
The coefficients 
$c_{I,x,q}$, 
$c_{1,I,J,x,q}$, $c_{2,I,J,x,q}$, and $G_{x,q}$, respectively, 
correspond to 
$c_{I,x}$, $c_{1,I,J,x}$, $c_{2,I,J,x}$, and $G_{x}$ in eq. (\ref{V_SA}), 
where $\omega_x$ in these coefficients has been replaced with $\omega_q$. 

We now replace $\omega_q$ with $2 \pi f$ 
in order to investigate the $f$ dependence of the coefficients. 
We then can regard 
$c_{I,x,q}$, $c_{1,I,J,x,q}$, 
$c_{2,I,J,x,q}$, and $G_{x,q}$ 
as $c_{I,x}$, $c_{1,I,J,x}$, $c_{2,I,J,x}$, and $G_x$ 
in eq. (\ref{V_SA}), respectively, 
where $\omega_x$=$2 \pi f$. 
Such coefficients may be directly compared with 
those of the spin-phonon interaction 
of the below-mentioned 
eqs. (\ref{V_SP}), (\ref{Mattuck_c}), and (\ref{Hartmann_c}). 
In particular, 
we focus on large components in the respective coefficients 
in the case of 
the Fe ion (3d$^6$) in the crystal field of tetragonal symmetry, i.e., 
$\Lambda_{y,x}^{(1)}$, 
$\Lambda_{z,x}^k$, 
$G_x D$, $G_x F$, 
$\Lambda_{X,Z;x}'$, 
$\Lambda_{Y,Z;x}'$, 
and 
$\Lambda_{z;z,x}^{\pm}$ (see Figs. \ref{comp} and \ref{comp1}). 

\subsection{Spin-phonon interaction}
\label{subsub_SP}

The spin-phonon interaction 
originally represents an interaction between 
a single spin located at the origin 
and the phonon of a lattice system.\cite{Mattuck,Hartmann}  
Here, a unit cell of the lattice consists of 
a magnetic ion (with a single spin) and the surrounding ions. 
In addition, all the ions vibrate. 
Note that exchange interactions between the spins 
are neglected.

As a similar lattice system to that in \S \ref{V_SA_V_SF}, 
a one-dimensional lattice system is considered, 
in which each unit cell is aligned along the $x$ axis. 
In addition, all the ions vibrate in the $x$ direction. 
The dominant term in the spin-phonon interaction,~\cite{Mattuck,Hartmann} 
${V}_{\rm SP}$, 
is then expressed as
\begin{eqnarray}
\label{V_SP}
{V}_{\rm SP}=
\sum_{I,J=X,Y,Z} S_I S_J 
\frac{1}{\sqrt{N}}\sum_q \left( c_q a_q + c_q^* a_q^\dag \right), 
\end{eqnarray}
where $N$ is the number of unit cells, $q$ is the wave vector, 
and $c_q$ is the coefficient.

The coefficient of ${V}_{\rm SP}$ derived 
by Mattuck and Strandberg,\cite{Mattuck} 
$c_q$, is written as 
\begin{eqnarray}
\label{Mattuck_c}
c_q = \frac{1}{4\pi \epsilon_0}\sqrt{\frac{\hbar \omega_q}{2M_{\rm u} v^2}} 
\frac{6 e^2 r_0^2}{\Delta^2 R^3} \lambda^2, 
\end{eqnarray}
with $\omega_q$=$2\pi f_q$.\cite{coefficient} 
Here, 
$\lambda$ is the spin-orbit coupling constant of eq. (\ref{so_lam}), 
$\Delta$ is the energy splitting under the crystal field, 
$R$ is the distance between the magnetic ion and the surrounding ions, 
$r_0$ is the ionic radius, 
$M_{\rm u}$ is the mass per unit cell, 
and $v$ is the phonon velocity defined as 
$v$=$\omega_q/|q|$, where $v$ is constant. 
This spin-phonon interaction 
is essentially the same as the expression by Van Vleck,\cite{Van1} 
which corresponds to 
the third-order perturbation energy. 
Here, 
the crystal field potential energy is the unperturbed term, 
while 
${V}_{\mbox{\footnotesize so0}}$ of eq. (\ref{V_so0}) and 
the modulation of crystal field potential energy 
due to the lattice vibration 
are the perturbed terms.\cite{Van1,Stoneham_b} 
We mention that 
the second-order perturbation energy due to 
${V}_{\mbox{\footnotesize so0}}$ and 
the modulation of crystal field potential energy 
vanishes owing to the Van Vleck cancellation.\cite{Van1,Kondo} 

To estimate $c_q$, 
we assume $\Delta$=1 eV, $R$=1 nm, and $r_0$=0.1 nm. 
We also use 
$M_{\rm u}$=[(26+30)+5$\times$(29+34)+2$\times$(7+7)]$\times$(1.67$\times$10$^{-27}$)+(26+5$\times$29+2$\times$7)$\times$(9.11$\times$10$^{-31}$) kg 
on the basis of results of the STM experiments 
for the Fe ion on the CuN surface.\cite{Hirjibehedin} 
The unit cell has been assumed here to consist of 
one Fe ion, five Cu ions, and two N ions. 
The numbers of protons (the number of neutrons) of Fe is 26 (30), 
that of protons (that of neutrons) of Cu is 29 (34), 
and that of protons (that of neutrons) of N is 7 (7). 
The numbers of electrons of Fe, Cu, and N neutral atoms 
are 26, 29, and 7, respectively. 
Since the numerical value of $v$ is unknown for this system, 
we utilize $v$=1.45$\times$10$^3$ m/s,\cite{Leuenberger} 
which is the value evaluated for Mn$_{12}$.

On the other hand, the coefficient of ${V}_{\rm SP}$ derived 
by Hartmann-Boutron {\it et al}.\cite{Hartmann} $c_q$ is given by 
\begin{eqnarray}
\label{Hartmann_c}
c_q = {\rm i} \sqrt{\frac{\hbar \omega_q }{2M_{\rm u} v^2}} V_q, 
\end{eqnarray}
with $\omega_q$=$2\pi f_q$. 
Here, $q$=$\omega_q/v$ has been used, where $v$ is constant. 
The quantity $V_q$ is regarded as 
the coefficient with a unit of energy.\cite{Politi} 
This interaction originally arises from 
the spin-orbit interaction and 
the local strain at the spin site due to the phonon. 
To estimate $c_q$, we  use 
$|V_q|$=4.45 meV and the above-mentioned $M_{\rm u}$ and $v$. 
This $|V_q|$ is obtained from 
$C_{\rm pre}$=$3|V_q|^2 /(2 \pi \hbar^4 \rho v^5)$,\cite{Mn12_1} 
with $\rho$=$M_{\rm u}/a^3$, 
where 
the volume of the unit cell, $a^3$, 
is set to be $a^3$=1 nm$^3$. 
The quantity 
$C_{\rm pre}$ corresponds to a prefactor of the expression of 
the relaxation rate, and then 
$C_{\rm pre}$=10$^{4}$ Hz/K$^3$ 
is chosen 
on the basis of experimental results for magnetic ions.\cite{Mn12_1,Abragam}

\subsection{Comparison}
\label{comparison}

We first make a comparison of the $f$ dependence of the coefficients 
between ${V}_{\rm SP}$ and ${V}_{\rm SA}$. 
Figure \ref{sp_coef} shows the $f$ dependence of $|c_q|$ 
of eqs. (\ref{Mattuck_c}) and (\ref{Hartmann_c}), 
where $f$ is defined by $f$=$f_q$. 
We find that $|c_q|$ is proportional to $\sqrt{f}$. 
Such $f$ dependence (i.e., $\sqrt{f}$) is different from 
those of the coefficients of the $S_I S_J a_q$ and $S_I S_J a_q^\dag$ terms 
of ${V}_{\rm SA}$ (i.e., $1/\sqrt{f}$), 
but the same as that of 
the $S_I(- a_q + a_q^\dag)$ term of ${V}_{\rm SA}$. 

Next, 
as seen from Figs. \ref{comp} - \ref{sp_coef}, 
$|G_x D|$, $|G_x F|$, $|\Lambda_{X,Z:x}'|$, and $|\Lambda_{Y,Z:x}'|$ 
in ${V}_{\rm SA}$ 
are larger than $|c_q|$ in the low $f$ region. 
For example, when $\lambda$=$-$10, $-$25, and $-$40 meV, 
the former are larger than the latter in a region of $f \lesssim$ 1 THz. 
Also, the region of $f$, in which 
the former become larger than the latter, 
expands with increasing $\lambda$.



As the main reason 
why the magnitudes of the coefficients of $V_{\rm SA}$ can be larger than 
$|c_q|$ of 
Mattuck and Strandberg, 
we consider that 
$M$ 
in the coefficients 
of $V_{\rm SA}$ ($\propto M^{-1/2}$) 
is smaller than $M_{\rm u}$ 
in $c_q$ of $V_{\rm SP}$ ($\propto {M_u}^{-1/2}$). 
Originally, the decrease in $M$ 
increases the magnitude of vibration displacement 
$|\Delta \xi_{\rm n}|$ ($\xi$=$x$, $y$, $z$) 
(see eq. (\ref{Delta_rn}) and Appendix \ref{D_rn_ap}). 
Furthermore, 
the increase in $|\Delta \xi_{\rm n}|$ 
enhances 
the magnitude of $V_{\rm SA}$ 
(i.e., ${v}_{{\mbox{\footnotesize so0,so1}},2}^{(2)}$) 
(e.g., eqs. (\ref{so0so1^2}) and (\ref{V_so1})).

On the other hand, 
$c_q$ of Hartmann-Boutron {\it et al}. is rewritten as 
$c_q$=${\rm i}\sqrt{ \hbar^5 \omega_q v^3 C_{\rm pre}/(3 a^3)}$, 
by substituting 
the above-mentioned $|V_q|^2$=$(2 \pi \hbar^4/3)
(M_{\rm u}/a^3) v^5 C_{\rm pre}$ 
into eq. (\ref{Hartmann_c}). 
Since the related quantities of 
$v$, $a$, and $C_{\rm pre}$ in $c_q$ are not contained in $V_{\rm SA}$, 
it may be impossible to identify the cause of 
the larger magnitudes of the coefficients of $V_{\rm SA}$ 
than $|c_q|$. 


We anticipate that 
at a low $f$, 
the $1/\sqrt{f}$ terms in $V_{\rm SA}$ 
can be verified from the pressure effect of 
the spin relaxation time, and so on.

\section{Transition Probability per Unit Time}
\label{Dis_tran}

In this section, 
we discuss transition probabilities per unit time 
due to ${V}_{\rm SA}$ and ${V}_{\rm SF}$ 
for a simple model of the Fe ion with bistable states. 
Our interests are in how 
the expressions of $V_{\rm SA}$ (e.g., $S_X S_Z (a + a^\dag)$) 
and $V_{\rm SF}$ (e.g., $S_X S_Z$) 
influence the transition probabilities. 
Note also that the transition probabilities 
will be necessary for investigating 
spin relaxation phenomena.\cite{Kokado1} 



When ${\cal V}_i$ is considered to be 
a time-dependent perturbed Hamiltonian, 
we have the probability per unit time 
of a transition from a state $| u \rangle$ to $| v \rangle$, 
$W_{u \to v}^{i}$: 
\begin{eqnarray}
\label{WWW}
W_{u \to v}^{i}=\frac{2\pi}{\hbar}
\left| \langle v | {\cal V}_i | u \rangle \right|^2 
\frac{1}{\pi} \frac{ \hbar/\tau}{(E_u - E_v)^2 + (\hbar/\tau)^2 },
\end{eqnarray} 
with 
${\cal H}_0^{\rm AS} |u \rangle$=$E_u |u \rangle$, 
where 
${\cal H}_0^{\rm AS}$ is 
the unperturbed Hamiltonian for the atomic spin, 
and $|u \rangle$ is the eigenstate of ${\cal H}_0^{\rm AS}$.\cite{Mag_reso} 
Here, a correlation time for ${\cal V}_i$, $\tau$, 
is introduced 
by assuming that the correlation function 
for ${\cal V}_i$ has an exponential decay. 
In addition, the case of $E_u$=$E_v$ represents 
the conservation of energy 
for a transition from $|u \rangle$ to $|v \rangle$, 
while 
the case of $E_u \ne E_v$ represents 
the nonconservation of energy 
for the transition. 


\begin{figure}[ht]
\begin{center}
\includegraphics[width=0.4\linewidth]{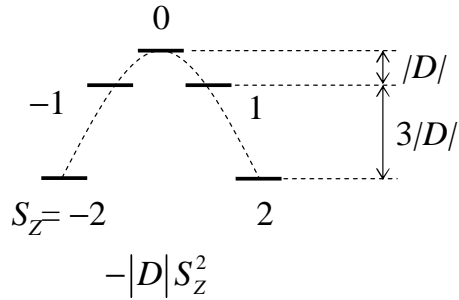}
\caption{
Energy levels of $-|D|S_Z^2$ with $S$=2. 
}
\label{bistable}
\end{center}
\end{figure}

We now propose ${\cal H}_0^{\rm AS}$ and ${\cal V}_i$ 
of a simple model for 
the Fe ion in the crystal field of tetragonal symmetry. 
Here, this ion is considered to vibrate in the $X$ direction. 
Note again that 
$x$=$X$, $y$=$Y$, and $z$=$Z$ are realized in this system, 
as described in Appendix \ref{app_theta}. 
When the representative 
terms in ${V}_{\rm SA}$ of eq. (\ref{V_SA}) 
and ${V}_{\rm SF}$ of eq. (\ref{V_SF}) 
(see Figs. \ref{a1_f} - \ref{ap_x_f}) are adopted, 
the Hamiltonian is written as follows: 
\begin{eqnarray}
&&\hspace*{-0.3cm}{\cal H}_0^{\rm AS}=DS_Z^2 + \hbar \omega a^\dag a, \\
\label{model_VSA}
&&\hspace*{-0.3cm}{\cal V}_{\rm SA1}=\Lambda_Z^k S_Z(-a + a^\dag), \\
&&\hspace*{-0.3cm}{\cal V}_{\rm SA2}=
\left[ \Lambda_{X,Z}' ( S_X S_Z + S_Z S_X) + 
\Lambda_{Y,Z}' (S_Y S_Z + S_Z S_Y) \right] (a + a^\dag), \\
&&\hspace*{-0.3cm}
{\cal V}_{\rm SA3}=G \left[ DS_Z^2 + FS(S+1)\right] (a + a^\dag), \\
\label{V_SF_Fe}
\label{model_VSF}
&&\hspace*{-0.3cm}{\cal V}_{\rm SF}=
\Lambda_{X,Z}^{\rm SF} (S_X S_Z + S_Z S_X) 
+ \Lambda_{Y,Z}^{\rm SF} (S_Y S_Z + S_Z S_Y), 
\end{eqnarray}
where $a$ ($a^\dag$) is the annihilation (creation) operator 
of the atomic vibration in the $X$ direction, 
and $\omega$ ($>$0) is the angular frequency in this direction. 
The coefficients $D$, $F$, $G$, and $\Lambda_Z^k$ are 
eqs. (\ref{appen_D}), (\ref{appen_F}), (\ref{G_xi1}), and (\ref{A^k}), 
respectively. 
In addition,  $\Lambda_{X,Z}'$ and $\Lambda_{Y,Z}'$ correspond to 
eq. (\ref{Lam_IJi_D}), 
while $\Lambda_{X,Z}^{\rm SF}$ and $\Lambda_{Y,Z}^{\rm SF}$ to 
eq. (\ref{Lambda_IJ^SF/D}). 
The state $|u \rangle$ is written as $|S_Z, n \rangle$, 
where $n$ is the vibrational quantum number.

For this model, we choose $S$=2 (see \S \ref{Application}) and $D$=$-|D|$ 
using examples from the experimental and theoretical studies 
of the Fe ion on the CuN surface.\cite{Hirjibehedin} 
As shown in Fig. \ref{bistable}, 
the model exhibits bistability between $S_Z$=$-2$ and $2$. 
The energy difference between 
the ground states and the first excited states is $3|D|$, 
while that between the first excited states and the second excited state 
is $|D|$.

By using eqs. (\ref{WWW}) - (\ref{V_SF_Fe}), 
the respective $W_{u \to v}^i$'s are obtained as follows: 
\begin{eqnarray}
&&\hspace*{-.5cm}W_{S_Z,n \to S_Z,n-1}^{\rm SA1}= \frac{2 \pi}{\hbar}
|\Lambda_Z^k|^2 S_Z^2 n \frac{1}{\pi}
\frac{\hbar/\tau}
{(\hbar \omega)^2 + (\hbar/\tau)^2}, \\
&&\hspace*{-.5cm}W_{S_Z,n \to S_Z,n+1}^{\rm SA1}= \frac{2 \pi}{\hbar}
|\Lambda_Z^k|^2 S_Z^2 (n+1) \frac{1}{\pi}
\frac{\hbar/\tau}
{(\hbar \omega)^2 + (\hbar/\tau)^2}, \\
&&\hspace*{-.5cm}W_{S_Z,n \to S_Z \pm 1,n-1}^{\rm SA2}= \frac{2 \pi}{\hbar}
\frac{\left[ (\Lambda_{X,Z}')^2+(\Lambda_{Y,Z}')^2 \right]}{4} (2S_Z \pm 1)^2 
(S \mp S_Z) (S \pm S_Z +1)n \nonumber \\
&&\hspace*{2.7cm} \times \frac{1}{\pi}
\frac{\hbar/\tau}
{\left[ -|D| ( \pm 2 S_Z +1) -\hbar \omega \right]^2 + (\hbar/\tau)^2}, 
\\
&&\hspace*{-.5cm}W_{S_Z,n \to S_Z \pm 1,n+1}^{\rm SA2}= \frac{2 \pi}{\hbar}
\frac{\left[ (\Lambda_{X,Z}')^2+(\Lambda_{Y,Z}')^2 \right]}{4} (2S_Z \pm 1)^2 
(S \mp S_Z) (S \pm S_Z +1) (n+1) \nonumber \\
&& \hspace*{2.7cm} \times \frac{1}{\pi}
\frac{\hbar/\tau}
{\left[ -|D| ( \pm 2 S_Z +1) +\hbar \omega \right]^2 + (\hbar/\tau)^2},\\
&&\hspace*{-.5cm}W_{S_Z,n \to S_Z,n-1}^{\rm SA3}= \frac{2 \pi}{\hbar}
G^2 \left[ -|D| S_Z^2 + FS(S+1) \right]^2 n \frac{1}{\pi}
\frac{\hbar/\tau}
{(\hbar \omega)^2 + (\hbar/\tau)^2}, \\
&&\hspace*{-.5cm}W_{S_Z,n \to S_Z,n+1}^{\rm SA3}= \frac{2 \pi}{\hbar}
G^2 \left[ -|D| S_Z^2 + FS(S+1) \right]^2 (n+1) \frac{1}{\pi}
\frac{\hbar/\tau}
{(\hbar \omega)^2 + (\hbar/\tau)^2}, \\
&&\hspace*{-.5cm}W_{S_Z,n \to S_Z \pm 1,n}^{\rm SF}= \frac{2 \pi}{\hbar}
\frac{\left[ (\Lambda_{X,Z}^{\rm SF})^2+(\Lambda_{Y,Z}^{\rm SF})^2 \right]}{4} (2S_Z \pm 1)^2 
(S \mp S_Z) (S \pm S_Z +1) 
\frac{1}{\pi}
\frac{\hbar/\tau}
{D^2 ( \pm 2 S_Z +1)^2 + (\hbar/\tau)^2}, \nonumber \\
\end{eqnarray}
where $\Lambda_Z^k$ is an imaginary number, 
whereas the other coefficients are real number (see \S \ref{Application}). 
The transition probability 
$W_{S_Z,n \to S_Z \pm 1,n-1}^{\rm SA2}$ 
($W_{S_Z,n \to S_Z \pm 1,n+1}^{\rm SA2}$) 
corresponds to a probability for a transition with 
the conservation of energy 
in the case of 
$\hbar \omega$=$-|D|(\pm 2S_Z + 1)$ 
($\hbar \omega$=$|D|(\pm 2S_Z + 1)$). 
The other $W_{u \to v}^{i}$'s 
are probabilities for transitions with the nonconservation of energy, 
irrespective of $\omega$ ($>0$).



\begin{figure}[ht]
\begin{center}
\includegraphics[width=0.45\linewidth]{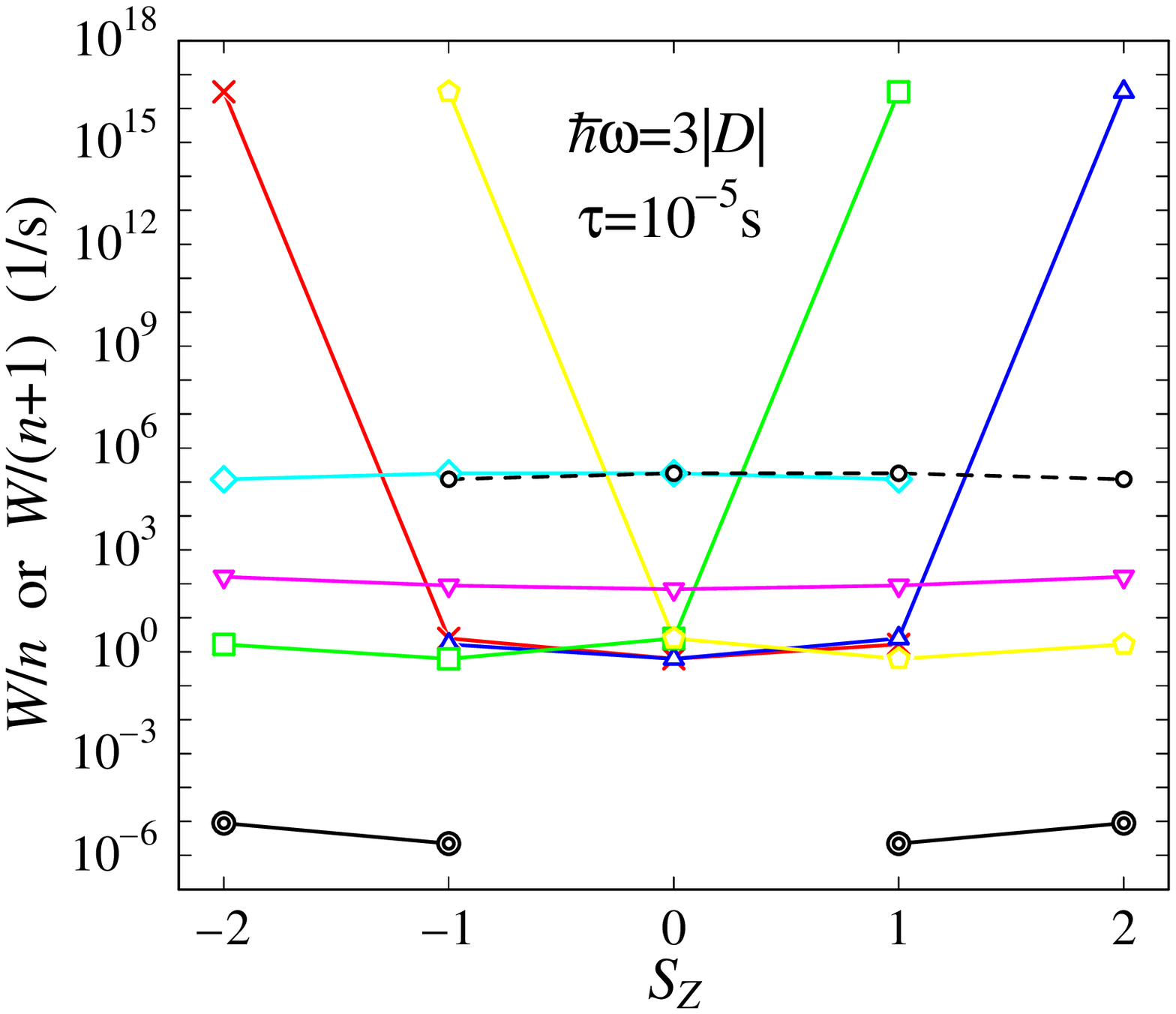}\\
\includegraphics[width=0.45\linewidth]{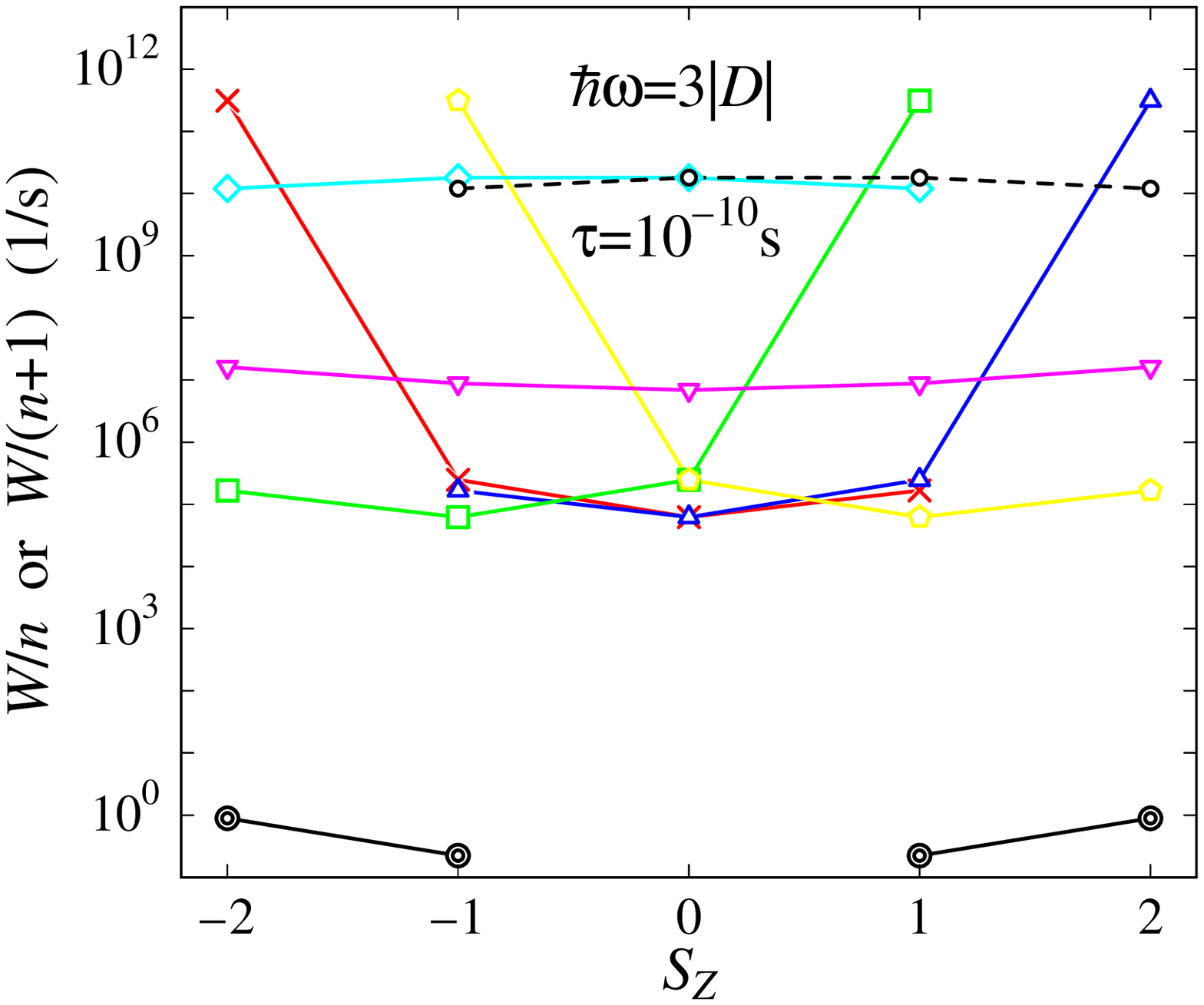}\\
\includegraphics[width=0.45\linewidth]{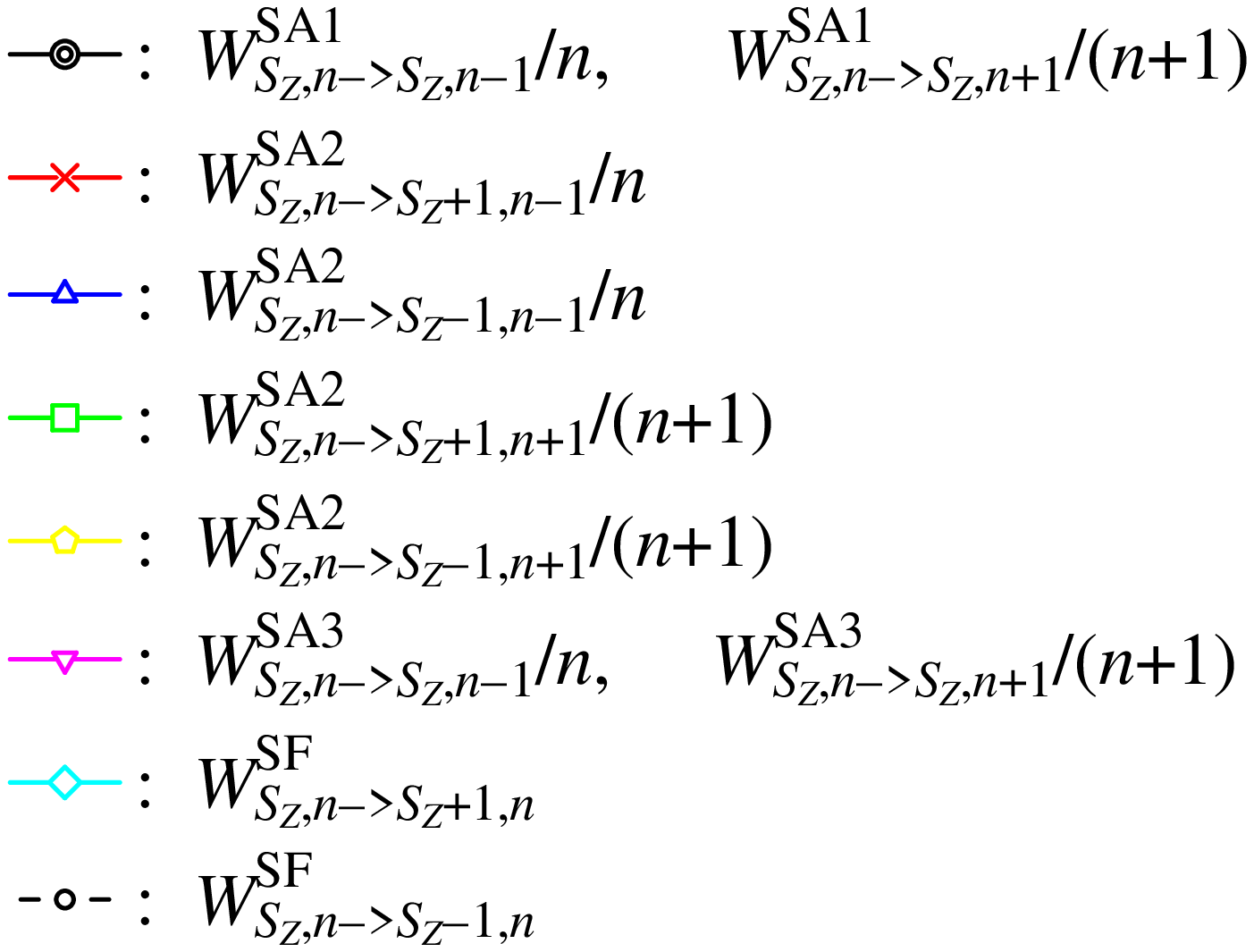}
\caption{
$Z$ component of the spin 
$S_Z$ dependences of 
$W_{S_Z,n \to S_Z,n-1}^{\rm SA1}/n$, 
$W_{S_Z,n \to S_Z \pm 1,n-1}^{\rm SA2}/n$, 
$W_{S_Z,n \to S_Z \pm 1,n+1}^{\rm SA2}/(n+1)$, 
$W_{S_Z,n \to S_Z,n-1}^{\rm SA3}/n$, 
$W_{S_Z,n \to S_Z,n+1}^{\rm SA3}/(n+1)$, 
$W_{S_Z,n \to S_Z \pm 1,n}^{\rm SF}$ of the $S$=2 system 
with $\hbar \omega$=3$|D|$. 
Upper panel: $\tau$=10$^{-5}$ s. 
Lower panel: $\tau$=10$^{-10}$ s. 
The energy levels of this system are shown in Fig. \ref{bistable}. 
}
\label{tran1}
\end{center}
\end{figure}


\begin{figure}[ht]
\begin{center}
\includegraphics[width=0.45\linewidth]{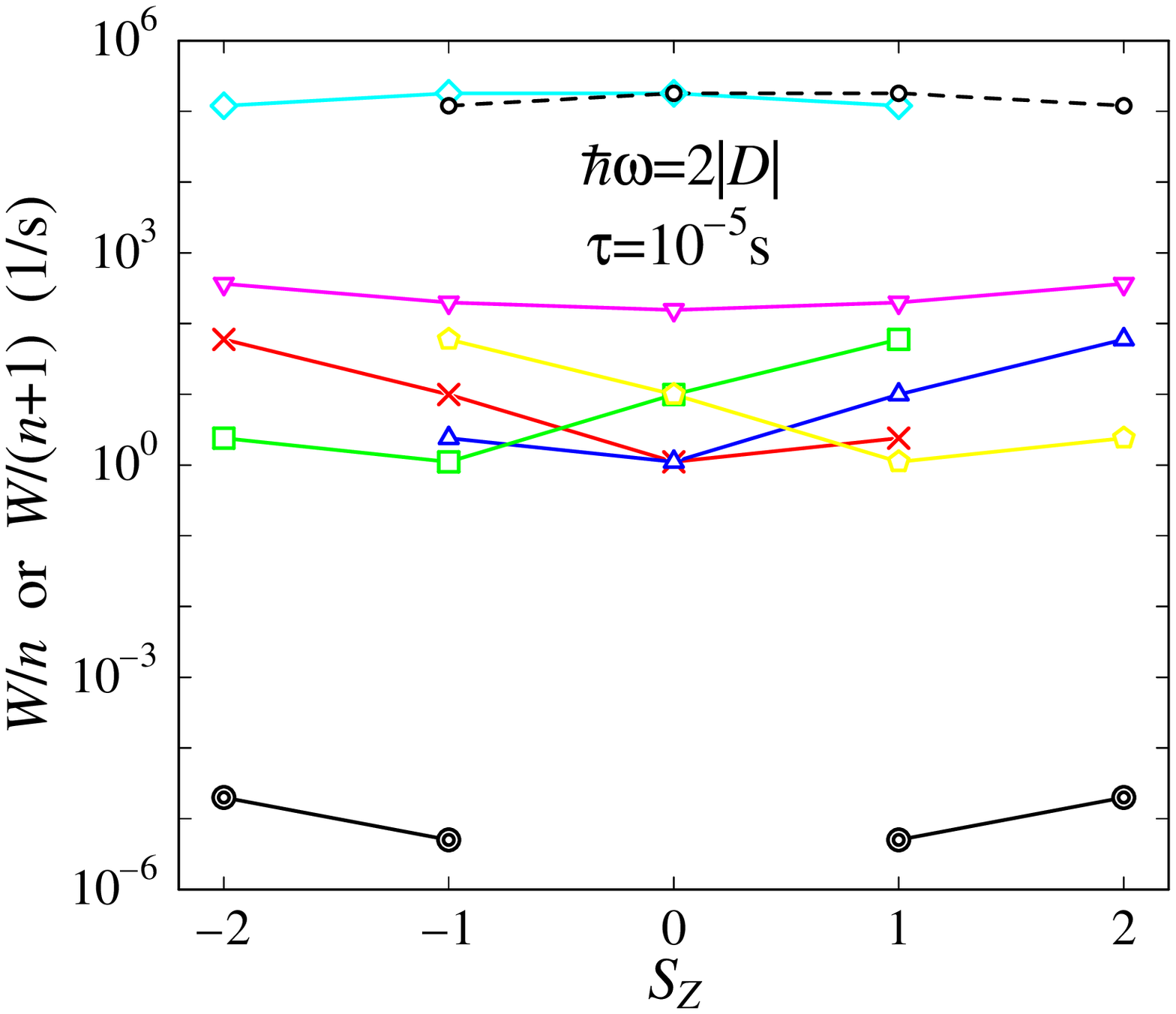}\\
\includegraphics[width=0.45\linewidth]{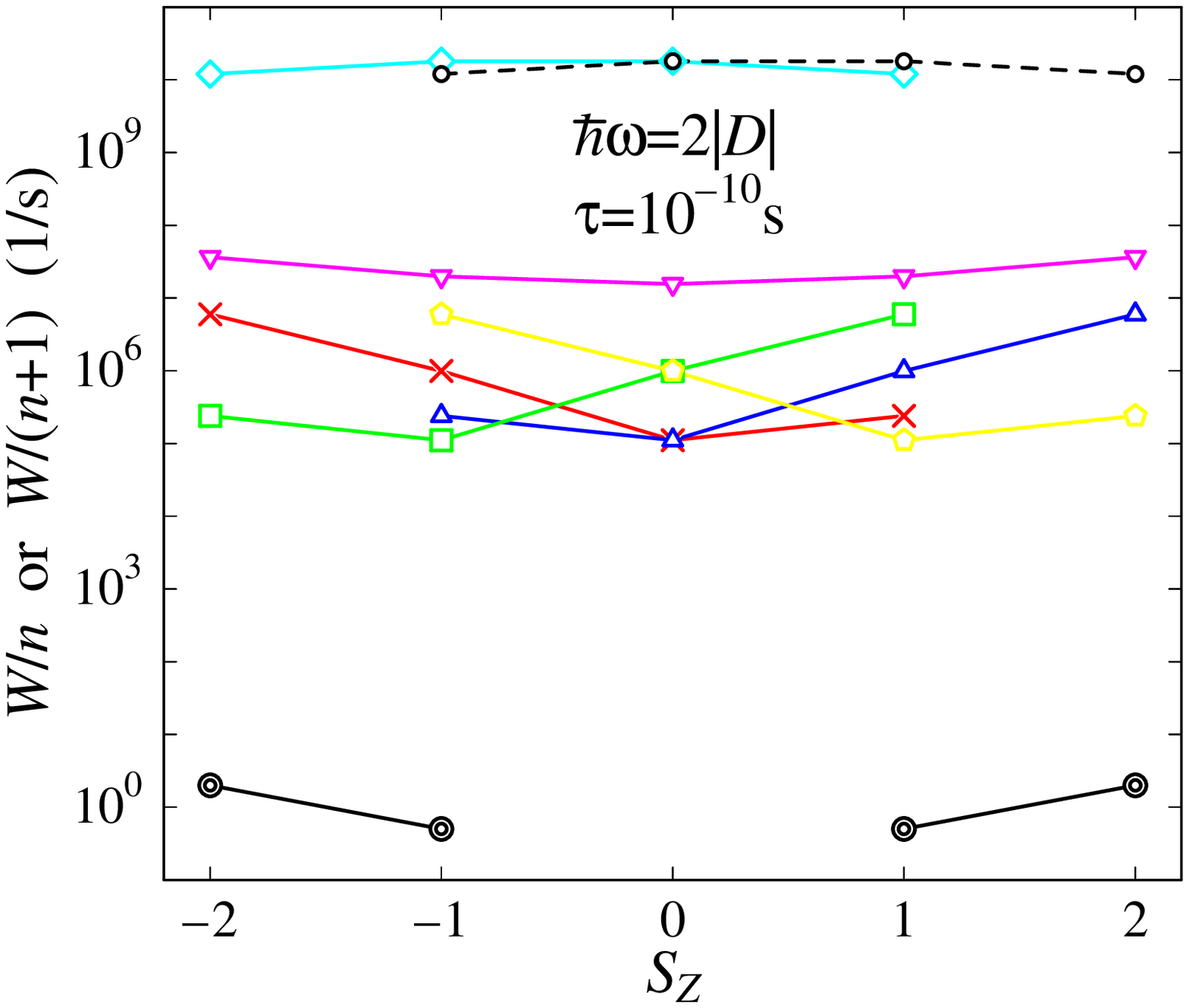}\\
\includegraphics[width=0.45\linewidth]{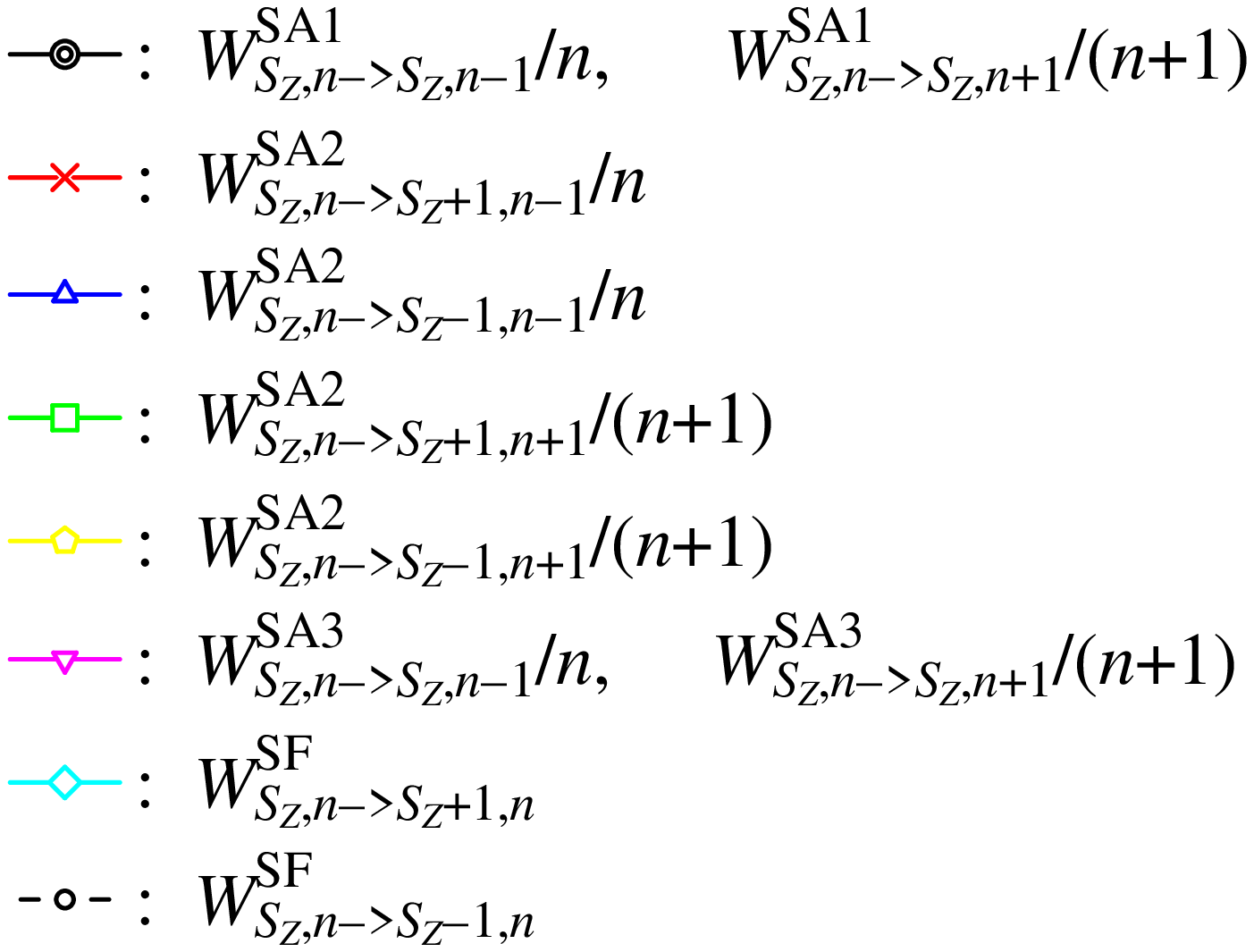}
\caption{
$Z$ component of the spin 
$S_Z$ dependences of 
$W_{S_Z,n \to S_Z,n-1}^{\rm SA1}/n$, 
$W_{S_Z,n \to S_Z \pm 1,n-1}^{\rm SA2}/n$, 
$W_{S_Z,n \to S_Z \pm 1,n+1}^{\rm SA2}/(n+1)$, 
$W_{S_Z,n \to S_Z,n-1}^{\rm SA3}/n$, 
$W_{S_Z,n \to S_Z,n+1}^{\rm SA3}/(n+1)$, 
$W_{S_Z,n \to S_Z \pm 1,n}^{\rm SF}$ of the $S$=2 system 
with $\hbar \omega$=2$|D|$. 
Upper panel: $\tau$=10$^{-5}$ s. 
Lower panel: $\tau$=10$^{-10}$ s. 
The energy levels of this system are shown in Fig. \ref{bistable}. 
}
\label{tran2}
\end{center}
\end{figure}

The parameters are now set to be 
$|D|$=1.5 meV,\cite{Hirjibehedin} 
$\eta$=0.05 (see Appendix \ref{ele_pol}), 
$\Lambda_Z^k$=$-$(1.00$\times$10$^{-5}$)$|D|{\rm i}$, 
$\Lambda_{X,Z}'$=(7.02$\times$10$^{-3}$)$|D|$, 
$\Lambda_{Y,Z}'$=$-$(1.94$\times$10$^{-3}$)$|D|$, 
$G$=9.40$\times$10$^{-3}$, 
$F$=$-$1.25$|D|$, 
$\Lambda_{X,Z}^{\rm SF}$=(7.47$\times$10$^{-1}$)$|D|$, and 
$\Lambda_{Y,Z}^{\rm SF}$=$-$(2.07$\times$10$^{-1}$)$|D|$. 
Here, 
$\Lambda_{X,Z}'$, $\Lambda_{Y,Z}'$, $G$, 
$\Lambda_{X,Z}^{\rm SF}$, and $\Lambda_{Y,Z}^{\rm SF}$ have been 
respectively evaluated from 
$\Lambda_{X,Z;x}'$, $\Lambda_{Y,Z;x}'$, $G_x$, 
$\Lambda_{X,Z}^{\rm SF}$, and $\Lambda_{Y,Z}^{\rm SF}$
at $f$=1 THz in Figs. \ref{sf_f} - \ref{ad_x_f}, 
while 
$\Lambda_Z^k$ has been roughly estimated from 
$\Lambda_{z,x}^k$ at $f$=1 THz 
in the upper panel of Fig. \ref{comp1}. 
In addition, 
$F$ has been evaluated from the above $|D|$ 
and eqs. (\ref{appen_F}) and (\ref{appen_D}) 
with $\Delta_s$=0.45$\Delta_m$.

In Fig. \ref{tran1}, we show 
the $S_Z$ dependence of 
$W_{u \to v}^{i}$ 
in the case of $\hbar \omega$=3$|D|$. 
When $\tau$=10$^{-5}$ s, 
$W_{-2,n \to -1,n-1}^{\rm SA2}/n$, 
$W_{-1,n \to -2,n+1}^{\rm SA2}/(n+1)$, 
$W_{1,n \to 2,n+1}^{\rm SA2}/(n+1)$, 
and 
$W_{2,n \to 1,n-1}^{\rm SA2}/n$ 
take the extremely large values 
because of the transition with the conservation of energy. 
When $\tau$=10$^{-10}$ s, however, 
such probabilities 
get closer to $W_{S_Z,n \to S_Z \pm 1,n}^{\rm SF}$. 
For each $\tau$, 
when a ratio in $W_{u \to v}^{i}$ 
is defined as the largest $W_{u \to v}^{i}$ divided by the smallest one, 
the ratio of $\tau$=10$^{-5}$ s 
is larger than that of $\tau$=10$^{-10}$ s 
because of the sharpness of peaks of $W_{u \to v}^{i}$. 
Figure \ref{tran2} shows 
the $S_Z$ dependence of $W_{u \to v}^i$ 
in the case of $\hbar \omega$=2$|D|$. 
All the $W_{u \to v}^{i}$'s are 
the probabilities for the transitions with the nonconservation of energy. 
The probability 
$W_{S_Z,n \to S_Z \pm 1,n}^{\rm SF}$ 
thus becomes dominant for $\tau$=10$^{-5}$ and 10$^{-10}$ s, 
indicating that the magnitudes of the coefficients of 
${\cal V}_{\rm SF}$ are larger than those of the other coefficients.

\section{Conclusions}
\label{Conc}
Bearing in mind a single magnetic ion on a substrate or 
the ion trapped between a tip and a substrate, 
we derived 
a spin-atomic vibration interaction 
and 
a spin-flip Hamiltonian 
of a single atomic spin in a crystal field due to the surrounding ions. 
Here, 
the spin-atomic vibration interaction meant 
the interaction between the spin of the magnetic ion and 
the vibration of this ion. 
The spin-flip Hamiltonian represented terms 
containing spin-flip operators and no atomic vibration operators. 

In the derivation, 
we applied the perturbation theory to the following model: 
the crystal field potential energy in the equilibrium state 
was contained in the unperturbed Hamiltonian, 
while the spin-orbit interaction and 
the modulation of crystal field potential energy 
due to the vibration displacement were the perturbed terms. 
The model also took into account 
the difference in vibration displacement 
between the effective nucleus and the electrons, $|\eta \Delta r_{\rm n}|$, 
where $\Delta r_{\rm n}$ is the displacement of the nucleus, 
and $\eta$ ($0 \le \eta < 1$) 
is a dimensionless quantity characterizing the difference. 
The unperturbed wave function $\psi_\zeta$ ($\zeta$=1 - 5) was 
roughly set to be 
$\psi_\zeta$=
$d_\zeta + c_d \sum_{m=1 (\ne \zeta)}^5 \overline{d_m} 
+ c_p \sum_{n=1}^3 \overline{p_n}$. 
Here, $d_\zeta$ was the dominant 3d orbital, 
while $\overline{d_m}$ and 
$\overline{p_n}$ were the other 3d orbital 
and 4p orbital in the same magnetic ion, respectively. 
The coefficient for the d orbital, $c_d$, 
(the coefficient for p orbital, $c_p$,) 
represented the degree of d-d hybridization 
(d-p hybridization). 

As results, the spin-atomic vibration interaction 
was obtained as 
$\sum_I c_{1,I} S_I (-a + a^\dag) + \sum_{I,J} (c_{2,I,J} S_I S_J a
+ c_{3,I,J} S_I S_J a^\dag)$, 
and the spin-flip Hamiltonian was expressed as 
$\sum_{I,J} c_{4,I,J} S_I S_J$, 
with $I$, $J$=$X$, $Y$, $Z$. 
Here, $S_I$ was the $I$ component of the spin ${\mbox{\boldmath $S$}}$=$(S_X,S_Y,S_Z)$, and 
$a$ ($a^\dag$) was 
the annihilation operator (creation operator) of 
the vibration of the magnetic ion. 
The coefficients of 
the spin-atomic vibration interaction $c_{1,I}$, $c_{2,I,J}$, and $c_{3,I,J}$ 
contained $\eta$, $c_p$, $c_d$, and $f$, 
while that of the spin-flip Hamiltonian $c_{4,I,J}$ had $c_d$, 
where $f$ was a vibration frequency. 

The respective details about the spin-atomic vibration interaction, 
spin-flip Hamiltonian, and transition probability per unit time 
are written as follows: 
\begin{enumerate}
\item Spin-atomic vibration interaction 

\begin{enumerate}
\item 
The spin-atomic vibration interaction appeared owing to 
the difference in vibration displacement 
and the d-p hybridization. 
Concretely speaking, 
this interaction was proportional to $\eta$ 
and also present for $c_p$$\ne$0. 
The condition of $c_p$$\ne$0 
was attributed to the result that 
the coefficients of this interaction 
contained the similar formula to 
the matrix element of the electric dipole transition. 
\item 
Most of the coefficients of 
the $S_I S_J a_\xi$ and $S_I S_J a_\xi^\dag$ terms 
(i.e., $c_{2,I,J}$ and $c_{3,I,J}$) 
showed the $1/\sqrt{f}$ dependence, 
indicating that 
these coefficients were proportional to 
the vibration displacement of the nucleus, 
$\Delta {\mbox{\boldmath $r$}}_n$ 
($\propto 1/\sqrt{f}$). 
In contrast, 
the coefficient of the $S_I (-a_\xi + a_\xi^\dag)$ term 
(i.e., $c_{1,I}$) 
had the $\sqrt{f}$ dependence 
because 
it was proportional to 
the momentum of the nucleus, 
$\Delta {\mbox{\boldmath $p$}}_n$ ($\propto \sqrt{f}$). 
Note that 
the coefficient of the conventional spin-phonon interaction 
was proportional to $\sqrt{f}$, 
where 
this spin-phonon interaction meant an interaction between a single spin and 
the phonon of a lattice system, 
\item 
\label{conc_1c}
In the case of the Fe ion in the crystal field of tetragonal symmetry, 
the magnitudes of the coefficients of the spin-atomic vibration interaction 
became larger than that of the spin-phonon interaction 
by Mattuck and Strandberg in the specific region of $f$. 
The main reason for this was that the mass of the oscillator 
of the spin-atomic vibration interaction 
was smaller than that of the spin-phonon interaction. 
Here, the oscillator of the spin-atomic vibration interaction 
was the single magnetic ion, 
while that of the spin-phonon interaction consisted of 
all the ions in the unit cell. 
Note that the decrease in the mass of the oscillator 
increased the magnitude of vibration displacement, 
and furthermore 
the increase in the magnitude of the displacement 
enhanced 
the magnitude of the dominant term of the spin-atomic vibration interaction. 
\end{enumerate}
\item 
Spin-flip Hamiltonian 
\begin{enumerate}
\item The spin-flip Hamiltonian stemmed from only the d-d hybridization, 
and was independent of the atomic vibration. 
In other words, the spin-flip Hamiltonian existed only for $c_d$$\ne$0, 
and did not contain $f$ and $\eta$. 
\item 
\label{conc_1d}
In the case of the Fe ion in the crystal field of tetragonal symmetry, 
the magnitudes of the coefficients of the spin-flip Hamiltonian 
tended to be larger than those of the spin-atomic vibration interaction. 
This tendency was attributed to the result that 
the coefficients of the spin-flip Hamiltonian did not contain $\eta$, 
whereas 
those of the spin-atomic vibration interaction 
were proportional to $\eta$, 
where $\eta$=0.05 was set in this paper. 
\end{enumerate}
\item Transition probability per unit time 

The transition probability per unit time 
was investigated for a simple model of the Fe ion. 
The unperturbed Hamiltonian was $-|D|S_Z^2 + \hbar \omega a^\dag a$, 
with $S$=2, 
where $D$ was an anisotropy constant, 
and $\omega$ (=$2\pi f$) was the angular frequency. 
As to $-|D|S_Z^2$, 
the energy difference between 
the ground states and the first excited states was $3|D|$, 
while that between the first excited states and the second excited state 
was $|D|$. 
The perturbed Hamiltonians were composed of 
the representative terms in the spin-atomic vibration interaction 
and spin-flip Hamiltonian. 
Note here that 
the magnitudes of the coefficients of the spin-atomic vibration interaction 
were set to be smaller than those of the spin-flip Hamiltonian 
(see \ref{conc_1d}). 
The transition probability due to the spin-atomic vibration interaction 
(e.g., $S_X S_Z (a+a^\dag)$) 
corresponded to the probability 
for the transition with the conservation of energy, 
depending on $\hbar \omega$ and the initial spin state (i.e., $S_Z$). 
In contrast, 
the transition probability due to the spin-flip Hamiltonian (e.g., $S_X S_Z$) 
became the probability 
for the transition with the nonconservation of energy. 
In the cases of $\hbar \omega$=3$|D|$ and specific $S_Z$'s, 
the transition probability due to the spin-atomic vibration interaction 
took the largest value in all the probabilities 
because of the transition with the conservation of energy. 
In the case of 
$\hbar \omega$=2$|D|$, all the transition probabilities were 
the probabilities 
for the transitions with the nonconservation of energy. 
The transition probability due to the spin-flip Hamiltonian 
thus became dominant, 
indicating that 
the magnitudes of the coefficients of 
the spin-flip Hamiltonian were larger than those of the other coefficients. 
\end{enumerate}

\begin{acknowledgments}
We would like to thank Prof. Toshiharu Hoshino of Shizuoka University 
for useful discussion. 
This work has been supported by 
a Grant-in-Aid for Young Scientists (B) 
(No. 20710076) from 
the Japan Society for the Promotion of Science. 
\end{acknowledgments}

\appendix

\section{Difference in Displacement between Nucleus and Electron}
\label{ele_pol}



Let us consider a model in which 
a single electron is present in the $\sigma$-spin shell 
(e.g., Fe$^{2+}$: an electron exists in the down-spin shell). 
For this model, we evaluate 
the difference in vibration displacement 
between an effective nucleus and the electron 
in the $\sigma$-spin shell. 
The effective nucleus consists of a nucleus and core electrons, 
where the core electrons correspond to 
the electrons other than those in the $\sigma$-spin shell. 
We here use the theory of electronic polarizability in ref. \citen{Kittel}. 




\begin{figure}[ht]
\begin{center}
\includegraphics[width=0.5\linewidth]{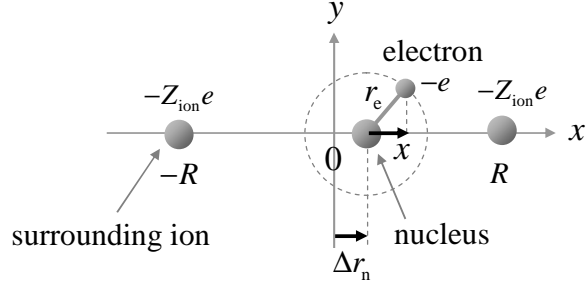}
\caption{
Schematic illustration of the vibrating magnetic ion. 
The effective nucleus has 
the displacement $\Delta r_{\rm n}$ in the $x$ direction. 
The electron then moves in a circular orbit around the nucleus 
fixed at $\Delta r_{\rm n}$, 
and simultaneously experiences 
the crystalline electric field due to the surrounding ions. 
The $x$ component of the position of the electron is 
written as $\Delta r_{\rm n}+x$. 
}
\label{classic}
\end{center}
\end{figure}


We consider a situation in which 
only the magnetic ion vibrates in the $x$ direction, 
while the surrounding ions are rigid. 
Such a situation
is schematically illustrated in Fig. \ref{classic}. 
In particular, this ion 
acts as a harmonic oscillator 
in the vicinity of 
a stable equilibrium position. 
This position corresponds to 
a local minimum point of a potential energy of the system. 
Here, the energy consists of ``a Coulomb potential energy 
between the magnetic ion and the surrounding ions'' 
and ``a repulsive potential energy 
due to the Pauli exclusion principle and the overlap of electrons between 
the magnetic ion and the surrounding ions''.\cite{Kittel1} 
Now, the magnetic ion is regarded as a mass point 
with the mass of this ion, $M$. 
The position of the mass point corresponds to 
the gravity point of the magnetic ion, 
which may be, in other words, 
the position of the effective nucleus. 
When this mass point has 
the displacement from the equilibrium position, 
$\Delta r_{\rm n}$, 
the equation of motion of the mass point is written as 
\begin{eqnarray}
\label{eq_mo}
M \Delta \ddot{r}_{\rm n} = - M \omega_{\rm n}^2 \Delta r_{\rm n}, 
\end{eqnarray}
where $\omega_{\rm n}$ is the vibration frequency of the magnetic ion.

When 
the effective nucleus is located at 
$\Delta r_{\rm n}$, 
the electron is considered to move in a circular orbit 
around ``the nucleus fixed at $\Delta r_{\rm n}$''. 
In particular, we focus on 
the $x$ component of 
the circular motion of the electron 
in the $xy$ plane (see Fig. \ref{classic}). 
The $x$ component of the position of the electron is 
written by $\Delta r_{\rm n} + x$. 
Here, $x$ ($-r_{\rm e} \le x \le r_{\rm e}$) is 
the $x$ component of a vector that points 
from the nucleus to the electron, 
with $r_{\rm e}$ being the electron radius. 
In addition, the electron experiences 
the crystalline electric field due to 
the surrounding ions. 
For this electron, 
we have the following equation of motion 
in the $x$ direction:  
\begin{eqnarray}
\label{eq_mo0}
m (\Delta \ddot{r}_{\rm n} + \ddot{x}) 
= - m \omega_{\rm e}^2 x - e E_{\rm ion}, 
\end{eqnarray}
where $\omega_{\rm e}$ is the angular velocity of the circular motion 
of the electron, $-e$ ($<0$) is the electronic charge, 
and 
$E_{\rm ion}$ is the $x$ component of the crystalline electric field. 

On the left-hand side of eq. (\ref{eq_mo0}), 
we note that 
$|\Delta \ddot{r}_{\rm n}|$ (=$|\omega_{\rm n}^2 \Delta r_{\rm n}|$) 
is estimated to be much smaller than 
$|\ddot{x}|$ ($\sim|\omega_{\rm e}^2 x|$), 
because of $10^{-12} \lesssim (\omega_{\rm n}/\omega_{\rm e})^2 
\lesssim 10^{-4}$ 
and $ 9.47 \times 10^{-3} \lesssim |\Delta r_{\rm n}/x| \lesssim 
9.47 \times 10^{-1}$. 
We here use 
$\omega_{\rm n}$=$2 \pi f$ 
with 0.01 THz $\le f \le$ 100 THz, and 
$\omega_{\rm e}$ of eq. (\ref{omega}). 
The magnitude of the displacement 
$|\Delta r_{\rm n}|$ 
has been considered to be 
9.47$\times$10$^{-4}$ nm 
$\lesssim \sqrt{ \overline{\Delta r_{\rm n}^2}}\lesssim$
9.47$\times$10$^{-2}$ nm 
in Appendix \ref{D_rn_ap}. 
In addition, $|x|$ has been roughly replaced with 
the square root of the time average of $x^2$, i.e., 
$r_{\rm e}/\sqrt{2}$ ($\sim$0.1 nm), 
where $r_{\rm e}$ corresponds to $r_n$ of eq. (\ref{r_n}). 

Thus, by neglecting $\Delta \ddot{r}_{\rm n}$, 
eq. (\ref{eq_mo0}) is rewritten as
\begin{eqnarray}
\label{eq_mo1}
m \ddot{x} = - m \omega_{\rm e}^2 x - e E_{\rm ion}. 
\end{eqnarray}
Regarding $E_{\rm ion}$, 
we consider a case in which 
the surrounding ions with the minus charge $-Z_{\rm ion}e$ ($<0$) 
are respectively 
located at $-R$ and $R$ in the $x$ axis as the point charges. 
Under the conditions of 
$r_{\rm e}/R$$\ll$1, $|x|/R$$\ll$1, and $|\Delta r_{\rm n}|/R\ll$1, 
$E_{\rm ion}$ is obtained as
\begin{eqnarray}
\label{ele_field}
&&\hspace*{-0.4cm}
E_{\rm ion} = \frac{-Z_{\mbox{\footnotesize ion}}e}{4\pi \epsilon_0} 
\left\{
\frac{-(R-\Delta r_{\rm n} -x)}
{\left[ (R-\Delta r_{\rm n} -x)^2 + r_{\rm e}^2 -x^2 \right]^{3/2}} + 
\frac{R+\Delta r_{\rm n} + x}
{\left[(R+\Delta r_{\rm n} + x)^2 + r_{\rm e}^2 -x^2 \right]^{3/2}}
\right\} \nonumber \\
&&\hspace*{0.35cm}\approx 
\frac{Z_{\mbox{\footnotesize ion}}e }{\pi \epsilon_0 R^3} 
(x + \Delta r_{\rm n}), 
\end{eqnarray}
where 
terms higher than the first order 
of $r_{\rm e}/R$, $x/R$, and $\Delta r_{\rm n}/R$ 
have been neglected. 
Note that, 
in eq. (\ref{ele_field}), 
the effect of the repulsive potential energy 
due to the overlap of electrons 
has not been 
taken into account 
for simplicity. 
Substituting eq. (\ref{ele_field}) into eq. (\ref{eq_mo1}), 
we obtain 
\begin{eqnarray}
\label{eq_mo2}
m \ddot{x} = -  \left( m \omega_{\rm e}^2  
+ \frac{Z_{\mbox{\footnotesize ion}}e^2 }{\pi \epsilon_0 R^3}\right) x
- \frac{Z_{\mbox{\footnotesize ion}}e^2 }{\pi \epsilon_0 R^3} 
\Delta r_{\rm n}. 
\end{eqnarray}
Here, by using 
the speed of electron, $v$=$r_{\rm e} \omega_{\rm e}$, 
the expression for the equilibrium of force, 
$mv^2/r_{\rm e}$=$Z_{\rm eff} e^2/(4\pi \epsilon_0 r_{\rm e}^2)$, 
and the Bohr's quantization condition $r_{\rm e} mv$=$n\hbar$ 
with $n$=1, 2, 3, $\cdot \cdot \cdot$, 
$\omega_{\rm e}$ is expressed as 
\begin{eqnarray}
\label{omega}
\omega_{\rm e} = \frac{Z_{\rm eff}^2 h}{2 \pi m n^3 
a_{\mbox{\tiny B}}^2}, 
\end{eqnarray}
with $a_{\mbox{\tiny B}}$=$\epsilon_0 h^2/(\pi m e^2)$, 
where $a_{\mbox{\tiny B}}$ is the Bohr radius.

The center position of the oscillation of the electron, 
$x_c$, corresponds to 
the difference in vibration displacement between the nucleus and the electron. 
This $x_c$ is obtained 
by putting $\ddot{x}$=0 and $x$=$x_c$ in eq. (\ref{eq_mo2}):
\begin{eqnarray}
&&x_c = - \eta \Delta r_{\rm n}, \\
\label{eta_ex}
&&\eta= \frac{4 (Z_{\mbox{\footnotesize ion}}/Z_{\rm eff}) 
\left(  r_n /R \right)^3
}{1+4 (Z_{\mbox{\footnotesize ion}}/Z_{\rm eff})
\left( r_n/R \right)^3}, \\
\label{r_n}
&& r_n= n^2 a_{\mbox{\tiny B}}/Z_{\rm eff}, 
\end{eqnarray}
with $0 \le \eta < 1$. 
Here, $x_c$=0 is confirmed 
for $Z_{\mbox{\footnotesize ion}}$=0.

\begin{figure}[ht]
\begin{center}
\includegraphics[width=0.5\linewidth]{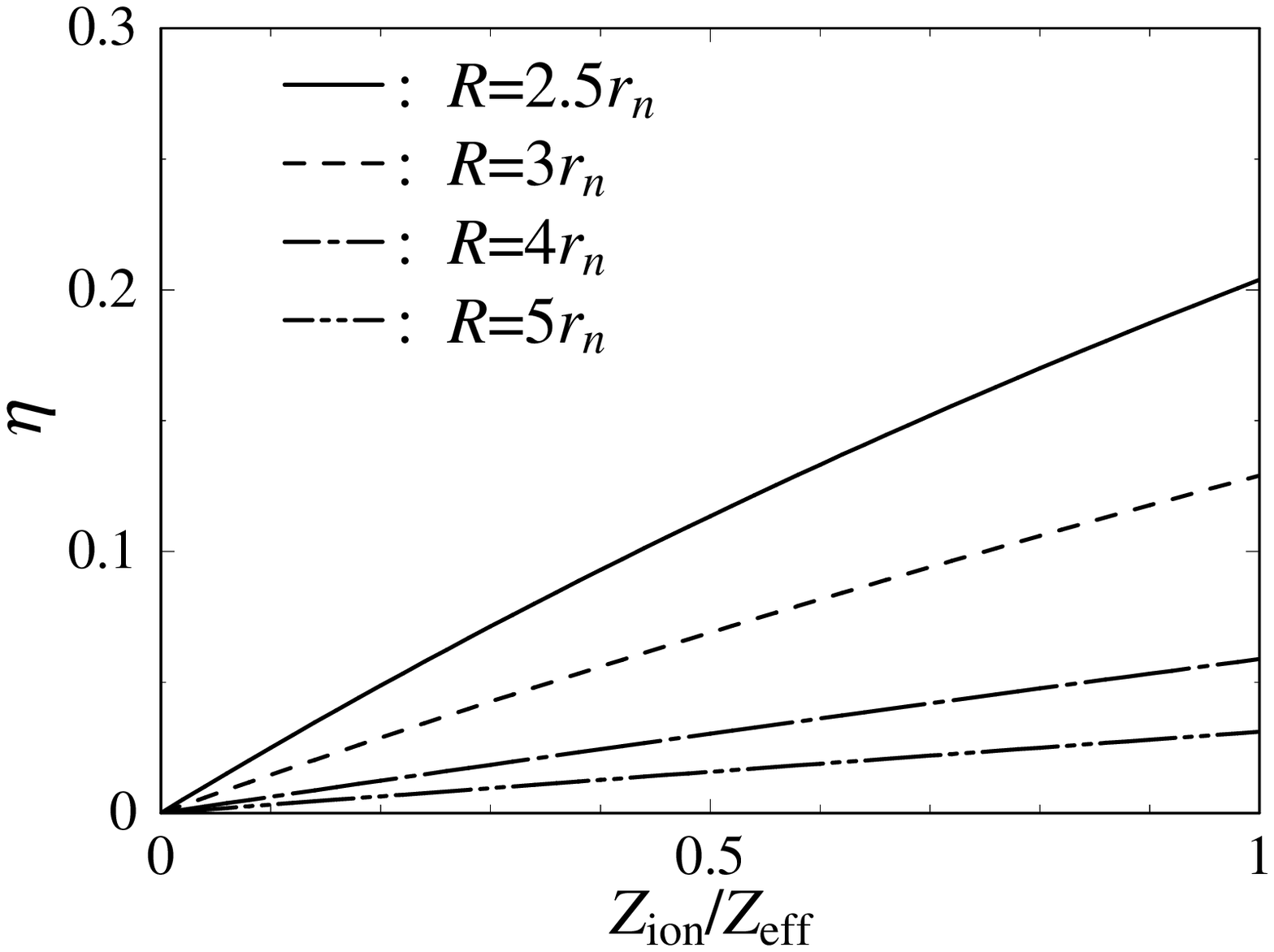}
\caption{
Quantity $Z_{\rm ion}/Z_{\rm eff}$ dependence of 
$\eta$ for $R$=2.5$r_n$, 3$r_n$, 4$r_n$, and 5$r_n$. 
}
\label{eta}
\end{center}
\end{figure}

In Fig. \ref{eta}, we show the $Z_{\mbox{\footnotesize ion}}/Z_{\rm eff}$ 
dependence of $\eta$ for 
$R$=2.5$r_n$, 3$r_n$, 4$r_n$, and 5$r_n$. 
The quantity $\eta$ increases with increasing 
$Z_{\mbox{\footnotesize ion}}/Z_{\rm eff}$. 
In the case of $Z_{\mbox{\footnotesize ion}}/Z_{\rm eff}$=0.2, 
we obtain 
$\eta$=0.05 for $R$=2.5$r_n$, 
$\eta$=0.03 for $R$=3$r_n$, 
$\eta$=0.01 for $R$=4$r_n$, and 
$\eta$=0.006 for $R$=5$r_n$.

As an application, we consider $\eta$ 
of the single Fe ion on the CuN surface.\cite{Hirjibehedin} 
The effective nuclear charge of the 3d orbital of Fe, $Z_{\rm eff}$, 
is obtained as $Z_{\rm eff}$=6.25 
according to Slater's rules,\cite{Slater} 
while $Z_{\mbox{\footnotesize ion}}$ arising from 
the nearest-neighbor N is set to be 
$Z_{\mbox{\footnotesize ion}}$=1.4 
on the basis of 
ref. \citen{Hirjibehedin}. 
From $Z_{\mbox{\footnotesize ion}}/Z_{\rm eff}$=0.224, 
the largest $\eta$ is evaluated to be about 0.05.


\section{Magnitude of Displacement of Nucleus $|\Delta r_{\rm n}|$}
\label{D_rn_ap}

On the basis of eq. (\ref{eq_mo}), 
we evaluate 
the magnitude of 
vibration displacement of the effective nucleus, $|\Delta r_{\rm n}|$. 
Here, $|\Delta r_{\rm n}|$ is defined as 
the square root of the time average of $\Delta r_{\rm n}^2$.

The solution of eq. (\ref{eq_mo}) is first obtained as
\begin{eqnarray}
\label{disp_ap}
\Delta r_{\rm n} = A \cos (\omega_{\rm n} t) 
+ B \sin (\omega_{\rm n} t), 
\end{eqnarray}
where
$A$ and $B$ are coefficients. 
The total energy is then written as 
\begin{eqnarray}
\label{energy_ap}
&&E=\frac{1}{2} M \Delta \dot{r}_{\rm n}^2 
+ \frac{1}{2} M \omega_{\rm n}^2 \Delta r_{\rm n}^2 \nonumber \\
&&\hspace*{0.4cm}=\frac{1}{2} M \omega_{\rm n}^2 ( A^2 + B^2 )
\nonumber \\
&&\hspace*{0.4cm}=M \omega_{\rm n}^2 \overline{\Delta r_{\rm n}^2}. 
\end{eqnarray}
Here, $\overline{\Delta r_{\rm n}^2}$ 
is the time average of $\Delta r_{\rm n}^2$, 
given by
\begin{eqnarray}
\label{time_ave}
\overline{\Delta r_{\rm n}^2} = 
\lim_{t' \to \infty} \frac{1}{t'}
\int_0^{t'} \Delta r_{\rm n}^2 {\rm d}t 
= \frac{1}{2} (A^2 + B^2).
\end{eqnarray}


On the other hand, 
the energy of a quantum harmonic oscillator 
is expressed as 
$( \langle n \rangle + 1/2 ) \hbar \omega_{\rm n}$. 
Here, 
$\langle n \rangle$ is the thermal average of the vibrational quantum number, 
given by 
$\langle n \rangle$
=$[ \exp(\hbar \omega_{\rm n}/k_{\mbox{\tiny B}} T) -1]^{-1}$, 
where $T$ is the temperature 
and $k_{\mbox{\tiny B}}$ is the Boltzmann constant. 

We now equate 
$E$ of eq. (\ref{energy_ap}) 
with the above energy;\cite{Dove} that is, 
\begin{eqnarray}
\label{time_ph}
M \omega_{\rm n}^2 \overline{\Delta r_{\rm n}^2} = 
\left( \langle n \rangle + \frac{1}{2} \right) \hbar \omega_{\rm n}.
\end{eqnarray}
From eq. (\ref{time_ph}), 
we can obtain 
$\sqrt{ \overline{\Delta r_{\rm n}^2} }$. 
At $T$=0 K, 
$\sqrt{ \overline{\Delta r_{\rm n}^2} }$ is written as
\begin{eqnarray}
\label{rmsd1}
\sqrt{ \overline{\Delta r_{\rm n}^2} } = 
\sqrt{ \frac{\hbar}{2 M \omega_{\rm n}}}.
\end{eqnarray}
The quantity $\sqrt{\overline{\Delta r_{\rm n}^2}}$ decreases 
with increasing $M$ and $\omega_{\rm n}$.

We also rewrite eq. (\ref{eq_mo}) as 
$M \Delta \ddot{r}_{\rm n} = - C \Delta r_{\rm n}$, 
with $C \equiv M \omega_{\rm n}^2$. 
Here, $C$ is a force constant 
that 
corresponds to 
the second derivative of the interatomic potential 
with respect to the displacement of the magnetic ion.\cite{Ziman} 
Substituting $\omega_{\rm n}$=$\sqrt{C/M}$ into eq. (\ref{rmsd1}), 
we obtain
\begin{eqnarray}
\label{rmsd2}
\sqrt{ \overline{\Delta r_{\rm n}^2} } = 
\sqrt{ \frac{\hbar}{2 \sqrt{M C}}}. 
\end{eqnarray}
The quantity $\sqrt{\overline{\Delta r_{\rm n}^2}}$ decreases 
with increasing $M$ and $C$. 

As an application, we consider 
$\sqrt{ \overline{\Delta r_{\rm n}^2} }$ 
at $T$=0 K for Fe$^{2+}$, 
where $M$ of Fe$^{2+}$ is 
$M$$\approx$(26+30)$\times$(1.67$\times$10$^{-27}$) kg. 
By using eq. (\ref{rmsd1}), 
$\sqrt{ \overline{\Delta r_{\rm n}^2} }$ is evaluated 
to be 9.47$\times$10$^{-4}$ nm $\lesssim 
\sqrt{ \overline{\Delta r_{\rm n}^2} } \lesssim$9.47$\times$10$^{-2}$ nm 
for 0.01 THz $\le f \le$ 100 THz, 
where $\omega_{\rm n} \equiv 2 \pi f$.

\section{Expression of Coefficients for F\lowercase{e} Ion}
\label{Appendix}

We describe the expressions of coefficients of 
the spin-atomic vibration interaction and the spin-flip Hamiltonian 
of the Fe ion (Fe$^{2+}$) 
in the crystal field of tetragonal symmetry. 
For convenience, 
the coefficients are represented by the following dimensionless quantities: 
$\Lambda_{\mu,\xi}^{(1)}/\lambda$, 
$\Lambda_{\mu,\xi}^k/\lambda$, 
$G_\xi$, $F/D$, 
$\Lambda_{I,J;\xi}'/D$, 
$\Lambda_{\mu;\nu,\xi}^{\pm}/D$, 
$\Lambda_{\mu,\xi;\nu}^{\pm}/D$, and 
$\Lambda_{I,J}^{\rm SF}/D$.

We here focus on one down-spin electron 
as described in \S \ref{Application}. 
The energy levels are assumed to be $E_j$ of Fig. \ref{energy}. 
In addition, we use the notation of 
$|\Psi_1)$=$|xy)$, 
$|\Psi_2)$=$|yz)$, 
$|\Psi_3)$=$|xz)$, 
$|\Psi_4)$=$|x^2-y^2)$, 
$|\Psi_5)$=$|3z^2 -r^2)$, 
$|d_1 \rangle$=$|xy \rangle$, 
$|d_2 \rangle$=$|xz \rangle$, 
$|d_3 \rangle$=$|yz \rangle$, 
$|d_4 \rangle$=$|x^2 -y^2 \rangle$, 
$|d_5 \rangle$=$|3z^2 -r^2 \rangle$, 
$|p_1 \rangle$=$|x \rangle$, 
$|p_2 \rangle$=$|y \rangle$, and 
$|p_3 \rangle$=$|z \rangle$. 
As to $E_j$, $e_j$, $c_{d_m}^{(j)}$, $c_{p_n}^{(j)}$, and $C_j$, 
we have the notation and parameter setup in \S \ref{notation}. 

In the below-mentioned calculations, 
we will use the following matrix elements: 
\begin{eqnarray}
\label{xz_Lx_xy}
&&\langle xz | L_x | xy \rangle={\rm i}, \\
&&\langle x^2 - y^2 |L_x | yz \rangle={\rm i}, \\
&&\langle 3z^2 -r^2 |L_x |yz \rangle=\sqrt{3}{\rm i}, \\
&&\langle yz | L_y | xy \rangle=-{\rm i}, \\
&&\langle x^2 - y^2 |L_y |xz \rangle={\rm i}, \\
&&\langle 3z^2 - r^2 |L_y |xz \rangle=-\sqrt{3}{\rm i}, \\
&&\langle x^2 - y^2 | L_z | xy \rangle=-2 {\rm i}, \\
\label{xz_Lz_yz}
&&\langle xz | L_z | yz \rangle=- {\rm i}, \\
&&\langle z |L_x |y \rangle={\rm i}, \\
&&\langle x |L_y |z \rangle={\rm i}, \\
&&\langle y |L_z |x \rangle={\rm i}, 
\end{eqnarray}
with $\langle d_{j'} |{\mbox{\boldmath $L$}} |d_j \rangle$=
$- \langle d_j |{\mbox{\boldmath $L$}} |d_{j'} \rangle$ 
and $\langle p_{j'} |{\mbox{\boldmath $L$}} |p_j \rangle$=
$- \langle p_j |{\mbox{\boldmath $L$}} |p_{j'} \rangle$.\cite{LLLLL}

\subsection{Principal axis coordinate system}
\label{app_theta}
When an electron in the crystal field of tetragonal symmetry 
has $E_j$ in Fig. \ref{energy}, 
we find that 
$X$, $Y$, and $Z$ are written as $X$=$x$, $Y$=$y$, and $Z$=$z$, respectively.

Using eqs. (\ref{xz_Lx_xy}) - (\ref{xz_Lz_yz}), 
we obtain ${\cal L}_{\mu,\nu}^{(0)}$ of eq. (\ref{lam_0}) as
\begin{eqnarray}
{\mbox{\boldmath ${\cal L}$}}^{(0)}=
\left(
  \begin{array}{ccc}
   {\cal L}_{x,x}^{(0)} & 0 & 0 \\
   0 & {\cal L}_{y,y}^{(0)} & 0  \\
   0 & 0 & {\cal L}_{z,z}^{(0)}  
  \end{array}
\right), 
\end{eqnarray}
where
\begin{eqnarray}
\label{lam_xx^0}
&&\hspace*{-0.5cm} {\cal L}_{x,x}^{(0)}=C^4
\frac{\langle xy | L_x | xz \rangle 
\langle xz | L_x | xy \rangle }{ e_{xy} - e_{xz} }=-\frac{C^4}
{\Delta_s}, 
\\
\label{lam_yy^0}
&&\hspace*{-0.5cm} {\cal L}_{y,y}^{(0)}= C^4
\frac{\langle xy | L_y | yz \rangle 
\langle yz | L_y | xy \rangle }{ e_{xy} - e_{yz} }= -\frac{C^4}
{\Delta_s}, 
\\
\label{lam_zz^0}
&&\hspace*{-0.5cm} {\cal L}_{z,z}^{(0)}=C^4
\frac{\langle xy | L_z | x^2 -y^2 \rangle 
\langle x^2 -y^2 | L_z | xy \rangle }{ e_{xy} - e_{x^2 -y^2} } 
= -\frac{4C^4}{\Delta_m}, 
\\
&&\hspace*{-0.5cm} C=\left( 1 + 4 c_d^2 + 3 c_p^2 \right)^{-1/2},
\end{eqnarray}
with
${\cal L}_{\mu,\nu}^{(0)}$=0 for $\mu$$\ne$$\nu$. 
Here, we have roughly set 
$e_1-e_j$=$E_1-E_j$, 
taking into account $|\Delta e_1-\Delta e_j|$$\ll$$|e_1-e_j|$ 
for $j \ne 1$ (see eq. (\ref{E_j})).

Since ${\cal L}^{(0)}$ is a diagonal matrix, 
we obtain 
$\theta_{X,x}$=$\theta_{Y,y}$=$\theta_{Z,z}$=0 
in eqs. (\ref{Unitary}) and (\ref{Unitary^T}), 
and the other $\theta_{I,\mu}$'s are $\pi$/2. 
Namely, 
$X$, $Y$, and $Z$ become $X$=$x$, $Y$=$y$, and $Z$=$z$, respectively.

\subsection{$D$, $E$, and $F$}
\label{app_DE}
Using eqs. 
(\ref{DDDDD}) - (\ref{FFFFF}) and 
(\ref{lam_xx^0}) - (\ref{lam_zz^0}), 
we obtain 
$D$, $E$, and $F$ as
\begin{eqnarray}
\label{appen_D}
&&\hspace*{-0.5cm} 
D= 
\lambda^2 
\left( {\cal L}_{Z,Z}^{(0)} - 
\frac{{\cal L}_{X,X}^{(0)} + {\cal L}_{Y,Y}^{(0)}}{2} \right) 
=\lambda^2 
\frac{ C^4}{\Delta_m} 
\left( \frac{\Delta_m}{\Delta_s} - 4 \right), 
\\
\label{E=0}
&&\hspace*{-0.5cm} 
E= 
\lambda^2 
\frac{ {\cal L}_{X,X}^{(0)} - {\cal L}_{Y,Y}^{(0)}}{2}=0, \\
\label{appen_F}
&&\hspace*{-0.5cm} 
F=
\lambda^2 
\frac{ {\cal L}_{X,X}^{(0)} + {\cal L}_{Y,Y}^{(0)}}{2}
= - \lambda^2 \frac{C^4}{\Delta_s}, 
\end{eqnarray}
with 
${\cal L}_{X,X}^{(0)}$=${\cal L}_{x,x}^{(0)}$, 
${\cal L}_{Y,Y}^{(0)}$=${\cal L}_{y,y}^{(0)}$, 
and ${\cal L}_{Z,Z}^{(0)}$=${\cal L}_{z,z}^{(0)}$. 
The bistability (i.e., $D <$ 0) 
is obtained for $\Delta_m/\Delta_s < 4$. 
The coefficient $D$ comes close to $-3\lambda^2 C^4/\Delta_m$ 
as $\Delta_m/\Delta_s$ comes to 1. 
In addition, we find that the conditions of $\Delta_s \sim \Delta_m$ 
and a small $\Delta_m$ as well as 
a large $|\lambda|$ enhance $|D|$. 
The coefficient $E$ 
becomes 0 
owing to ${\cal L}_{X,X}^{(0)}$=${\cal L}_{Y,Y}^{(0)}$.

\subsection{$\Lambda_{\mu,\xi}^{(1)}/\lambda $}
\label{app_first}
Using eqs. (\ref{L_mxi^(1)}), (\ref{so_lam}), and $|\Psi_1 )$=$|xy)$, 
we have 
\begin{eqnarray}
\label{A_mn^pm}
\frac{\Lambda_{\mu,\xi}^{(1)}}{\lambda} = 
-\eta
\frac{1}{\hbar} 
{\rm i} \frac{m}{M} \sqrt{\frac{M \hbar \omega_\xi}{2}}
( xy | \mu | xy), 
\end{eqnarray}
with $\mu$, $\xi$=$x$, $y$, $z$. 
By using $| xy)$ in eq. (\ref{dj}), 
$( xy | \mu | xy)$ is obtained as follows: 
\begin{eqnarray}
&&\hspace*{-0.3cm} 
( xy | x | xy ) =
C^2 c_p \left[ 
\left( \langle xy |x \overline{| y \rangle} 
+
\overline{ \langle y |}x | xy \rangle \right) 
+c_d \left( 
\overline{\langle xz |}x \overline{| z \rangle} 
+ \overline{\langle z |}x \overline{| xz \rangle} 
+ \overline{\langle x^2 -y^2 |}x \overline{| x \rangle} 
+ \overline{\langle x |}x \overline{| x^2 -y^2 \rangle} \right. \right. \nonumber \\
&& \hspace*{1.7cm} 
\left. \left. + \overline{\langle 3z^2 -r^2 |}x \overline{| x \rangle} 
+ \overline{\langle x |}x \overline{| 3z^2 -r^2 \rangle} 
\right) \right], \\
&&\hspace*{-0.3cm} 
( xy | y | xy ) =
C^2 c_p \left[ 
\left( \langle xy |y \overline{| x \rangle} 
+
\overline{ \langle x |}y | xy \rangle \right) 
+c_d \left( 
\overline{\langle yz |}y \overline{| z \rangle} 
+ \overline{\langle z |}y \overline{| yz \rangle} 
+ \overline{\langle x^2 -y^2 |}y \overline{| y \rangle} 
+ \overline{\langle y |}y \overline{| x^2 -y^2 \rangle} \right. \right. \nonumber \\
&&\hspace*{1.7cm} 
\left. \left. + \overline{\langle 3z^2 -r^2 |}y \overline{| y \rangle} 
+ \overline{\langle y |}y \overline{| 3z^2 -r^2 \rangle} 
\right) \right], \\
&&\hspace*{-0.3cm} 
( xy | z | xy ) =
C^2 c_d c_p \left( 
\overline{\langle xz |} z \overline{| x \rangle} 
+
\overline{ \langle x |}z \overline{| xz \rangle} 
+ 
\overline{\langle yz |}z \overline{| y \rangle} 
+ \overline{\langle y |}z \overline{| yz \rangle} 
+ \overline{\langle 3z^2 -r^2 |}z \overline{| z \rangle} 
+ \overline{\langle z |}z \overline{| 3z^2 -r^2 \rangle} 
\right), \nonumber \\
\end{eqnarray}
where $( xy | \mu | xy)$
vanishes at $c_p$=0. 
The matrix element $\langle p_{j'} |\mu| d_j \rangle$ 
is obtained using eqs. (\ref{xyf32}) - (\ref{zf31}):
\begin{eqnarray}
\label{x_y_xy}
&&\langle x | y | xy \rangle=-8.31 \times 10^{-12}, \\
&&\langle x | x |3z^2 -r^2 \rangle=4.80 \times 10^{-12}, \\
&&\langle z | z |3z^2 -r^2 \rangle=-9.59 \times 10^{-12}, \\
&&\langle  x| x  |x^2 -y^2 \rangle=-8.31 \times 10^{-12}, \\
\label{y_y_x2}
&&\langle  y| y  |x^2 -y^2 \rangle=8.31 \times 10^{-12}, 
\end{eqnarray}
where 
$\langle d_j |\mu| p_{j'} \rangle$=$\langle p_{j'} |\mu| d_j \rangle$.

\subsection{$\Lambda_{\mu,\xi}^k/\lambda $}
\label{app_first}
Using eqs. (\ref{Lam^k}) and (\ref{so_lam}), 
we obtain 
$\Lambda_{\mu,\xi}^k/\lambda$ as
\begin{eqnarray}
\label{A^k}
\frac{\Lambda_{\mu,\xi}^k}{\lambda} = 
-2 \eta \frac{m}{M} \frac{1}{\hbar}
\sqrt{\frac{M\hbar \omega_\xi}{2}}
\sum_{j (\ne 1)} (xy |L_\mu|\Psi_j )(\Psi_j |\xi|xy), 
\end{eqnarray}
for $\mu$, $\xi$=$x$, $y$, $z$. 
The matrix element $(xy | {\mbox{\boldmath $L$}} | \Psi_j )$ 
is expressed by 
\begin{eqnarray}
\label{xy_Lx_yz}
&&(xy|L_x|yz)=C^2 \left[ c_d \langle xy |L_x \overline{|xz \rangle} 
+ c_d \left( \overline{\langle x^2 -y^2 |} L_x |yz \rangle 
+ \overline{ \langle 3z^2 -r^2 |} L_x |yz \rangle \right) 
+c_p^2 \left( \overline{ \langle y |} L_x \overline{|z \rangle}
+ \overline{ \langle z |} L_x \overline{|y \rangle} \right) 
\right. \nonumber \\
&& \hspace*{2cm} 
\left. +c_d^2 \left( \overline{ \langle xz |} L_x \overline{|xy \rangle}
+ \overline{ \langle yz |} L_x \overline{|x^2 -y^2 \rangle} 
+ \overline{ \langle yz |} L_x \overline{|3z^2 -r^2 \rangle} \right) \right] 
= \sqrt{3}c_d(1-c_d)C^2 {\rm i}, \\
&&(xy|L_x|xz)=C^2 \left[ \langle xy |L_x|xz \rangle +
c_p^2 \left( \overline{ \langle y |} L_x \overline{|z \rangle}
+ \overline{ \langle z |} L_x \overline{|y \rangle} \right) 
+c_d^2 \left( \overline{\langle xz |} L_x \overline{|xy \rangle } 
+ \overline{\langle x^2 -y^2 |} L_x \overline{|yz \rangle } \right. \right. 
\nonumber \\
&&\hspace*{2.cm}
\left. \left. + \overline{\langle yz |} L_x \overline{|x^2 -y^2 \rangle } 
+ \overline{\langle 3z^2 -r^2 |} L_x \overline{|yz \rangle } 
+ \overline{\langle yz |} L_x \overline{|3z^2-r^2 \rangle } \right) \right] 
= (c_d^2 -1 )C^2 {\rm i}, \\
&&(xy|L_x |x^2 -y^2)= 
C^2 \left[ 
c_d \left( \langle xy |L_x \overline{ |xz \rangle} 
+ \overline{ \langle yz |} L_x | x^2 -y^2 \rangle \right) 
+c_p^2 \left( \overline{ \langle y |} L_x \overline{|z \rangle}
+ \overline{ \langle z |} L_x \overline{|y \rangle} \right) \right. 
\nonumber \\
&&\hspace*{2.8cm} 
\left. + c_d^2 \left( \overline{ \langle xz |} L_x \overline{|xy \rangle } 
+
\overline{ \langle x^2 -y^2 |} L_x \overline{|yz \rangle } 
+
\overline{ \langle 3z^2 -r^2 |} L_x \overline{|yz \rangle } 
+
\overline{ \langle yz |} L_x \overline{|3z^2 -r^2 \rangle }  \right) \right] 
\nonumber \\
&&\hspace*{2.5cm} = 2c_d(c_d -1)C^2 {\rm i}, \\
&&(xy|L_x|3z^2 -r^2) =
C^2 \left[ 
c_d \left( \langle xy |L_x \overline{ | xz \rangle } + 
\overline{ \langle yz |} L_x | 3z^2 -r^2 \rangle \right) 
+c_p^2 \left( \overline{ \langle y |} L_x \overline{|z \rangle}
+ \overline{ \langle z |} L_x \overline{|y \rangle} \right) \right. 
\nonumber \\
&&\hspace*{2.95cm} 
\left. 
+c_d^2 \left( 
\overline{ \langle xz |} L_x \overline{|xy \rangle } 
+
\overline{ \langle x^2 -y^2 |} L_x \overline{|yz \rangle } 
+
\overline{ \langle yz |} L_x \overline{|x^2 -y^2 \rangle } 
+
\overline{ \langle 3z^2 -r^2 |} L_x \overline{|yz \rangle } 
\right) \right] \nonumber \\
&&\hspace*{2.65cm} =(1 + \sqrt{3}) c_d(c_d -1)C^2 {\rm i},  \\
&&(xy|L_y|yz)=
C^2 \left[ 
\langle xy |L_y |yz \rangle 
+c_p^2 \left( \overline{ \langle x |} L_y \overline{|z \rangle}
+ \overline{ \langle z |} L_y \overline{|x \rangle} \right)
+ c_d^2 \left( 
\overline{\langle yz |} L_y \overline{|xy \rangle} 
+ 
\overline{\langle x^2 -y^2 |} L_y \overline{|xz \rangle} \right. \right. 
\nonumber \\
&&\hspace*{2cm} 
\left. \left. 
+ 
\overline{\langle xz |} L_y \overline{|x^2 -y^2 \rangle} 
+ 
\overline{\langle 3z^2 -r^2 |} L_y \overline{|xz \rangle} 
+\overline{\langle xz |} L_y \overline{|3z^2 -r^2 \rangle} 
\right) \right] 
= (1-c_d^2 )C^2 {\rm i}, \\
&&(xy|L_y | xz)=
C^2 \left[ 
c_d \left( 
\langle xy | L_y \overline{|yz \rangle} 
+
\overline{\langle x^2 -y^2 |} L_y |xz \rangle 
+ 
\overline{\langle 3z^2 -r^2 |} L_y |xz \rangle
\right) 
+c_p^2 \left( \overline{ \langle x |} L_y \overline{|z \rangle}
+ \overline{ \langle z |} L_y \overline{|x \rangle} \right) \right. 
\nonumber \\
&&\hspace*{2cm} 
\left. 
+ c_d^2 \left( 
\overline{\langle yz |} L_y \overline{|xy \rangle} 
+ 
\overline{\langle xz |} L_y \overline{|x^2 -y^2 \rangle} 
+ 
\overline{\langle xz |} L_y \overline{|3z^2 -r^2 \rangle} 
\right) \right]=(2 - \sqrt{3}) c_d(1-c_d)C^2 {\rm i}, \\
&&(xy|L_y|x^2 -y^2)=
C^2 \left[ 
c_d \left( 
\langle xy | L_y \overline{|yz \rangle} 
+
\overline{\langle xz |} L_y |x^2 -y^2 \rangle  \right) 
+c_p^2 \left( \overline{ \langle x |} L_y \overline{|z \rangle}
+ \overline{ \langle z |} L_y \overline{|x \rangle} \right) \right. 
\nonumber \\
&&\hspace*{2.8cm} \left. 
+ c_d^2 \left(
\overline{\langle yz |} L_y \overline{|xy \rangle} 
+ 
\overline{\langle x^2 -y^2 |} L_y \overline{|xz \rangle} 
+ 
\overline{\langle 3z^2 -r^2 |} L_y \overline{|xz \rangle} 
+ 
\overline{\langle xz |} L_y \overline{|3z^2 -r^2 \rangle} \right) \right] 
= 0, \nonumber \\\\
&&(xy|L_y|3z^2 -r^2)=
C^2 \left[ 
c_d \left(
\langle xy | L_y \overline{|yz \rangle}
+ 
\overline{\langle xz |} L_y |3z^2 -r^2 \rangle \right)
+c_p^2 \left( \overline{ \langle x |} L_y \overline{|z \rangle}
+ \overline{ \langle z |} L_y \overline{|x \rangle} \right) \right. 
\nonumber \\
&&\hspace*{2.95cm} \left. 
+ c_d^2 \left( 
\overline{\langle yz |} L_y \overline{|xy \rangle} 
+ 
\overline{\langle x^2 -y^2 |} L_y \overline{|xz \rangle} 
+ 
\overline{\langle xz |} L_y \overline{|x^2 -y^2 \rangle} 
+ 
\overline{\langle 3z^2 -r^2 |} L_y \overline{|xz \rangle} 
\right) \right]\nonumber \\
&&\hspace*{2.65cm}=(1+\sqrt{3}) c_d(1-c_d)C^2 {\rm i}, \\
&&(xy|L_z|yz)= 
C^2 \left[ 
c_d \left(
\langle xy | L_z \overline{|x^2 -y^2 \rangle} 
+
\overline{\langle xz |} L_z |yz \rangle \right) 
+c_p^2 \left( \overline{ \langle x |} L_z \overline{|y \rangle}
+ \overline{ \langle y |} L_z \overline{|x \rangle} \right) \right. 
\nonumber \\
&&\hspace*{2cm} \left. 
+ c_d^2 \left( 
\overline{\langle x^2 -y^2 |} L_z \overline{|xy \rangle} 
+ 
\overline{\langle yz |} L_z \overline{|xz \rangle} 
\right) \right]=c_d(1-c_d)C^2 {\rm i}, \\
&&(xy|L_z|xz)=
C^2 \left[ 
c_d \left( 
\langle xy | L_z \overline{|x^2 -y^2 \rangle}
+ 
\overline{\langle yz |} L_z |xz \rangle  \right)
+c_p^2 \left( \overline{ \langle x |} L_z \overline{|y \rangle}
+ \overline{ \langle y |} L_z \overline{|x \rangle} \right) \right. 
\nonumber \\
&&\hspace*{2cm} \left. 
+ c_d^2 \left( 
\overline{\langle xz |} L_z \overline{|yz \rangle} 
+ 
\overline{\langle x^2-y^2 |} L_z \overline{|xy \rangle} 
\right) \right]=3c_d(1-c_d)C^2 {\rm i}, \\
&&(xy|L_z|x^2 -y^2)= C^2 \left[ \langle xy | L_z |x^2 -y^2 \rangle 
+c_p^2 \left( \overline{ \langle x |} L_z \overline{|y \rangle}
+ \overline{ \langle y |} L_z \overline{|x \rangle} \right) \right. 
\nonumber \\
&&\hspace*{2.8cm} \left. 
+ c_d^2 \left( 
\overline{\langle x^2 -y^2 |} L_z \overline{|xy \rangle} 
+ \overline{\langle xz |} L_z \overline{|yz \rangle} 
+ \overline{\langle yz |} L_z \overline{|xz \rangle} 
\right) \right]=2(1-c_d^2)C^2 {\rm i}, \\
\label{xy_Lz_3z}
&&(xy|L_z|3z^2 -r^2)=
C^2 \left[ c_d \langle xy | L_z \overline{|x^2 -y^2 \rangle} 
+c_p^2 \left( \overline{ \langle x |} L_z \overline{|y \rangle}
+ \overline{ \langle y |} L_z \overline{|x \rangle} \right)\right. 
\nonumber \\
&&\hspace*{2.9cm} \left. 
+ c_d^2 \left( 
\overline{\langle x^2 -y^2 |} L_z \overline{|xy \rangle} 
+ 
\overline{\langle xz |} L_z \overline{|yz \rangle} 
+ 
\overline{\langle yz |} L_z \overline{|xz \rangle} 
\right) \right]=2c_d(1-c_d)C^2 {\rm i}, 
\end{eqnarray}
where 
$(\Psi_j|${\mbox{\boldmath $L$}}$|\Psi_j)$=0 and 
$(\Psi_{j'}|${\mbox{\boldmath $L$}}$|\Psi_j)$=$-(\Psi_j|${\mbox{\boldmath $L$}}$|\Psi_{j'})$ for $j$$\ne$$j'$.\cite{LLLLL}
The matrix element $(\Psi_{j} |\xi |xy)$ is obtained as 
\begin{eqnarray}
\label{xz_x_xy}
&&\hspace*{-0.3cm} ( xz | x | xy ) =
C^2 c_p \left[ 
\left( \langle xz |x \overline{| z \rangle} 
+\overline{ \langle y |}x | xy \rangle \right) 
+ c_d \left( \overline{\langle z |}x \overline{| xz \rangle} 
+ \overline{\langle xy |}x \overline{| y \rangle }
+\overline{\langle x |}x \overline{| x^2 -y^2 \rangle} 
 \right. \right. \nonumber \\
&&\hspace*{1.5cm}\left.\left. 
+\overline{\langle x^2 -y^2 |}x \overline{| x \rangle}
+\overline{\langle x |}x \overline{| 3z^2 -r^2 \rangle} +\overline{\langle 3z^2 -r^2 |}x \overline{| x \rangle} \right) \right], \\
&&\hspace*{-0.3cm} ( xz | y | xy )=C^2 c_p 
\left[ \overline{\langle x |} y | xy \rangle 
+ c_d \left( 
\overline{\langle xy |} y \overline{| x \rangle} 
+  \overline{\langle z |} y \overline{| yz \rangle} 
+ \overline{\langle yz |} y \overline{| z \rangle} 
+ \overline{\langle y |} y \overline{| x^2 -y^2 \rangle} 
+ \overline{\langle x^2 -y^2 |} y \overline{| y \rangle} 
\right. \right. \nonumber \\
&&\hspace*{1.5cm}
\left. \left. 
+ \overline{\langle y |} y \overline{| 3z^2 -r^2 \rangle} 
+ \overline{\langle 3z^2 -r^2 |} y \overline{| y \rangle} \right) \right], \\
&&\hspace*{-0.3cm} ( xz | z | xy )=C^2 c_p 
\left[ \langle xz |z \overline{| x \rangle }
+ c_d \left( \overline{\langle x |}z \overline{| xz \rangle} 
+ \overline{\langle y |}z \overline{| yz \rangle} 
+ \overline{\langle yz |}z \overline{| y \rangle} 
+ \overline{\langle z |}z \overline{| 3z^2 -r^2 \rangle} 
+ \overline{\langle 3z^2 -r^2 |}z \overline{| z \rangle} \right) \right], 
\nonumber \\\\
&&\hspace*{-0.3cm} ( yz | x | xy )=C^2 c_p \left[ 
\overline{\langle y |}x | xy \rangle + c_d
\left( \overline{\langle xy |}x \overline{| y \rangle} 
+ \overline{\langle z |} x \overline{| xz \rangle} 
+ \overline{\langle xz |}x \overline{| z \rangle }
+\overline{\langle x |}x \overline{| x^2 -y^2 \rangle} 
+\overline{\langle x^2 -y^2 |}x \overline{| x \rangle} \right. \right.
\nonumber \\
&&\hspace*{1.5cm}\left. \left.
+\overline{\langle x |}x \overline{| 3z^2 -r^2 \rangle} 
+\overline{\langle 3z^2 -r^2 |}x \overline{| x \rangle} 
\right)\right], \\
&&\hspace*{-0.3cm} ( yz | y | xy )=C^2 c_p \left[ 
\left( \langle yz |y \overline{| z \rangle} 
+ \overline{\langle x |}y | xy \rangle \right)
+c_d \left( \overline{\langle z |} y \overline{| yz \rangle} 
+ \overline{\langle xy |} y \overline{| x \rangle }
+ \overline{\langle y |} y \overline{| x^2 -y^2 \rangle }
+ \overline{\langle x^2 -y^2 |} y \overline{| y \rangle }
 \right. \right. 
\nonumber \\
&& \hspace*{1.5cm}\left. \left. 
+ \overline{\langle y |} y \overline{| 3z^2 -r^2 \rangle}  
+ \overline{\langle 3z^2 -r^2 |} y \overline{| y \rangle }
\right) \right], \\
&&\hspace*{-0.3cm} ( yz | z | xy )=
C^2 c_p \left[ \langle yz |z \overline{| y \rangle }
+ c_d \left( \overline{\langle y |}z \overline{| yz \rangle} 
+ \overline{\langle xz |}z \overline{| x \rangle }
+ \overline{\langle x |}z \overline{| xz \rangle}
+ \overline{\langle z |}z \overline{| 3z^2 -r^2 \rangle }
+ \overline{\langle 3z^2 -r^2 |}z \overline{| z \rangle} \right) \right], 
\nonumber \\\\
&&
\hspace*{-0.3cm}
( x^2 -y^2 | x | xy )=C^2 c_p \left[ 
\left( \langle x^2 -y^2 |x \overline{| x \rangle} 
+ \overline{\langle y |} x | xy \rangle \right) 
+ c_d \left( 
\overline{\langle x |}x \overline{| x^2 -y^2 \rangle }
+\overline{\langle xy |}x \overline{| y \rangle }
+\overline{\langle xz |}x \overline{| z \rangle }
+\overline{\langle z |}x \overline{| xz \rangle}
\right. \right. \nonumber \\
&&\hspace*{2.25cm}
\left.\left. 
+\overline{\langle x |}x \overline{| 3z^2 -r^2 \rangle} 
+\overline{\langle 3z^2 -r^2 |}x \overline{| x \rangle }
\right) \right], \\
&&
\hspace*{-0.3cm}
( x^2 -y^2 | y | xy )=C^2 c_p \left[ 
\left(
\langle x^2 -y^2 | y \overline{| y \rangle} 
+ \overline{\langle x |}y|xy \rangle 
\right) 
+ c_d \left( 
\overline{\langle y |} y \overline{| x^2 -y^2 \rangle} + 
\overline{\langle xy |} y \overline{| x \rangle} 
+ \overline{\langle z |} y \overline{| yz \rangle }
+ \overline{\langle yz |} y \overline{| z \rangle}
\right.\right. \nonumber \\
&&\hspace*{2.25cm}
\left. \left. 
+ \overline{\langle y |} y \overline{| 3z^2 -r^2 \rangle} 
+ \overline{\langle 3z^2 -r^2 |} y \overline{| y \rangle }
\right) \right], \\
&&
\hspace*{-0.3cm}
( x^2 -y^2 | z | xy )= C^2 
c_d c_p \left( \overline{\langle x |}z \overline{| xz \rangle} 
+ \overline{\langle xz |}z \overline{| x \rangle} 
+ \overline{\langle y |}z \overline{| yz \rangle}
+ \overline{\langle yz |}z \overline{| y \rangle} 
+ \overline{\langle z |}z \overline{| 3z^2 -r^2 \rangle} 
+ \overline{\langle 3z^2 -r^2 |}z \overline{| z \rangle} \right), 
\nonumber \\\\
&&
\hspace*{-0.3cm} 
( 3z^2 -r^2 | x | xy )=C^2 c_p \left[ 
\left( \langle 3z^2 -r^2 |x \overline{| x \rangle} +
\overline{\langle y |}x | xy \rangle \right) + c_d \left(
\overline{\langle x |}x \overline{| 3z^2 -r^2 \rangle}
+\overline{\langle xy |}x \overline{| y \rangle}
+\overline{\langle xz |}x \overline{| z \rangle} 
\right. \right. \nonumber \\
&&\hspace*{2.4cm}
\left. 
\left. 
+\overline{\langle z |}x \overline{| xz \rangle}
+\overline{\langle x^2 -y^2 |}x \overline{| x \rangle}
+\overline{\langle x |}x \overline{| x^2 -y^2 \rangle}
\right) \right],\\
&&
\hspace*{-0.3cm}
( 3z^2 -r^2 | y | xy )=C^2 c_p \left[ 
\left( \langle 3z^2 -r^2 |y \overline{| y \rangle} +
\overline{\langle x |}y | xy \rangle \right) 
+ c_d \left(
\overline{\langle y |}y \overline{| 3z^2 -r^2 \rangle} 
+\overline{\langle xy |}y \overline{| x \rangle} 
+ \overline{\langle yz |}y \overline{| z \rangle} 
\right. \right. \nonumber \\
&&\hspace*{2.4cm}
\left. \left. 
+\overline{\langle z |}y \overline{| yz \rangle}
+\overline{\langle x^2 -y^2 |}y \overline{| y \rangle}
+\overline{\langle y |}y \overline{| x^2 -y^2 \rangle}
\right) \right], \\
\label{3z_z_zy}
&&
\hspace*{-0.3cm}
( 3z^2 -r^2 | z | xy )=C^2 c_p \left[ 
\langle 3z^2 -r^2 |z \overline{| z \rangle}  
+ c_d \left(\overline{\langle z |}z \overline{| 3z^2 -r^2 \rangle}
+\overline{\langle xz |}z \overline{| x \rangle}
+\overline{\langle x |}z \overline{| xz \rangle}
+\overline{\langle yz |}z \overline{| y \rangle}
+\overline{\langle y |}z \overline{| yz \rangle} \right) \right], 
\nonumber \\
\end{eqnarray}
where $(\Psi_{j'} |\xi | \Psi_j)$=$(\Psi_j |\xi | \Psi_{j'})$. 
Here, 
$\langle p_{j'} |\xi| d_j \rangle$ 
is given by eqs. (\ref{x_y_xy}) - (\ref{y_y_x2}).

\subsection{$\Lambda_{I,J}^{\rm SF}/D $}
\label{Lam_IJ_D}
Using eqs. (\ref{Lam^SF}) and (\ref{DDDDD}), 
we obtain 
$\Lambda_{I,J}^{\rm SF}/D$ as
\begin{eqnarray}
\label{Lambda_IJ^SF/D}
\displaystyle{ \frac{ \Lambda_{I,J}^{\rm SF} }{ D  } }=
\frac{{\cal L}_{I,J}'}
{  {\cal L}_{Z,Z}^{(0)} - 
( {\cal L}_{X,X}^{(0)} + {\cal L}_{Y,Y}^{(0)} )/2},
\end{eqnarray}
with $I$, $J$=$X$, $Y$, $Z$. 
Here, 
${\cal L}_{Z,Z}^{(0)} - ({\cal L}_{X,X}^{(0)} + {\cal L}_{Y,Y}^{(0)})/2$ 
(see eq. (\ref{appen_D})) becomes 
\begin{eqnarray}
\label{denominator}
{\cal L}_{Z,Z}^{(0)} 
- \frac{{\cal L}_{X,X}^{(0)} + {\cal L}_{Y,Y}^{(0)}}{2} 
= \left( \frac{1}{\Delta_s} - \frac{4}{\Delta_m} \right) C^4. 
\end{eqnarray}
Using eqs. (\ref{xy_Lx_yz}) - (\ref{xy_Lz_3z}), 
we obtain ${\cal L}_{I,J}'$ 
of eqs. (\ref{L_I,J'}) (or ${\cal L}_{\mu,\nu}'$ of eq. (\ref{L_mu,nu'})) as 
\begin{eqnarray}
&&\hspace*{-1cm}{\cal L}_{X,Z}'={\cal L}_{x,z}'=
\frac{( xy | L_x | yz ) 
( yz | L_z | xy ) }{ E_{xy} - E_{yz} } 
+ 
\frac{( xy | L_x | xz ) 
( xz | L_z | xy ) }{ E_{xy} - E_{xz} } 
+ 
\frac{( xy | L_x | x^2 -y^2 ) 
( x^2 -y^2 | L_z | xy ) }{ E_{xy} - E_{x^2 -y^2} }  
\nonumber \\
&&\hspace*{1.4cm}
+ 
\frac{( xy | L_x | 3z^2 -r^2 ) 
( 3z^2 -r^2 | L_z | xy ) }{ E_{xy} - E_{3z^2 -r^2} } 
- {\cal L}_{x,z}^{(0)} \nonumber \\
&&\hspace*{0.9cm}= 
C^4 c_d(c_d-1) \left[
\frac{(3-\sqrt{3})c_d^2 + \sqrt{3}c_d -3}{\Delta_s} 
+\frac{4(c_d^2-1)}{\Delta_m} 
+\frac{2(1+\sqrt{3})c_d(c_d-1)}{\Delta_l} 
\right], \\
&&\hspace*{-1cm}{\cal L}_{Y,Z}'={\cal L}_{y,z}'=
\frac{( xy | L_y | yz ) 
( yz | L_z | xy ) }{ E_{xy} - E_{yz} } 
+ 
\frac{( xy | L_y | xz ) 
( xz | L_z | xy ) }{ E_{xy} - E_{xz} } 
+ 
\frac{( xy | L_y | 3z^2 -r^2 ) 
( 3z^2 -r^2 | L_z | xy ) }{ E_{xy} - E_{3z^2 -r^2} } 
- {\cal L}_{y,z}^{(0)}\nonumber \\
&&\hspace*{0.9cm}
=
C^4 c_d(c_d-1) \left[
\frac{(-7 + 3\sqrt{3})c_d^2 + (6-3\sqrt{3})c_d + 1}{\Delta_s}
+ \frac{-(2+2\sqrt{3}) c_d^2 + (2+2 \sqrt{3})c_d}{\Delta_l} \right], \\
&&\hspace*{-1cm}{\cal L}_{X,Y}'={\cal L}_{x,y}'=
\frac{( xy | L_x | yz ) 
( yz | L_y | xy ) }{ E_{xy} - E_{yz} } 
+ 
\frac{( xy | L_x | xz ) 
( xz | L_y | xy ) }{ E_{xy} - E_{xz} } 
+ 
\frac{( xy | L_x | 3z^2 -r^2 ) 
( 3z^2 -r^2 | L_y | xy ) }{ E_{xy} - E_{3z^2 -r^2} } 
- {\cal L}_{x,y}^{(0)} \nonumber \\
&&\hspace*{0.9cm}
=C^4 c_d(c_d-1) \left[
\frac{ (2 - 2\sqrt{3})c_d^2 + 2\sqrt{3}-2}{\Delta_s} 
+ \frac{ (1+\sqrt{3})^2 (c_d^2 -c_d)}{\Delta_l} 
\right], \\
&&\hspace*{-1cm}{\cal L}_{X,X}'={\cal L}_{x,x}'=
\frac{( xy | L_x | yz ) 
( yz | L_x | xy ) }{ E_{xy} - E_{yz} } 
+ 
\frac{( xy | L_x | xz ) 
( xz | L_x | xy ) }{ E_{xy} - E_{xz} } 
+ 
\frac{( xy | L_x | x^2 -y^2 ) 
( x^2 -y^2 | L_x | xy ) }{ E_{xy} - E_{x^2 -y^2} } 
\nonumber \\
&&\hspace*{1.4cm}
+ 
\frac{( xy | L_x | 3z^2 -r^2 ) 
( 3z^2 -r^2 | L_x | xy ) }{ E_{xy} - E_{3z^2 -r^2} } 
- {\cal L}_{x,x}^{(0)} \nonumber \\
&&\hspace*{0.9cm}
= 
C^4 c_d \left\{ 
- \frac{4 c_d + 2}{\Delta_s} - \frac{4 c_d}{\Delta_m}
- \frac{(1 + \sqrt{3})^2 c_d}{\Delta_l}
+ (c_d-2) \left[ 
-\frac{ 4 c_d^2 +2 c_d +1}{\Delta_s} 
- \frac{4c_d^2}{\Delta_m} 
- \frac{(1+ \sqrt{3})^2 c_d^2}{\Delta_l} 
\right] \right\}, \nonumber \\\\
&&\hspace*{-1cm}{\cal L}_{Y,Y}'={\cal L}_{y,y}'=
\frac{( xy | L_y | yz ) 
( yz | L_y | xy ) }{ E_{xy} - E_{yz} } 
+ 
\frac{( xy | L_y | xz ) 
( xz | L_y | xy ) }{ E_{xy} - E_{xz} } 
+
\frac{( xy | L_y | 3z^2 -r^2 ) 
( 3z^2 -r^2 | L_y | xy ) }{ E_{xy} - E_{3z^2 -r^2} } 
- {\cal L}_{y,y}^{(0)} \nonumber \\
&&\hspace*{0.9cm}
= 
C^4 c_d \left\{
- \frac{( 8 - 4\sqrt{3})c_d + 2}{\Delta_s}
- \frac{(1+\sqrt{3})^2 c_d}{\Delta_l} \right. \nonumber \\
&&\hspace*{1.2cm} \left. + (c_d -2)
\left[
- \frac{(8 -4 \sqrt{3})c_d^2 + 2c_d + 1}{\Delta_s} 
- \frac{(1+\sqrt{3})^2 c_d^2 }{\Delta_l} 
\right] \right\}, \\
&&\hspace*{-1cm}{\cal L}_{Z,Z}'={\cal L}_{z,z}'=
\frac{( xy | L_z | yz ) 
( yz | L_z | xy ) }{ E_{xy} - E_{yz} } 
+ 
\frac{( xy | L_z | xz ) 
( xz | L_z | xy ) }{ E_{xy} - E_{xz} } 
+ 
\frac{( xy | L_z | x^2 -y^2 ) 
( x^2 -y^2 | L_z | xy ) }{ E_{xy} - E_{x^2 -y^2} } 
\nonumber \\
&&\hspace*{1.4cm}
+ 
\frac{( xy | L_z | 3z^2 -r^2 ) 
( 3z^2 -r^2 | L_z | xy ) }{ E_{xy} - E_{3z^2 -r^2} } 
- {\cal L}_{z,z}^{(0)}\nonumber \\
&&\hspace*{0.9cm}
= 
C^4 c_d \left\{
- \frac{10 c_d}{\Delta_s} - \frac{4 (c_d + 2)}{\Delta_m} 
- \frac{4 c_d}{\Delta_l}
+ (c_d -2) 
\left[ 
- \frac{10c_d^2}{\Delta_s}
- \frac{4(c_d +1)^2}{\Delta_m}
- \frac{4c_d^2}{\Delta_l}
\right] \right\}, 
\end{eqnarray}
with 
${\cal L}_{x,z}^{(0)}$=${\cal L}_{y,z}^{(0)}$=${\cal L}_{x,y}^{(0)}$=0. 
In addition, we have 
${\cal L}_{Z,X}'$=${\cal L}_{z,x}'$=${\cal L}_{x,z}'$, 
${\cal L}_{Z,Y}'$=${\cal L}_{z,y}'$=${\cal L}_{y,z}'$, 
${\cal L}_{Y,X}'$=${\cal L}_{y,x}'$=${\cal L}_{x,y}'$.\cite{LL} 
Note that ${\cal L}_{I,J}'$ becomes 0 at $c_d$=0.

\subsection{$G_\xi$ and $F/D$}
\label{sub_G_i}
The coefficient $G_\xi$ is given by eq. (\ref{G_xi}):
\begin{eqnarray}
\label{G_xi1}
G_\xi=2 \eta
\frac{ \left\langle 3 \xi r^{-5} \right\rangle}
{\left\langle r^{-3} \right\rangle } 
\sqrt{\frac{ \hbar}{2 M \omega_\xi}}, 
\end{eqnarray}
with $\xi$=$x$, $y$, $z$. 
Here, $\langle r^{-3} \rangle$ and 
$\langle 3 \xi r^{-5} \rangle$ 
are regarded as the respective ground-state expectation values: 
\begin{eqnarray}
\label{r^-3_xy}
&&\langle r^{-3} \rangle = ( xy | r^{-3} | xy ) \nonumber \\
&&\hspace*{0.92cm}= C^2 \left[ \langle xy | r^{-3} | xy \rangle 
+ c_p^2 
\left( \overline{\langle x |} r^{-3} \overline{| x \rangle} 
+ \overline{\langle y |} r^{-3} \overline{| y \rangle} 
+ \overline{\langle z |} r^{-3} \overline{| z \rangle} \right) 
+ c_d^2 \left( \overline{\langle yz |} r^{-3} \overline{| yz \rangle} 
+\overline{\langle xz |} r^{-3} \overline{| xz \rangle} \right. \right. 
\nonumber \\
&& \hspace*{1.2cm}
\left.\left.
+ \overline{\langle x^2 -y^2 |} r^{-3} \overline{| x^2 -y^2 \rangle} 
+\overline{\langle 3z^2 -r^2 |} r^{-3} 
\overline{| 3z^2 -r^2 \rangle} 
\right) \right], \\
\label{3xr^-5}
&&\langle 3 x r^{-5} \rangle =
(xy | 3 x r^{-5} | xy) \nonumber \\
&&\hspace*{1.3cm}=
C^2 c_p \left[ 
\left( \langle xy | 3 x r^{-5} \overline{| y \rangle} 
+ \overline{\langle y |} 3 x r^{-5} | xy \rangle \right) 
+ c_d \left( 
\overline{\langle x^2 -y^2 |} 3 x r^{-5} \overline{|x \rangle} 
+ 
\overline{\langle x |} 3 x r^{-5} \overline{| x^2 -y^2 \rangle} 
\right. \right. 
\nonumber \\
&&\hspace*{1.6cm} \left. \left. + 
\overline{\langle 3z^2 -r^2 |} 3 x r^{-5} \overline{| x \rangle} 
+ \overline{\langle x |} 3 x r^{-5} \overline{| 3z^2 -r^2 \rangle}  
+ \overline{\langle xz |} 3 x r^{-5} \overline{| z \rangle} 
+ \overline{\langle z |} 3 x r^{-5} \overline{| xz \rangle} \right) \right], \\
\label{3yr^-5}
&&\langle 3 y r^{-5} \rangle =
(xy | 3 y r^{-5} | xy) \nonumber \\
&&\hspace*{1.28cm}=
C^2 c_p \left[ 
\left( \langle xy | 3 y r^{-5} \overline{| x \rangle} 
+ \overline{\langle x |} 3 y r^{-5} | xy \rangle \right) 
+ c_d \left( 
\overline{\langle x^2 -y^2 |} 3 y r^{-5} \overline{|y \rangle} 
+ 
\overline{\langle y |} 3 y r^{-5} \overline{| x^2 -y^2 \rangle} 
\right. \right. 
\nonumber \\
&&\hspace*{1.6cm} \left. \left. + 
\overline{\langle 3z^2 -r^2 |} 3 y r^{-5} \overline{| y \rangle} 
+ \overline{\langle y |} 3 y r^{-5} \overline{| 3z^2 -r^2 \rangle}  
+ \overline{\langle yz |} 3 y r^{-5} \overline{| z \rangle} 
+ \overline{\langle z |} 3 y r^{-5} \overline{| yz \rangle} \right) \right], \\
\label{3zr^-5}
&&\langle 3 z r^{-5} \rangle =
(xy | 3 z r^{-5} | xy) \nonumber \\
&&\hspace*{1.28cm}=
C^2 
c_d c_p \left( \overline{\langle xz |} 3 z r^{-5} \overline{|x \rangle} 
+ \overline{\langle x |} 3 z r^{-5} \overline{| xz \rangle} 
+
\overline{\langle yz |} 3 z r^{-5} \overline{| y \rangle} 
+ 
\overline{\langle y |} 3 z r^{-5} \overline{| yz \rangle}  
\right. 
\nonumber \\
&&\hspace*{1.6cm}\left.  +  
\overline{\langle 3z^2 -r^2 |} 3 z r^{-5} \overline{| z \rangle} 
+ \overline{\langle z |} 3 z r^{-5} \overline{| 3z^2 -r^2 \rangle} \right), 
\end{eqnarray}
with 
\begin{eqnarray}
\label{r^-3}
&&\langle x | r^{-3} | x \rangle 
= 2.42 \times 10^{30}, \\
&&\langle xy | r^{-3} | xy \rangle= 4.08 \times 10^{30}, \\
&&\langle 3z^2 -r^2 | r^{-3} | 3z^2 -r^2 \rangle
=  4.06 \times 10^{30}, \\
&&\langle x^2 -y^2 | r^{-3} | x^2 -y^2 \rangle
=  4.06 \times 10^{30}, \\
&&\langle xy | 3x r^{-5} | y \rangle=1.26 \times 10^{41}, \\
&& \langle x^2 -y^2 | 3x r^{-5} |x \rangle 
=  1.26 \times 10^{41}, \\
&& \langle x^2 -y^2 | 3y r^{-5} |y \rangle 
=   -1.26 \times 10^{41}, \\
&& \langle 3z^2 -r^2 | 3x r^{-5} |x \rangle 
=  -7.26 \times 10^{40}, \\
\label{3zr^-5}
&& \langle 3z^2 -r^2 | 3z r^{-5} |z \rangle 
=  1.45 \times 10^{41}, 
\end{eqnarray}
where eqs. (\ref{r^-3}) - (\ref{3zr^-5}) have been obtained by using 
eqs. (\ref{xyf32}) - (\ref{zf31}).

\subsection{$\Lambda_{I,J;\xi}'/D$}
\label{subsub_d}
Using eqs. (\ref{Lam_IJxi}) and (\ref{DDDDD}), 
we have 
\begin{eqnarray}
\label{Lam_IJi_D}
&&\frac{ \Lambda_{I,J;\xi}' }{D}=2 \eta
\frac{ \left\langle 3 \xi r^{-5} \right\rangle}
{\left\langle r^{-3} \right\rangle } 
\sqrt{\frac{ \hbar}{2 M \omega_\xi}} 
\frac{{\cal L}_{I,J}'}
{  {\cal L}_{Z,Z}^{(0)} - 
( {\cal L}_{X,X}^{(0)} + {\cal L}_{Y,Y}^{(0)} )/2}, 
\end{eqnarray}
for $I$, $J$=$X$, $Y$, $Z$ and $\xi$=$x$, $y$, $z$. 
This $\Lambda_{I,J;\xi}'/D$ is also represented by 
$\Lambda_{I,J;\xi}'/D$=$G_\xi \Lambda_{I,J}^{\rm SF}/D$ 
(see eqs. (\ref{G_xi}) and (\ref{Lambda_IJ^SF/D})).

\subsection{$\Lambda_{\mu;\nu,\xi}^{\pm}/D$} 
\label{A/A}
Using eqs. (\ref{xy_Lx_yz}) - (\ref{xy_Lz_3z}), 
we express $\Lambda_{\mu;\nu,\xi}^{\pm}/D$ of eq. (\ref{lam_d}) 
with $\mu$, $\nu$, $\xi$=$x$, $y$, $z$ as 
\begin{eqnarray}
&&\hspace*{-0.8cm}
\frac{\Lambda_{x;\nu,\xi}^{\pm}}{D}=
\left[ 
\frac{ \sqrt{3}c_d(1-c_d) {\rm i}(yz|\nu|xy)}
{-\Delta_s} \kappa_{s,\xi}^{\pm} 
+ \frac{ (c_d^2 -1) {\rm i}(xz|\nu|xy)}{-\Delta_s} \kappa_{s,\xi}^{\pm}  
+ \frac{ 2c_d(c_d-1) {\rm i}(x^2 -y^2|\nu|xy)}
{-\Delta_m} \kappa_{m,\xi}^{\pm}  
\right. \nonumber \\
&& \left. \left. 
\hspace*{0.5cm} 
+ \frac{ (1+\sqrt{3})c_d(c_d-1) {\rm i}(3z^2 -r^2|\nu|xy)}{-\Delta_l} 
\kappa_{l,\xi}^{\pm} \right] \right/ 
\left( \frac{1}{\Delta_s} - \frac{4}{\Delta_m} \right) C^2, \\
&&\hspace*{-0.8cm}
\frac{\Lambda_{y;\nu,\xi}^{\pm}}{D}=
\left[ 
\frac{ (1-c_d^2) {\rm i}(yz| \nu | xy)}{-\Delta_s} \kappa_{s,\xi}^{\pm}  
+ \frac{ (2-\sqrt{3}) c_d(1-c_d) {\rm i}(xz| \nu | xy)}{-\Delta_s} 
\kappa_{s,\xi}^{\pm} \right. 
\nonumber \\
&& \left. \left.  \hspace*{0.5cm}
+ 
\frac{ (1+\sqrt{3}) c_d(1-c_d) {\rm i} (3z^2-r^2| \nu | xy)}{-\Delta_l} 
\kappa_{l,\xi}^{\pm} \right] \right/ 
\left( \frac{1}{\Delta_s} - \frac{4}{\Delta_m} \right) C^2, 
\\
&&\hspace*{-0.8cm}
\frac{\Lambda_{z;\nu,\xi}^{\pm}}{D}=
\left[ 
\frac{ c_d(1-c_d) {\rm i}(yz| \nu | xy)}{-\Delta_s} \kappa_{s,\xi}^{\pm}  
+ 
\frac{ 3c_d(1-c_d) {\rm i}(xz| \nu | xy)}{-\Delta_s} \kappa_{s,\xi}^{\pm}  
+
\frac{ 2(1-c_d^2) {\rm i}
(x^2 -y^2| \nu | xy)}{-\Delta_m} \kappa_{m,\xi}^{\pm} \right. 
 \nonumber \\
&& \left. \left. \hspace*{0.5cm}
+ 
\frac{ 2c_d(1-c_d) {\rm i}
(3z^2 -r^2| \nu | xy)}{-\Delta_l} \kappa_{l,\xi}^{\pm} \right] \right/
\left( \frac{1}{\Delta_s} - \frac{4}{\Delta_m} \right) C^2, 
\end{eqnarray}
with 
\begin{eqnarray}
\label{kappa}
\kappa_{u,\xi}^{\pm} =
\eta
\frac{{\rm i}}{\hbar} 
\left(
\frac{m}{M} \sqrt{ \frac{M \hbar \omega_\xi}{2}}
\pm 
\sqrt{\frac{\hbar}{2 M \omega_\xi}} \frac{m}{\hbar} \Delta_u 
\right), 
\end{eqnarray}
for $u$=$s$, $m$, and $l$. 
The matrix element $(\Psi_{j'} | \nu | \Psi_j)$ is given by 
eqs. (\ref{xz_x_xy}) - (\ref{3z_z_zy}). 
We also note that 
$\Lambda_{\mu;\nu,\xi}^{\pm}$ 
satisfies 
$\Lambda_{\mu;\nu,\xi}^{\pm}$=$-\Lambda_{\nu,\xi;\mu}^{\mp}$
=$(\Lambda_{\mu;\nu,\xi}^{\pm} )^*$ 
for $\mu$, $\nu$, $\xi$=$x$, $y$, $z$,\cite{Lambda} 
where 
$\Lambda_{\nu,\xi;\mu}^{\mp}$ is eq. (\ref{lam_dd}).


\begin{thebibliography}{10}
\bibitem{ex_negle}
In the spin-phonon interaction,\cite{Mattuck,Hartmann} 
each magnetic ion (or each magnetic molecule) in the crystal 
has a single spin, 
where the exchange interactions between the spins are neglected. 
The spin-phonon interaction 
originally represents an interaction between 
the spin located at the origin and the phonon of a lattice system.
\bibitem{Z_eff}
According to Slater's rule,\cite{Slater} 
$Z_{\rm eff}$ of Fe is evaluated to be 6.25.
\bibitem{disp_ref}
For example, see Appendix \ref{D_rn_ap}. 
As found from eq. (\ref{rmsd2}), 
the magnitude of vibration displacement decreases with increasing mass 
and force constant. 
When the vibration of the surrounding ions is considered, 
the magnetic ion (the surrounding ions) 
in this Appendix should be 
replaced with the surrounding ions (the other ions adjacent to them 
in the substrate).
\bibitem{LLLLL}
We here use 
the wave functions of 3d orbitals, i.e., 
$d_1$=$xy f_{3,2}(r)$, 
$d_2$=$xz f_{3,2}(r)$, $d_3$=$yz f_{3,2}(r)$, 
$d_4$=$[(x^2 -y^2)/2] f_{3,2}(r)$, 
and $d_5$=$[(3z^2 -r^2)/2\sqrt{3}] f_{3,2}(r)$ 
(see eqs. (\ref{xyf32}) - (\ref{3z2f32})), 
where $d_\zeta$ ($\zeta$=1 - 5) is a real function. 
We then have $\langle d_\zeta | {\mbox{\boldmath $L$}} |d_\zeta \rangle$=0 
and $\langle d_{\zeta'}| {\mbox{\boldmath $L$}} |d_\zeta \rangle$=
$-\langle d_\zeta | {\mbox{\boldmath $L$}} |d_{\zeta'} \rangle$ 
for $\zeta$$\ne$$\zeta'$, 
because ${\mbox{\boldmath $L$}}$ is purely imaginary and a Hermitian operator. 
Next, 
we consider $\Psi_j$=$\sum_{\zeta} c_{\zeta}^{(j)} d_\zeta$, 
where $\Psi_j$ is chosen to be a real function. 
We have $( \Psi_j | {\mbox{\boldmath $L$}} |\Psi_j )$=
$\sum_{\zeta'} \sum_{\zeta}$$ c_{\zeta'}^{(j)} c_{\zeta}^{(j)} 
\langle d_{\zeta'}| {\mbox{\boldmath $L$}} |d_{\zeta} \rangle$
= $\sum_{\zeta'>\zeta} c_{\zeta'}^{(j)} c_{\zeta}^{(j)}$
$\left( \langle d_{\zeta'}| {\mbox{\boldmath $L$}} |d_{\zeta} \rangle 
+ \langle d_{\zeta}| {\mbox{\boldmath $L$}} |d_{\zeta'} \rangle \right)$=0. 
Furthermore, when $j$$\ne$$j'$, we obtain $( \Psi_{j'} | {\mbox{\boldmath $L$}} | \Psi_j )$=$\sum_{\zeta'} \sum_{\zeta}$$ c_{\zeta'}^{(j')} c_{\zeta}^{(j)}  $$ \langle d_{\zeta'}| {\mbox{\boldmath $L$}} |d_{\zeta} \rangle$
= 
$\sum_{\zeta} \sum_{\zeta'} $$ c_{\zeta}^{(j)} c_{\zeta'}^{(j')}$$ \left(-\langle d_{\zeta}| {\mbox{\boldmath $L$}} |d_{\zeta'} \rangle \right)$
=$-( \Psi_j | {\mbox{\boldmath $L$}} |\Psi_{j'} )$.
\bibitem{LL}
Regarding ${\cal L}_{\mu,\nu}$ of eq. (\ref{cal L}), 
we obtain ${\cal L}_{\mu,\nu}$=${\cal L}_{\nu,\mu}$ using 
$(\Psi_{j'} | L_\mu | \Psi_j)$ = $-(\Psi_j | L_\mu | \Psi_{j'})$.\cite{LLLLL} 
In a similar way, 
${\cal L}_{\mu,\nu}^{(0)}$=
${\cal L}_{\nu,\mu}^{(0)}$ 
is obtained using 
$ \langle \Psi_{j'} | L_\mu | \Psi_j \rangle$ 
= $- \langle \Psi_j | L_\mu | \Psi_{j'} \rangle$.\cite{LLLLL} 
From ${\cal L}_{\mu,\nu}'$=${\cal L}_{\mu,\nu} - {\cal L}_{\mu,\nu}^{(0)}$, 
we have 
${\cal L}_{\mu,\nu}'$=${\cal L}_{\nu,\mu}'$.
\bibitem{Lambda}
Using 
$(\Psi_{j'} | L_\mu | \Psi_j) = -(\Psi_j | L_\mu | \Psi_{j'})$ 
(see ref. \citen{LLLLL})
and 
$(\Psi_{j'} | F_{\nu,\xi}^{\pm} | \Psi_j)$ 
= 
$(\Psi_j | F_{\nu,\xi}^{\mp} | \Psi_{j'})$ of eq. (\ref{jFi}),
we obtain 
$\Lambda_{\mu;\nu,\xi}^{\pm}$=
$\eta
\lambda^2 
(1/\hbar)$
$\sum_{j (\ne 1)}  
(\Psi_1 | L_\mu | \Psi_j)$
$(\Psi_j | F_{\nu,\xi}^{\pm} | \Psi_1)$$/( E_1 - E_j)$
=
$-\eta
\lambda^2 
(1/\hbar)$
$\sum_{j (\ne 1)}  
(\Psi_1 | F_{\nu,\xi}^{\mp} | \Psi_j)$
$(\Psi_j | L_\mu | \Psi_1)$$/( E_1 - E_j)$
=
$-\Lambda_{\nu,\xi;\mu}^{\mp}$. 
In addition, 
we have 
$(\Lambda_{\mu;\nu,\xi}^{\pm})^*$=
$\eta
\lambda^2 (1/\hbar)$
$\sum_{j (\ne 1)}  
(\Psi_1 | (F_{\nu,\xi}^{\pm})^\dag | \Psi_j)$
$(\Psi_j | (L_\mu)^\dag | \Psi_1)$$/( E_1 - E_j)$
=
$\eta
\lambda^2 (1/\hbar)$
$\sum_{j (\ne 1)}  
(\Psi_1 | (-F_{\nu,\xi}^{\mp}) | \Psi_j)$
$(\Psi_j | L_\mu | \Psi_1)$$/( E_1 - E_j)$
=$-\Lambda_{\nu,\xi;\mu}^{\mp}$ = $\Lambda_{\mu;\nu,\xi}^{\pm}$.
\bibitem{coefficient}
On the basis of the coefficient in eq. (7) in ref. \citen{Tucker}, 
$c_q$ has been written as eq. (\ref{Mattuck_c}).
\end{thebibliography}
\end{document}